\documentclass[a4paper,11pt,twoside]{ThesisStyle}

\usepackage{amsmath,amssymb}             % AMS Math
\usepackage[latin1]{inputenc}
\usepackage{epigraph}
\usepackage[T1]{fontenc}
\usepackage[left=1.5in,right=1.3in,top=1.1in,bottom=1.1in,includefoot,includehead,headheight=13.6pt]{geometry}

\usepackage[nottoc, notlof, notlot]{tocbibind}
\usepackage{minitoc}
\usepackage{aecompl}

\usepackage[intoc]{nomencl}

\makenomenclature

\usepackage{ifpdf}

\ifpdf
  \usepackage[pdftex]{graphicx}
  \DeclareGraphicsExtensions{.jpg}
  \usepackage[a4paper,pagebackref,hyperindex=true]{hyperref}
\else
  \usepackage{graphicx}
  
  \DeclareGraphicsExtensions{.ps,.eps}
  \usepackage[a4paper,dvipdfm,pagebackref,hyperindex=true]{hyperref}
\fi

\graphicspath{{.}{images/}}

\usepackage{color}
\definecolor{linkcol}{rgb}{0,0,0.4} 
\definecolor{citecol}{rgb}{0.5,0,0} 

\usepackage{rotating}                    % Sideways of figures & tables
\usepackage{fancyhdr}                    % Fancy Header and Footer

\let\headruleORIG\headrule
\renewcommand{\headrule}{\color{black} \headruleORIG}

\usepackage{colortbl}
\arrayrulecolor{black}

\fancypagestyle{plain}{
  \fancyhead{}
  \fancyfoot{}
  
}

\usepackage{algorithm}
\usepackage[noend]{algorithmic}

\makeatletter

\def\cleardoublepage{\clearpage\if@twoside \ifodd\c@page\else%
  \hbox{}%
  \thispagestyle{empty}%              % Empty header styles
  \newpage%
  \if@twocolumn\hbox{}\newpage\fi\fi\fi}

\makeatother
 
{%

\hrulefill
\vspace*{0.5cm}%
\end{minipage}
}

\let\minitocORIG\minitoc
\renewcommand{\minitoc}{\minitocORIG \vspace{1.5em}}

\usepackage{multirow}
\usepackage{slashbox}

{ \begin{list}%
	{$\bullet$}%
	{\setlength{\labelwidth}{25pt}%
	 \setlength{\leftmargin}{30pt}%
	 \setlength{\itemsep}{\parsep}}}%
{ \end{list} }

\renewcommand{\epsilon}{\varepsilon}

\begin{document}

\begin{titlepage}
 % \vspace*{\stretch{1}}
  %\begin{center}
   % \textbf{\huge Two-particle azimuthal correlations at forward rapidity in STAR}
 % \end{center}
%\vspace*{\stretch{4}}
%\newpage
%\thispagestyle{empty}
% --------------------achterkant frans titelblad

%\vspace*{1em}
%\vfill
%\noindent
%%\newline
%Cover: Ermes Braidot, Amsterdam~\copyright~2010~~~~braidot@rcf.rhic.bnl.gov

% -------------------- echte titelblad
\newpage

\thispagestyle{empty}
  \vspace*{1em}
  \begin{center}
    \textbf{\huge Two-particle azimuthal correlations at forward rapidity in STAR}

    \vspace{\stretch{1}}
    
    \thispagestyle{empty}
%  \vspace*{0.4em}
    \text{\Large Azimutale twee-deeltjescorrelaties bij grote rapiditeit in STAR}

    \vspace{\stretch{1}}

    \thispagestyle{empty}
  \vspace*{0.4em}
    \text{(met een samenvatting in het Nederlands)}

    \vspace{\stretch{5}}

    {\Large Proefschrift}

    \vspace{\stretch{1}}

    {ter verkrijging van de graad van doctor 
      aan de Universiteit Utrecht
     op gezag van de rector magnificus 
      prof.dr.~J.C.~Stoof, ingevolge het besluit van het college voor promoties 
    in het openbaar te verdedigen 
     op maandag 17 januari 2011 des middags te 12.45 uur}

    \vspace{\stretch{2}}

    door

    \vspace{\stretch{1}}

    {\Large Ermes Braidot}

    \vspace{\stretch{.5}}

    geboren op 9 november 1980 te Trieste, $\mathrm{Itali\ddot{e}}$

    \vspace*{\stretch{1.5}}
  \end{center}

%voor de pedel:
%\vspace{1cm}
%\begin{tabular}{p{8cm}l}
%      Datum: 		& Datum: \\[1ex]
%      Secretaris	& Voorzitter \\
%\end{tabular} 
%einde voor de pedel

%----------achterkant titel
  \newpage
  \thispagestyle{empty}
  \begin{flushleft}
%    \begin{large}
	\begin{tabular}{@{}ll}
      Promotoren: &  Prof.~dr.~T.~Peitzmann \\
      & Prof.~dr.~E. L. N. P. ~Laenen \\
       Co-promotor: & Dr.~A. ~Mischke \\
  	\end{tabular} \\
    %    \end{large}
  \end{flushleft}

  \vspace{\stretch{1}}

{Dit werk maakt deel uit van het onderzoekprogramma van de Stichting 
voor Fundamenteel Onderzoek der Materie (FOM), die financieel wordt 
gesteund door de Nederlandse Organisatie voor Wetenschappelijk Onderzoek (NWO).}

\newpage
\thispagestyle{empty}

\end{titlepage}

\dominitoc

\pagenumbering{roman}

\cleardoublepage

\setlength{\epigraphwidth}{.37\textwidth}
\epigraph{L\`{a}, tout n'est qu'ordre et beaut\'{e},\\Luxe, calme et volupt\'{e}.}{\textit{L'invitation au voyage}\\\textsc{Charles Baudelaire}}

\vspace*{1em}
\vfill
\noindent
%\newline
Cover: Ermes Braidot, Amsterdam \\
Bookmark: Amanda, ``Rainbow Pancakes''

\tableofcontents

\mainmatter

%%% Chapter heading commands %%%

\chapter{Introduction}
\label{chapter:introduction}

\index{Introduction}

%%% Abstract %%%

%\begin{Abstract}

%\end{Abstract}

%%% Chapter sections %%%

Quarks and gluons are the building blocks of the matter that populate our Universe. It is believed that, during the very first moments after the Big Bang, the newborn Universe was filled with a very hot and dense quark-gluon plasma (QGP). At these conditions, quarks were decoupled and behaved essentially as free particles. After a few microseconds, the Universe thermalised and quarks and gluons started to group themselves into heavier particles, in what is called hadronization. Presently, the plasma may still be present only in the very dense core of neutron stars. In ordinary energy regimes, instead, the strong nuclear force binds these elementary components into bound states so that it is not possible to isolate them. 

The interest of studying collisions between relativistic heavy ions comes from the possibility of recreating the conditions of the early Universe in laboratory. The Relativistic Heavy Ion Collider (RHIC) at Brookhaven National Laboratory (BNL) and, more recently, the Large Hadron Collider (LHC) at the Conseil Europ\'{e}en pour la Recherche Nucl\'{e}aire (CERN) facility, provide collisions between heavy nuclei with the purpose of creating the QGP and study its properties. Nuclei are accelerated up to ultra-relativistic speed, so that they appear Lorentz contracted along their longitudinal dimension, like thin ``pancakes''. When two of these pancakes collide, they mostly pass through each other, leaving behind a hot and dense plasma of interacting quarks and gluons. 

However, modification in the structure of the relativistic nucleus may happen independently of the collision. At these energies, the density of quarks that can be probed increases significantly, and this requires an all-new description for the nuclear structure. This can be achieved by using a framework of prescriptions and equations, known as the Color Glass Condensate (CGC). In this approach, the density of gluons in the nucleus is \emph{saturated} and strong collective behavior characterize its components, leading to a series of new phenomena. While the Quark-Gluon Plasma is the product of the collision of relativistic pancakes of nuclear matter, the Color Glass Condensate describes the pancakes themselves. A good understanding of the structure of the projectiles is therefore crucial for a correct description of the final state effects that characterize heavy ion interactions, especially with an eye on LHC, where saturation may be relevant for the proton structure as well.

\bigskip

\index{Introduction}

%%% Chapter heading commands %%%

\chapter{Theory: scattering off a glass}
\label{chapter:theoretical_background}

\index{Theory: scattering off a glass}

%%% Abstract %%%

%\begin{Abstract}

%\end{Abstract}

%%% Chapter sections %%%

\section{Deep inelastic scattering}
The cleanest way of studying hadronic and subnuclear matter is by probing it with structureless projectiles like an electron or a muon. The interaction between the lepton and a constituent of the proton (or, analogously, of the nucleons within a nucleus) goes through the exchange of a virtual photon. The hadronic structure is probed with a spatial resolution which depends on the momentum $Q$ transferred from the lepton to the hadron by the virtual photon, and can be quantified as the De Broglie wavelength of the photon $\lambda\sim 1/Q$. When $Q$ is large enough the partonic structure becomes manifest and the hadron itself breaks into some new hadronic final state. When this happens we talk about \emph{deep inelastic scattering} (DIS).

\begin{figure}[h]
\includegraphics[width=0.5\textwidth]{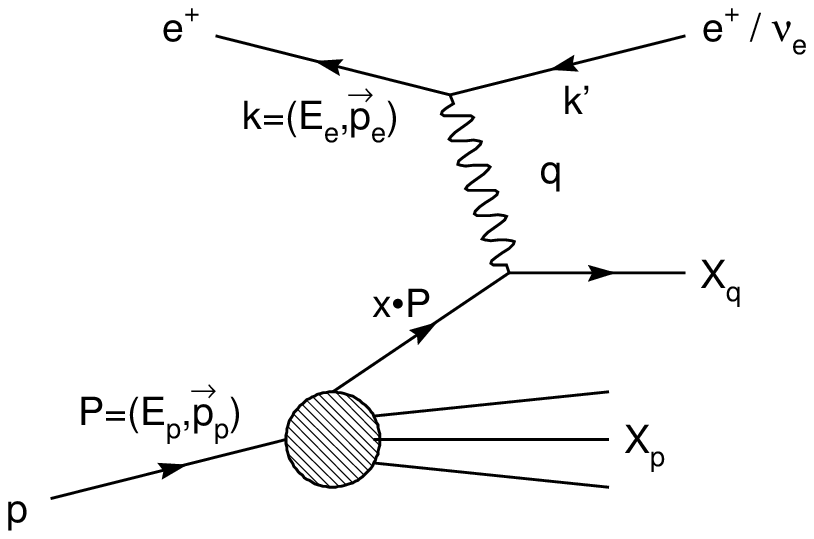} 
\includegraphics[width=0.5\textwidth]{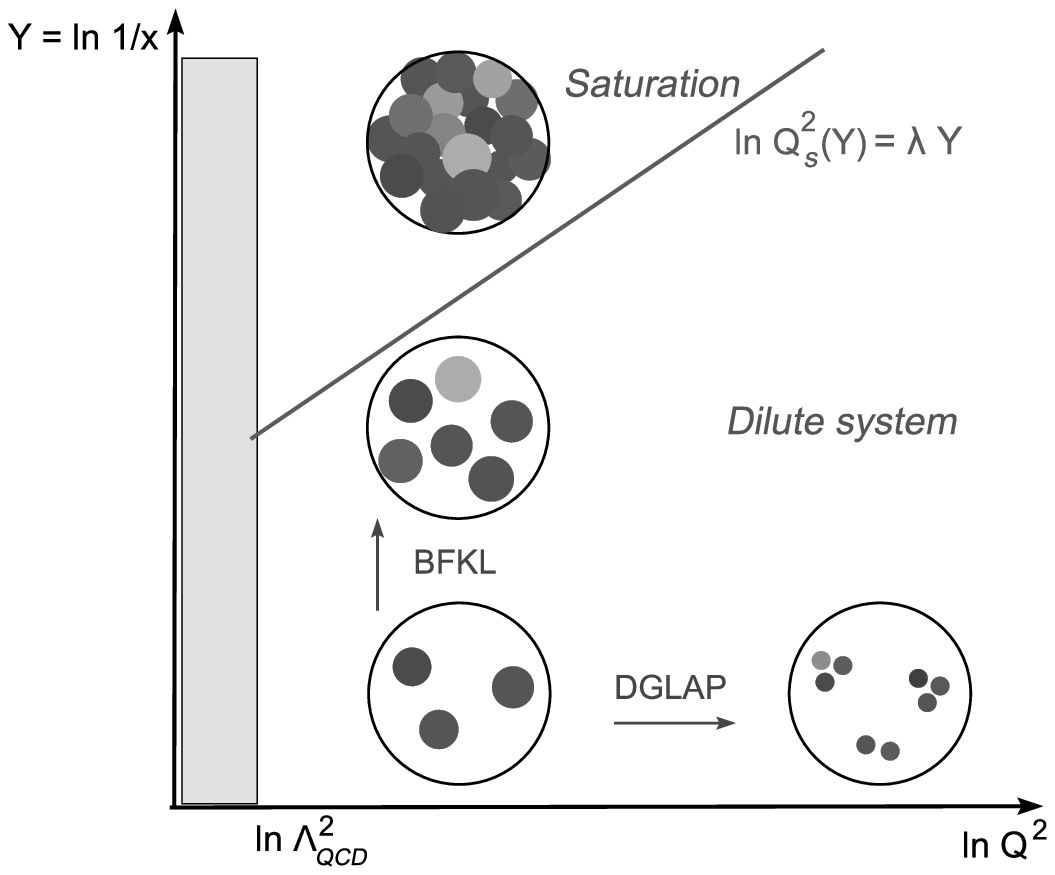}
\caption{Left: lowest order description of Deep Inelastic Scattering. Right: pictorial representation of parton density function evolution in $Q^2$ (DGLAP) and $x$ (BFKL), taken from \cite{2010arXiv10020333G}. Circles represent the resolved partons whose transverse size $1/Q^2$ decreases with $Q^2$ while density increases with $1/x$.}\label{dis}
\end{figure}

DIS cross sections can be factorized into a leptonic part $L_{\mu\nu}$, describing the radiation of the virtual photon by the electron, and a hadronic part $W^{\mu\nu}$, describing the interaction between the photon and the hadron. The structure of the hadron, as seen by the virtual photon, is parametrized by \emph{structure functions}  $F_{i}(x_B,Q^2)$ that depend on the transferred momentum $Q^2=-q^2$ and the Bjorken scaling variable $x_B=Q^2/(2p\cdot q)$. 

We can understand different properties of hadrons and nuclei by choosing a specific frame of reference and an appropriate gauge for DIS interactions: the \emph{Bjorken frame} has been fundamental for understanding the partonic nature of hadrons; on the other hand in the \emph{dipole frame} unitarity constraints become manifest.

\subsection{Bjorken frame}
\label{bjframe}

In the frame in which a proton is moving very fast along the $z$-axis (\emph{infinite-momentum frame}) the deep inelastic scattering variable $x_B$ assumes a finite value. In this so-called Bjorken limit (fixed $x_B$ and $Q^2 \to\infty$) the hadronic structure functions scale, i.e. they become independent of $Q^2$: $F_i(x_B,Q^2)\to F_i(x_B)$. Scaling turns out to be a result of the partonic structure of the hadron (structure functions depending on $x_B$-only can be recovered by an elementary quark-to-lepton treatment of the hadronic scattering). One of its consequences, the Callan-Gross relation for the structure functions $F_2(x_B)=2x_B F_1(x_B)$, relates to the fermionic nature of the quarks (the relation is theoretically recovered for particles with spin $s=1/2$).

In this Bjorken frame the proton is Lorentz contracted along its longitudinal direction. It is composed of pointlike fermions (essentially free, due to the vanishing of the strong coupling constant $\alpha_S(Q^2)$ in the limit $Q^2 \to \infty$) with a finite longitudinal momentum and a small transverse size. The scattering happens between the fermion and one of these free partons through the exchange of a virtual photon which is mainly transverse in the infinite-momentum limit. The photon is therefore a good probe of the partonic structure of the hadron, since its spacial (transverse) resolution $1/Q^2$ becomes very small in the limit $Q^2\to\infty$.

In the parton picture that follows, the DIS cross section is given by a convolution of the cross section of the elementary electron-quark scattering with the probability to find such a quark in the proton. The hadronic structure functions can be expressed through the \emph{parton distribution functions} (pdf's) $q_i(x)$ that represent the probability to find a parton of species $i$ with longitudinal momentum $p_i=xp$ within the hadron. The variable $x$ gives the fraction of the hadronic longitudinal momentum carried by the parton, i.e. the fraction of momentum at which the hadron is probed, and it is given by the Bjorken variable $x_B$. It is related to the center-of-mass energy $s$ ($x_B\sim Q^2/s$ when $s\gg Q^2$) so that high energy DIS means low $x_B$.

This is of course an approximation: quarks are not free particles even if they behave so in high $Q^2$ interactions with virtual photons; Bjorken scaling and Callan-Gross relation are in fact exact only in a leading order approximation in $\alpha_S$. In a perturbative QCD treatment of DIS that goes beyond the \emph{na\"ive} parton model, logarithmic scaling violations appear, driven by the possibility of a quark radiating a gluon, thus acquiring transverse momentum. Similarly, the pdf's acquire a logarithmic $Q^2$ dependence through higher order corrections in $\alpha_S(Q^2)$. Their $Q^2$ evolution is expressed through a set of equations (DGLAP: Dokshitzer-Gribov-Lipatov-Altarelli-Parisi equations) which include the probabilities of gluon emission. The DGLAP equations work well in the limit $Q^2\to\infty$ where dominant contributions to the perturbative radiation of gluons comes from the region in phase space where the transverse momenta of the gluons are strongly ordered: $Q^2\ge k^{2}_{1\perp}\gg k^{2}_{2\perp}\gg ...$ As $Q^2$ grows, the spatial resolution $1/Q^2$ with which the hadron is probed reduces; we become then sensitive to an increasing number of gluons with decreasing transverse momentum. But pdf's depend also on the longitudinal momentum fraction $x$. In regimes where $x$ is small but the transverse momentum transferred is still large ($Q^2\to\infty$) contributions of  $\mathrm{\ln{(1/x)}}$ cannot be neglected anymore. The behaviour of the distributions is here recovered by applying a double leading logarithmic approximation (DLLA) in terms of $\mathrm{\ln{(1/x)}}$ and $\mathrm{\ln{(Q^2)}}$ to the DGLAP equations.

 As $x$ becomes even smaller the DGLAP equations lose their applicability ($\mathrm{\ln{(1/x)}}$ contributions become larger than $\mathrm{\ln{(Q^2)}}$ ones) so one needs to consider evolution driven by the radiation of gluons strongly ordered in $x$. This is achieved by the Balitsky-Fadin-Kuraev-Lipatov (BFKL) equation. In the small-$x$ region the BFKL equation predicts a steep growth of the gluon density that leads eventually to an inconsistent infinite number of gluons (small-$x$ problem). At the same time the transverse size $1/Q^2$ of the probed partons stays constant within BFKL evolution. This means that at a high enough energies and densities (i.e. low enough $x$) gluons start to overlap. When this happens, non-linear effects have to be taken into account while computing the parton distribution function. Such effects are expected to tame the otherwise divergent growth of the gluon density. This goes under the name of \emph{saturation} of the gluon density.
 
One first attempt of including non-linear effects into a evolution equation of partons in a dense proton or nucleus has been made by Gribov, Levin and Ryskin in 1983 \cite{GLR}. Their GLR equation includes contributions from gluon recombination, which at these high densities are not negligible anymore. This happens when the gluon \emph{occupation number} $A\cdot xg(x,Q^2)/Q^2$, i.e the total transverse area they occupy (number of gluons in a nucleus A, times their transverse size $1/Q^2$), becomes of the order of the nuclear transverse area $S_{\perp}=A^{2/3}S_{0\perp}$. Since the probability of gluon interaction is proportional to $\alpha_S$, we can estimate the scale at which saturation will occur by defining the \emph{saturation scale} $Q_S^2(x)$ which indicates, at a given $x$, the (inverse of the) transverse size gluons need to have in order to feel each other:
\begin{equation}\label{Qscale}
Q^2_S(x)\simeq\alpha_S(Q_S)A^{1/3}\frac{xg(x,Q_S^2)}{S_{0\perp}}
\end{equation}
This definition leads to a straight line division between saturated and dilute matter in a $\mathrm{\ln{Q^2(x)}}$-$\mathrm{\ln{(x)}}$ plane, as depicted on the right panel of Figure \ref{dis}.

Saturation of the gluon density is expected also from a more general argument: the unitarity of the hadronic scattering amplitude. The total cross section of a hadronic scattering process is in fact expected to be bounded by the logarithm squared of the center of mass energy $s$, following what is know as the Froissart bound \cite{PhysRev.123.1053}: $\sigma_{tot}(s)\le\sigma_0 \ln^2(s)$. The rise of the gluon density, without non-linear correction, would lead eventually to a violation of the unitarity bound for the scattering amplitude and thus of the Froissart bound.

\subsection{Dipole frame}
\label{dipo}

Although the Bjorken frame is essential to give a description of the partonic structure of the hadron and it provides a first glimpse on saturation, it is also useful to consider the deep inelastic scattering process from a different point of view. In DIS the virtual photon scatters off a quark from the hadron. At high energies the quark is likely not a valence, but rather a sea quark emitted by a small-$x$ gluon. It is therefore convenient to disentangle the final quark emission from the rest of the partonic evolution in the hadron. In order to do so, we can perform a Lorentz boost on the system and strip the $\gamma^{\star}q\bar{q}$ vertex of the elementary interaction out of the hadron. As in the Bjorken frame, most of the energy is still carried by the hadron, but now the virtual photon has a longitudinal component and enough energy to split into a quark-antiquark pair (\emph{color dipole}) long before interacting with the hadron.

This picture was originally developed by Mueller \cite{Mueller1994373, Mueller1994gb} to describe the interactions between highly energetic color dipoles. It is more appropriated for high energy DIS, since in this regime the lifetime of the quark-antiquark pair is much larger than the interaction timescale. It is therefore possible to factorize the cross section of the scattering process into  the probability of the photon to fluctuate into a quark-antiquark pair and the dipole-hadron cross section $\sigma_{dipole} (r_\perp,x)$ of a dipole of transverse size $r_\perp$ that scatters off a small-$x$ gluon field. The former can be computed perturbatively, while the latter, which contains all the strong interaction physics, is modeled in a semiclassical way. Since $x$ is small, the gluon density in the hadron is large and quantum effects can be neglected. We can then describe a state with high occupation numbers as a classical gluon field. 
In addition, it is more convenient to describe the gluon field of the hadron through the \emph{unintegrated} (in the transverse space) \emph{gluon distribution} $\varphi_x(k_\perp)$ rather than the usual pdf, since in this frame the partonic description itself starts losing its meaning.
The dipole-hadron cross section $\sigma_{dipole} (r_\perp,x)=2\int d^2 b_\perp \mathcal{N}_x(r_\perp,b_\perp)$ can be obtained by integrating the forward scattering amplitude $\mathcal{N}_x(r_\perp,b_\perp)$ over the impact parameter $b_\perp$.

Since in the \emph{dipole frame} the gluon density is contained in the scattering amplitude, one considers the evolution in $x$ of the entire amplitude $\mathcal{N}_x(r_\perp,b_\perp)$. This leads to Mueller's form of the BFKL equation which is equivalent to the original BFKL equation but involving  $\mathcal{N}_x(r_\perp,b_\perp)$. It is interesting to note that this new version of the BFKL equation shows the same inconsistent behaviour at low $x$ seen in the Bjorken frame. At high energy, in fact,  the solutions for the scattering amplitude violates the unitarity bound $|\mathcal{N}_x(r_\perp,b_\perp)|\le1$ and more generally the Froissart bound $\sigma_{dipole}(s)\le\sigma_0 \ln^2(s)$ for the total dipole cross section.

As for the Bjorken frame, in order to correct this, non-linear contributions in the form of gluon recombination need to be added to the evolution equation. In the dipole picture, however, one can also  imagine to boost the system even further in order to accelerate the dipole, and study the evolution of the dipole itself. The color dipole will then have enough energy to radiate a gluon which can be described again as a (new) color dipole and eventually interact with the gluon field. When the energy is high enough one has to include in the evolution of the scattering amplitude also the probability of simultaneous interactions of two (or more) dipoles off the hadronic gluon field. These effects are encoded in a non-linear evolution equation, the Balitsky-Kovchegov (BK) equation \cite{Balitsky199699,PhysRevD.60.034008,PhysRevD.61.074018}, where contributions from multiple dipole-hadron scatterings damp the rise of the total amplitude at low $x$. Hence, in the ``boosted'' dipole picture, saturation of the gluon density translates into saturation of the scattering amplitude\footnote{This is true for a dipole frame where the perturbative evolution lies fully in the probe wave-function. In a more general (``less boosted'') dipole picture, non-linear effects are a mix of gluon recombination and multiple scattering effects.}.  
It is interesting to note that while in the Bjorken frame perturbative evolution (DGLAP, BFKL, GLR) is put entirely in the wave-function of the hadron and saturation arises from recombination of its components, in this ``boosted'' dipole frame it is the dipole (the probe) that evolves by emitting softer gluons. In this frame gluon recombination is seen from the opposite point of view as a splitting of the dipole into two dipoles interacting simultaneously with the target. Parton recombination and multiple (simultaneous) scattering are properties of the wave-functions of hadron and probe, respectively. Therefore they are both frame dependent description of the phenomenon saturation. Unitarity instead is a property of the scattering and it is therefore frame independent. In this context, the dipole picture provides a better frame to work with since it  shows how unitarity is restored using the non-linear BK equation. 
  
The BK equation is a useful tool to describe saturation effects in low-$x$ hadronic interactions. Unfortunately, there is no analytical solution to it and one needs to solve it numerically or by applying phenomenological models able to reproduce the scattering amplitude in different regimes.    
For DIS data at the HERA facility, a number of models for the dipole scattering amplitude have been proposed, mainly based on the Glauber model. The Golec-Biernat-W\"{u}sthoff (GBW) \cite{GBW} and the Iancu-Itakura-Munier (IIM) \cite{IIM} models both show good results in fitting HERA data, even if they apply to different (and limited) small-$x$ regions. In order to preserve unitarity, they need to follow two main conditions. Dipoles with low transverse resolution $1/Q$ (or ``small'' dipoles) are supposed to interact weakly with the system ($\mathcal{N}(r)\ll 1$), as expected in interactions with diluted systems (this goes under the name of \emph{color transparency}). On the contrary, dipoles with high transverse resolution (or ``large'' dipoles) need to be strongly absorbed ($\mathcal{N}(r)\approx 1$) in order to restore unitarity (\emph{blackening} of the cross section). Remarkably, the distinction between the two regimes is given by the same saturation scale $Q_{S}$ defined before (Eq.\ref{Qscale}). This means that the unitarity limit in the dipole frame corresponds to gluon saturation in the hadronic wave-function in the Bjorken frame.

\section{Hadronic interactions}

We have seen that saturation is a necessary feature to be included in order to describe interactions that involve a dense medium. In the previous chapter the concept of saturation has been applied to deep inelastic scatterings of an elementary probe off a dense hadron at very small $x$. The same idea can be implemented for hadronic interactions (interactions involving a hadronic probe) as well. In this case the hadronic projectile needs a higher energy in order to probe $x$ as low as in DIS and to reach the saturation regime. However, from Eq.\ref{Qscale}, we know that the saturation scale (i.e. the scale where saturations effects start becoming important) grows with the atomic number $A$ of the nuclear target. This means that hadronic interactions between a proton and a very large nucleus (say: a gold nucleus with atomic number $A=197$) should also provide the sufficient conditions to look for saturation effects. Moreover it is possible to select kinematical regions that privilege the interaction with lower $x$ gluons in the target, such as forward hadron production. 

High energy interactions between two colliding hadrons (protons or nucleons) can be described  via elementary QCD interaction by using the partonic description. When the energy is high enough or, more precisely, when the transfered momentum $Q^2$ is large enough ($Q\rightarrow\infty$ at fixed momentum fractions $x$), short and long range effects can be disentangled. In this way, the cross section can be \emph{factorized} into a perturbative component, representing the hard scattering between the constituents (quark and gluons) of the hadrons, and a non-perturbative quantity: the probability of finding such components within the hadrons. Hard interactions are selected by the scale variable $Q^2$ which discriminates between partons with large transverse momentum, which contribute to the hard scattering, and soft partons, which are instead absorbed in the parton distribution. These parton distribution functions (pdf's) are the same used for DIS. The cross section for a hard scattering process initiated by two hadrons can be then written as the convolution of the elementary cross section between two partons with the probability (encoded in the pdf's) to find such partons in the hadrons.
As in DIS, the total cross section is given by the incoherent sum of all these partonic contributions.
Examples of hadronic interactions underlying elementary hard scattering can be found in the production of two high-$p_T$ jets or in Drell-Yan lepto-production. 

The success of perturbative QCD and factorization theorems is however limited to the description of phenomena that present high $p_T$ particles production in the final state. More precisely, factorization theorems work well only if partons can be described as independent. As we saw for DIS, as the energy grows, interactions between the elementary components need to be taken into account. This eventually leads to saturation of the gluon density and to a description of the nucleus where quarks and gluons present a coherent, and more or less collective, behaviour. This clearly breaks the basis of consistency for factorization theorems.  Processes such as forward di-hadron production in p+p or p+A involve low-$x$ gluons in the wave-function of the target and are characterized by low transverse momentum of the produced particles. These processes access indeed the region where incoherence is not assured, and factorization theorems do not hold anymore. 

A possible solution is indicated by the comparison with DIS. 
In the dipole picture of the scattering, in fact, non linearities arise from contributions of coherent multiple gluon exchange between the probe and the target medium. In terms of the factorization approach, in p+p interactions this contribution is suppressed in comparison to the leading hard partonic process. However, when the target is a large relativistic nucleus, multiple scatterings are enhanced by the dense parton density of the nucleus. In high energy p+A interactions the longitudinal resolution of the probe from the proton can become larger than the size of the nucleons within the target. This makes multiple scattering between the probe and the component from different nucleons not negligible. This is clearly a non-linear effect and needs to be taken into account. It has been shown \cite{qiu-2006-632} that multiple interaction contributions can be added perturbatively and included in the cross section. This allows the extension of factorization approaches to relatively small transverse momentum regions, bridging the gap between a model with independent partons and the possible onset of gluon saturation. Nonetheless this is still an attempt of using perturbative QCD in a region where its validity is not completely assured. For a full description of non-linearities in the nuclear wave-function and in the scattering interaction, one will need to overcome the perturbative approach and try to solve non-linear contributions. The most famous attempt of doing this is know as the \emph{Color Glass Condensate} (CGC) model which is summarized in the next session.

\section{Color Glass Condensate}

In this section we will first describe a model (the \emph{McLerran-Venugopalan model}) for the different partonic components of a relativistic large nucleus in its infinite momentum frame. This will lead to a more general picture where the nucleus is described as a very dense system of color charges characterized by strong collective behaviour. We will see how non-linear effects can be naturally included in the evolution of the dipole-nucleus scattering amplitude by using the Color Glass Condensate framework. It is interesting to notice that this model, which was introduced in order to describe hadronic interactions, can also be applied to deep inelastic scatterings.
The fact that the same saturation model may describe DIS as well as hadronic interactions goes under the name of \emph{universality} of high energy scattering.

\subsection{The McLerran-Venugopalan model}
Let us consider a large nucleus in the Bjorken frame. In this frame the nucleus is moving relativistically with momentum $P\rightarrow\infty$ along the $z$ direction and its partonic structure is manifest. Fast partons, that are the hadron constituents  (like the valence quarks) who carry a large fraction of the momentum, move almost as free particles and act as sources for the sea of soft (i.e. slow) partons. The ``valence'' partons are Lorentz contracted to a distance $\sim 2R_{A}/\gamma = 2R_{A}m_{n}/P$, where $m_{n}$ is the mass of the nucleon. The cloud of slow (or ``wee'') partons with low momentum fraction ($x\ll 1$) is instead delocalized over larger distances. For momentum fractions small enough ($x\ll A^{-1/3}$) the longitudinal resolution $\lambda =h/(xP)$ of the slow partons becomes larger than the nuclear diameter and they are not able to resolve the longitudinal distribution of  ``valence'' partons anymore. From their point of view, fast partons are seen as a thin sheet of color charges. The same kinematical distinction can be applied to parton lifetime. Due to a different time dilation, the lifetime of slow partons is much shorter than that of the fast partons. Compared to slow parton timescale the fast ones appear to live forever. They are seen as \emph{static} (thus recoilless) \emph{sources of color charge}.    
This kinematic distinction is at the base of the \emph{McLerran-Venugopalan} (MV) \emph{model} for the structure of the nucleus. It is useful for what follows to consider this distinction between fast and slow partons to be sharp and to introduce a cutoff momentum $\Lambda$, of the order of the typical longitudinal momentum of the valence 
quarks, to distinguish fast and slow modes.
Since the dynamics of the two have very different time scales, we are entitled to model them separately. We will first derive a density distribution for the sources and then a field theory for the low-$x$ gluons emitted. 

We can describe the color distribution of the fast partons by considering an external probe (for example, a low-$x$ parton) traveling through the nucleus, and counting the color sources the probe is locally sensitive to. The MV model refers to nuclei so large to be considered nearly infinite in the transverse dimension.  If the parton density is high enough, we can assume partons to be uniformly distributed on the transverse space\footnote{However, this is not a strict requirement: the model can be generalized to finite size nuclei and extended including realistic nuclear density profiles.}. The transverse resolution $1/Q$ is given by the momentum transferred $Q^2$ in the process. At high energies, $Q^2\gg\Lambda_{QCD}^{2}$ and the probe can resolve transverse distances smaller than the nucleon size $\sim 1/\Lambda_{QCD}$. As a consequence, the probe sees ``valence'' quarks through the nucleon structure as sources of color charge. Moreover, if the density $n\equiv N_{C}A/\pi R^{2}_{A}\simeq\Lambda_{QCD}^{2}A^{1/3}$ of valence quarks is large, the number of color charges resolved per transverse area $n/Q^2$ is large (this translates into the condition $Q^{2}\ll\Lambda_{QCD}^{2}A^{1/3}$). On the contrary, along the longitudinal direction the resolution of the probe is too low to disentangle the partons. What the probe sees, instead, is a distribution of charges from different nucleons, thus uncorrelated. We can think of them as random sources of color charge and describe the charge distribution through a weight function $W_A\lbrack\rho\rbrack$ which is locally Gaussian with respect to the density $\rho$. 
In other words: in the MV model the color density is taken to be a stochastic random variable with a Gaussian distribution. The subscript $A$ in $W_A\lbrack\rho\rbrack$ refers to the cutoff momentum $\Lambda$ we introduced to distinguish between slow and fast partons. It reminds us that the prescriptions of the MV model are valid within some restricted kinematic ranges of momentum fraction $x\ll A^{-1/3}$ and transverse resolution $\Lambda^{2}_{QCD}\ll Q^{2}\ll\Lambda_{QCD}^{2}A^{1/3}$. 

Once the color density distribution at given time and $x$ is known, we can compute the gluon field radiated. The MV model was formulated as a means to describe the low $x$ component of the wave-function of nucleons within large nuclei. When the parton density is very large (i.e. in very large nuclei or at very low $x$ values) the coupling constant for strong interactions is in fact weak. It is then possible to use weak coupling methods to compute the gluon field at low $x$. To do so we can use an analog from QED, where the coupling constant is always weak. In QED it is possible to describe the soft photon dressing (mainly photons from bremsstrahlung) of an ultrarelativistic electron by boosting the classical Coulomb field radiated by an electric charge to a frame where the charge moves along the lightcone. This goes under the name of \emph{Weizs\"{a}cker-Williams} (WW) \emph{field}. In the same way, the MV model uses the WW prescription in a QCD system, namely by evolving the field of a color charge moving relativistically. It is possible to use here the classical Yang-Mills equations to compute the color field, since the system's high occupation numbers make quantum effects negligible. In the approximation of weak coupling, the gluons radiated are soft, and they leave the momentum of the valence parton (the fast color charge) practically unchanged. We recover here the picture described above: fast partons are depicted as recoilless sources for slow gluons, that can be described through classical fields.     

\subsection{An effective field theory}

The McLerran-Venugopalan (MV) model leads to a picture in which the relativistic nucleus is described by a \emph{semi-classical} field theory instead of by parton distribution functions. Under high density conditions we can use the weak coupling approximation and treat slow partons as a classic Yang-Mills field. The source for this field is represented by the fast partons, whose \emph{effective} distribution at a fixed $x=\Lambda/p$ is obtained from the weight function $W_{\Lambda}\lbrack\rho\rbrack$, which in the MV model is Gaussian in $\rho$. In order to compute physical observables (such as the gluon density function) the first step is to solve the classical Yang-Mills equation of motion for a given color configuration. As mentioned before this can be done using as a guideline the Weizs\"{a}cker-Williams (WW) approach for the Coulomb field radiated by a relativistic charge. The color field emitted by the fast partons forms the non-abelian counterpart of the WW field. It is important to stress here that, although the coupling constant is weak, the corresponding classical fields for gluons are very strong, precisely because the occupation numbers are large; the saturation regime remains then highly non-perturbative. This means that a perturbative expansion of the solution for the Yang-Mills equations is not sufficient. Instead one needs to solve the equation of motion \emph{exactly}, i.e. including interactions between radiated gluons which are obviously not considered in the QED analog. On the other hand it is the classical context which makes exact calculations possible. Once the solution is known as a function of the color density $\rho$ of the source, one needs to average it over all possible distribution of sources, weighting them with $W_{\Lambda}\lbrack\rho\rbrack$. 
This procedure of averaging over different configuration of the distribution $\rho$ is that of a \emph{Color Glass} as it resembles the approach used in the context of ``spin-glasses'' \cite{PhysRevLett.43.744}. For this reason, the above description of a relativistic nucleus is called \emph{Color Glass Condensate} (CGC). It is in fact a theory for gluons, which are ``colored''. It describes a disordered system which evolves very slowly and whose internal dynamics (given by the fast partons) appears frozen, in the same manner as a glass. And it is a condensate due to the high density of gluons involved, which is what allows us to use weak couplings methods. 

As suggested before, one of the major benefits of describing the high energetic nucleus in the CGC framework is that non-linearities in the hadronic wave-function (read: soft gluons recombination) are treated \emph{classically}. As we will see, a quantum evolution of the system is still needed in order to describe the evolution of the system with $x$. Even if the MV model does not include evolution in $x$, it provides a description for the gluon density that includes the saturated scale already at the tree level (leading order in the perturbative theory). It is in fact possible to express the unintegrated distribution $\varphi_{A}(k_{\perp})$ of gluons with transverse momentum $k_{\perp}\le Q$ by weighting the exact non-linear solution for the classical gluon field with the Gaussian approximation $W_{A}\lbrack\rho\rbrack$ of the weight function. This lead to an expression for $\varphi_{A}(k_{\perp})$ that presents different behaviour depending on the gluon transverse momentum $k_{\perp}$: at high momenta $k_{\perp}\gg Q_{S}(A)$ the distribution behaves the same as in the linear WW prediction, and it grows as $A^{1/3}$. Instead, when the transverse momentum is smaller than the saturation scale $k_{\perp}\ll Q_{S}(A)$, the distribution appears to saturate, showing a slower growth that goes like $\ln{( A/k_{\perp}^{2})}$. Here the saturation momentum, which separates linear and non-linear trends of the gluon distribution, grows as $A^{1/3}\ln{A}$ and, as well as all the observables computed within the MV model, it is independent of $x$. This reflects the fact that the MV model is built at a fixed separation scale, and it does not involve evolution in $x$.

The same approach can be used to obtain the forward scattering amplitude of a color dipole off the dense gluon field (CGC). While in the infinite momentum frame of the nucleus, we can use the Gaussian weight function $W_{A}\lbrack\rho\rbrack$ from the MV model to average the point-to-point interactions of the components of the dipole with the target. As already anticipated, when the dipole size is larger than the saturation scale $Q_{S}(A)$, non-linear effects (here: multiple simultaneous scattering) become significant and help to restore unitarity in the scattering amplitude. We finally note that in the MV model, as well as in the DGLAP evolution, the dipole scatters \emph{independently} off the color sources. This picture will change substantially when, in the next section, we will include quantum evolution towards lower $x$, which will have the effect of inducing correlations between the color sources.

\subsection{Quantum evolution in a glass}
The model developed so far depends on the momentum cutoff $\Lambda$, which defines the notion of small (and large) $x$ partons. 
However, when we reach lower $x$ regimes, new softer gluons become accessible. Some of the partons that were before considered slow are now acting like probes: they freeze out, i.e. they become part of the color glass and they need to be included in the effective theory. In other words, the momentum scale $\Lambda$, and so the weight function $W_{\Lambda}\lbrack\rho\rbrack$, varies with the momentum fraction $x$. The form of the classical small-$x$ fields, on the other side, does not suffer any consequences by the change of momentum cutoff. So all we need to do is to modify the weight function $W_{\Lambda}\lbrack\rho\rbrack$ to the new scale. In the Color Glass Condensate (CGC) this translates to a \emph{renormalization} of the effective color source. In the BFKL evolution, saturation arises from the competition of radiation of more and more soft gluons (linear effects that increase the gluon density) and recombination of gluons due to high occupation numbers (non-linear effects that tame the rise of the gluon density), which corresponds, in the ``boosted'' dipole frame, to multiple dipoles simultaneously scattering off the target. The CGC provides a natural framework for the description of both these effects. A linear contribution is given, for example, by the radiation of a soft gluon with momentum larger than the cutoff momentum $\Lambda$. This will still be considered fast and therefore included in the distribution of sources by renormalizing the effective color charge. At this point, non-linearities involving slow partons (with respect to the momentum cutoff) are included in the exact classical solution of the color field generated by this effective source. However, non-linearities affect fast partons as well; therefore they need to be taken into account while computing the color field and its weight function. Also, ``mixed'' non-linearities, such as propagation of a radiated ``semi-fast'' gluon through the color field (rescattering), have to be included. All this is achieved by using a  \emph{renormalization group equation} (RGE), also known as the Jalilian-Marian-Iancu-McLerran-Weigert-Leonidov-Kovner (JIMWLK) equations. This is a functional, \emph{non-linear}, evolution equation for the weight function $W_{\Lambda}\lbrack\rho\rbrack$ that, starting from a charge configuration at $\Lambda$ (for which the MV Gaussian weight function can be used), derives the effective charge distribution at a new (lower) value of the cutoff momentum $\Lambda^{\prime}$ by integrating out the quantum degrees of freedom with longitudinal momentum $\Lambda^{\prime}<k<\Lambda$, that now fall into the definition of ``fast'' partons. 

The McLerran-Venugopalan model was historically the first saturation model for the dipole forward amplitude. It was originally developed for very large nuclei, and for some particular kinematic conditions, for which the distribution of the color charges could be described as Gaussian. 
The evolution of the model at lower $x$, in general, does not preserve the Gaussian form of the weight function $W_{\Lambda}\lbrack\rho\rbrack$ so that the MV model does not incorporate any $x$ evolution (as we saw, the saturation scale is independent of $x$). Nevertheless, it represents a very good initial condition for the quantum evolution depicted before. For phenomenological predictions at low $x$ regimes we can use the fact that, remarkably, the JIMWLK equations reduces to the BK equation for the non-linear evolution of dipole amplitudes. The functional equation for $W_{\Lambda}\lbrack\rho\rbrack$ can in fact be converted into ordinary evolution equations which turn out to be equivalent to the BK equation in the weak field regime, for which approximate solutions at different $Q^2$ have been modeled: the Kharzeev- 
Kovchegov-Tuchin, KKT \cite{KKT}, and the Dumitru-Hayashigaki-Jalilian-Marian, DHJ \cite{DHJ}, models. 

\section{Phenomenology}
We have seen how the description of a dense system changes drastically when non-linear effects between gluons start to become relevant and to contribute to the hadronic wave-function. In these regimes, partons cannot be considered independent anymore and interactions with a dense medium cannot be limited to cases of incoherent hard scattering, as if quarks were free particles within a diluted system. The high occupation numbers characterizing dense systems lead to strong collective behaviour between gluons (and quarks). In this case, the description of the interaction is that of a probe scattering coherently off multiple partons from different nucleons or, following the CGC picture, off a color charge distribution which represents the effective color field of the low-x gluons the probe is sensitive to. We have seen how this is more easily achieved by identifying the probe with a color dipole and by describing its propagation through the medium as a dipole-CGC interaction. Although this picture requires phenomenological models in order to derive approximative solutions for the interaction, it provides nevertheless an economical description of a wide range of data with only a few parameters. In this section we will discuss some application of the CGC formalism that can be used to analyze and predict a series of different phenomena, from DIS involving a saturated target to hadro-production in relativistic nuclear interactions.       

\subsection{Geometric scaling}
\begin{figure}\begin{center}
\includegraphics[height=0.615\textwidth]{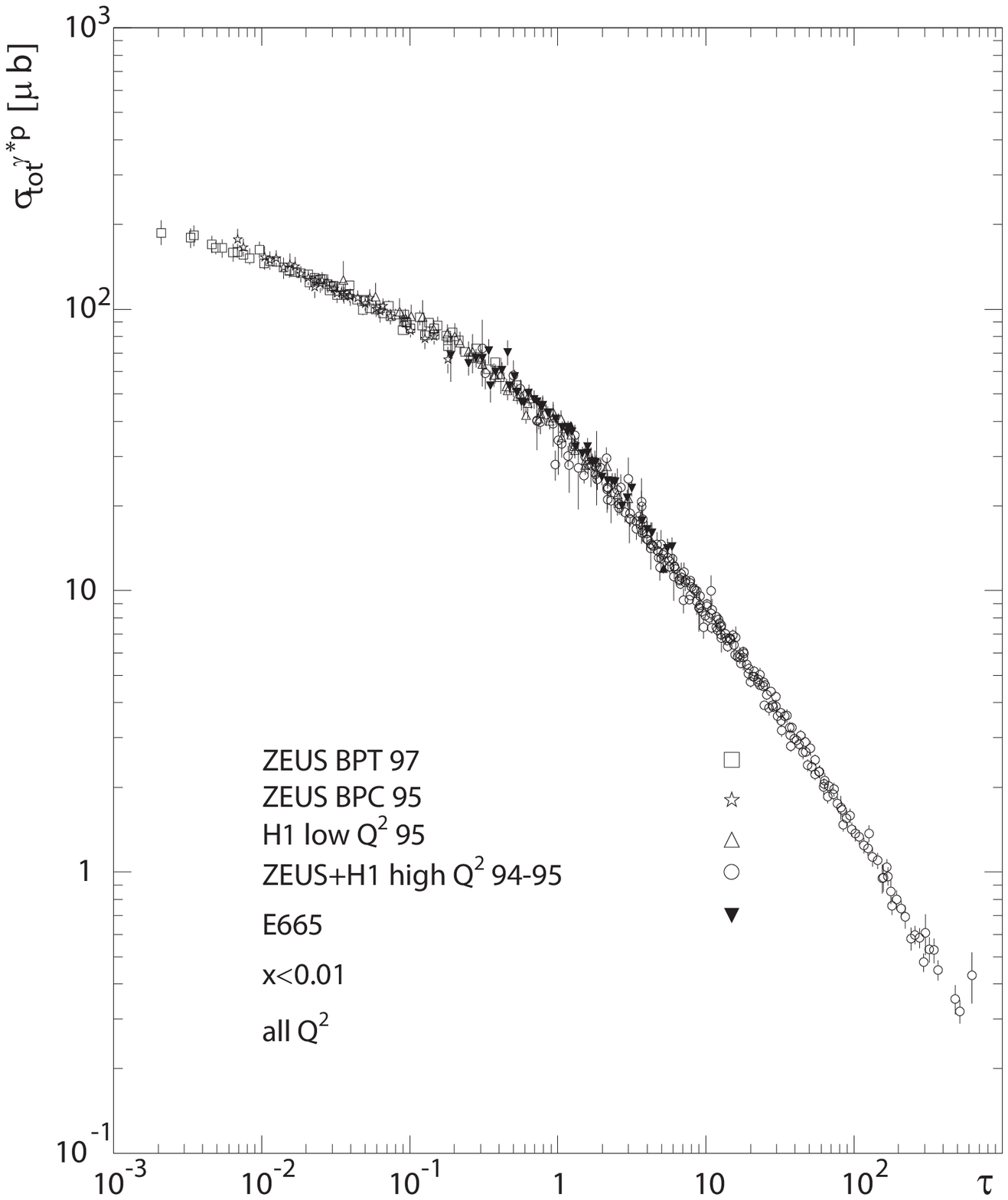} 
\includegraphics[height=0.62\textwidth]{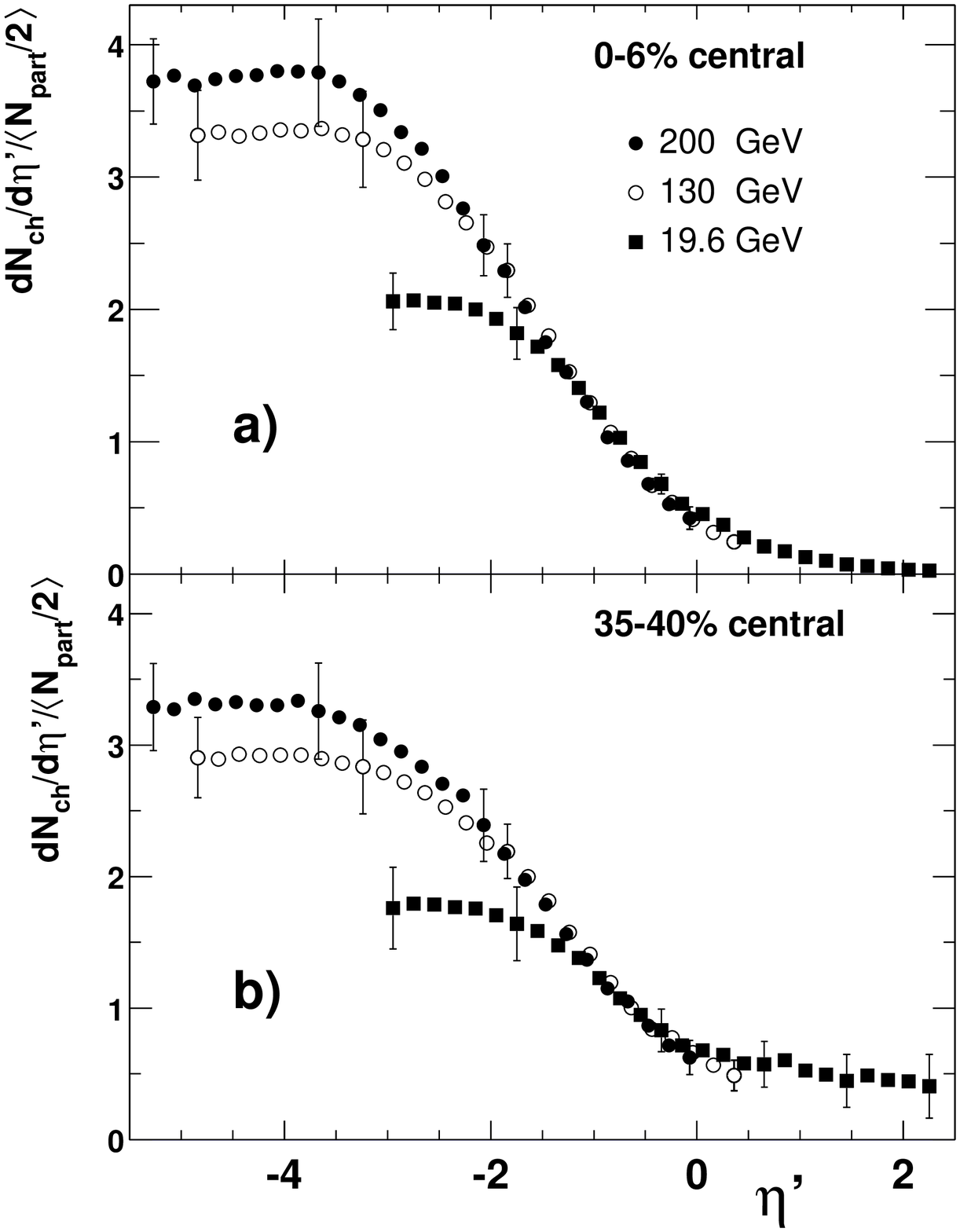}
\end{center}
\caption{Left: geometric scaling at HERA. Figure from \cite{Stasto:2000er}. Right: limiting fragmentation in Au-Au collisions at RHIC. Figure from \cite{PhysRevLett.91.052303}.}\label{proofs1}
\end{figure}

\emph{Geometric scaling} is a feature of the total cross section at small $x$, first observed at HERA. Measurements of DIS at small $x$ ($x<0.01$) showed that the inclusive virtual photon-proton cross section does not depend anymore on $Q^{2}$ and $x$ \emph{independently}, as expected for hard interactions. Instead, it \emph{scales} with the single variable $\tau=Q^{2}/Q_{s}^{2}(x)$, where $Q_{s}(x)$ turns out to be identifiable with the saturation scale (left panel of Figure \ref{proofs1}). For this reason, the property of geometric scaling is often seen as an indication of saturation. It is indeed possible to describe HERA data at low-$x$ using the Golec-Biernat-W\"{u}sthoff (GBW) \cite{GBW} model, a phenomenological model for the dipole amplitude which incorporates saturation. Within this picture, the scaling variable $\tau$ represents the ratio between the typical size of regions with strong color fields $1/Q_{s}^{2}(x)$ and the size of the dipole $1/Q^{2}$ into which the virtual photon fluctuates. It therefore provides an information on the probability of the color dipole to interact, either strongly (\emph{blackening} of the cross section) or weakly (\emph{color transparency}), following the argument illustrated in section \ref{dipo}. 

Even if geometric scaling may suggest the presence of a semi-hard dynamical scale in the proton wave-function, it is not clear if this property is indeed caused by saturation. While it is true that the GBW model is able to reproduce the $\tau$ dependence of the cross section, this can also be described by perturbative QCD, using next to leading order DGLAP evolution. Moreover, data seem to scale with $\tau$ in a region much larger than what expected from small-$x$ evolution involving saturation. For these reasons, more detailed comparison with data are necessary.  

Geometric scaling has also been observed at HERA in many other processes, such as inclusive diffraction, exclusive vector meson production ($\rho$, $J/\psi$) and deeply virtual Compton scattering. In particular, the phenomenon of hard diffraction in DIS, where the proton target remains intact and particles are produced almost exclusively in the fragmentation region, is particularly sensitive to saturation. When diffraction is described as a non-perturbative phenomenon, the diffractive cross section shows a stronger dependence of the energy than expected for the total inclusive cross section. This translates in a strong energy dependence of the ratio of the diffractive cross section and the total inclusive cross section at fixed $Q^2$. In the saturation picture, instead, diffractive cross sections are dominated by large size dipole contributions which have the effect of damping the growth of the diffractive cross section. In this scenario, the ratio between diffractive and total cross sections is expected to be nearly constant as a function of energy. Data from HERA seems to qualitatively support this last prediction \cite{PhysRevD.66.014001}. 

\subsection{Limiting fragmentation}
\emph{Limiting fragmentation} is the property of strong interactions that the rapidity distribution of particles becomes independent of the collision energy in the fragmentation region. When plotted as a function of  the rapidity gap $\eta^{\prime}\equiv\eta-y_{proj}$ between the produced particle and the projectile, particle multiplicities appear to be independent of the energy of the interaction for large values of the parameter $\eta^{\prime}$, i.e.: in the fragmentation region. Limiting fragmentation was observed in nuclear collisions at RHIC by the experiment PHOBOS \cite{PhysRevLett.91.052303} (right panel of Figure \ref{proofs1}); however it was postulated some decades ago \cite{BCY} following kinematical arguments. Saturation has been suggested as a possible explanation of this effect and calculations using the CGC framework proved to reproduce RHIC data at higher energies. The physical motivation beyond the use of saturation physics to explain limiting fragmentation becomes clear by noticing that $\eta^{\prime}\equiv\eta-y_{proj}\approx-\ln{(1/x)}$, with $x$ the longitudinal momentum fraction of the produced particle. In this case, limiting fragmentation implies that, with increasing energy, the fast (large $x$) degrees of freedom do not change much, while new modes populate the small $x$ region. This is what we expect from saturation models, where the low-$x$ component of the hadronic wave-function, selected in production of particles with very high rapidities, is increasingly populated by softer and softer gluons. One can see limiting fragmentation from another point of view: the rapidity distribution of produced particles, in the fragmentation region, becomes a function of $x$ alone and not of the energy anymore. This is similar to what we saw in Chapter \ref{bjframe} regarding the Bjorken scaling of parton distributions.

\subsection{Inclusive single particle production}
One of the most striking results achieved by the experiments at RHIC has been the characterization of the dense and hot plasma of partons created by colliding two relativistic heavy nuclei: the \emph{Quark-Gluon Plasma} (QGP). By comparing final state products in interactions between dense systems (nucleus-nucleus collisions) with final states in interactions between dilute systems (proton-proton collisions), a clear suppression of high-$p_{T}$ jets at mid-rapidity has been measured, as expected from energy loss of  particles traveling through a hot plasma. Suppression of high-$p_{T}$ jets has been considered one of the most significant indication of the creation of a dense medium in the final state of the interaction. Indeed, further measurements of \emph{inclusive particle production} in deuteron-nucleus interactions, where neither large final state effects nor QGP are expected, showed similar features as in p+p interactions, providing clear evidence that the strong high-$p_{T}$ suppression of jets at mid-rapidity is due to final state effects rather than initial state conditions. 

\begin{figure}\begin{center}
\includegraphics[width=0.99\textwidth]{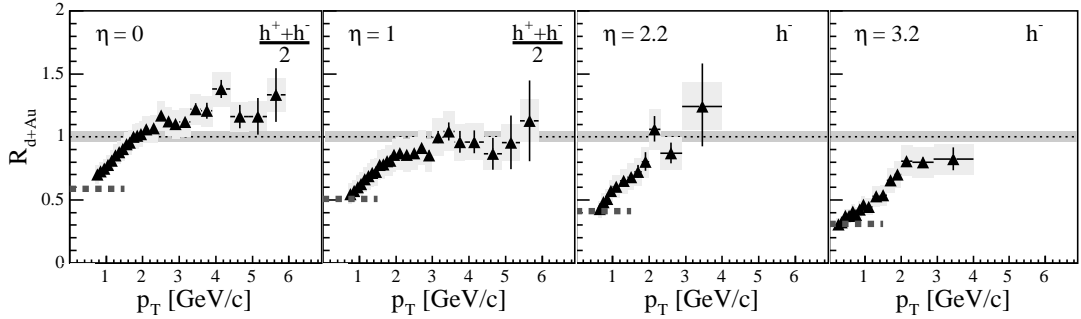} 
\end{center}
\caption{Nuclear modification factor for different pseudo-rapidities, from \cite{PhysRevLett.93.242303}.}\label{proofs2}
\end{figure}

However, a more complete set of measurements performed by the BRAHMS experiment \cite{PhysRevLett.93.242303} showed that at higher rapidities the suppression of high-$p_{T}$ jets starts to become significant in d+Au too (Figure \ref{proofs2}). Such suppression is not revealed at mid-rapidity, but it starts already at $\eta\approx1$ and it becomes more and more important as the rapidity grows. Since large final state effects are not expected in d+Au interactions, it has been proposed that this suppression is caused by modifications in the wave-function of the initial state participants. We can understand such a suppression at forward rapidity by admitting saturation effects in the initial state of the colliding nucleus. Large densities in the nucleus make multiple interactions between the probe and the dense gluon field of the target more probable, especially at higher rapidities where the lower $x$ gluons are selected. Scattering off a dense medium increases in this way the jet suppression in the final state. 

One way of quantifying such suppression is to compute the \emph{nuclear modification factor}  $R_{pA}$ (or $R_{dA}$ for deuteron-nucleus collisions). This is defined as the ratio between the cross sections of a particular process in a proton(deuteron)-nucleus interaction, and the correspondent proton-proton cross section multiplied by a factor which accounts for the different number of nucleons involved. The nuclear modification factor compares the effective nuclear composition with a crude description of the nucleus as a incoherent superposition of nucleons. This na\"{i}ve picture of the nucleus fails as soon as partons exhibit collective behaviour and the nuclear modification factor analysis shows a clear suppression of high-$p_T$ jets. 
Calculations using the CGC framework proved to be able to qualitatively predict inclusive single particle production and its transverse momentum dependence, as well as nuclear modification factors distributions in different rapidity regions \cite{PhysRevD.68.094013,Albacete:2010bs}. 

\subsection{Two particle correlations}\label{chap1:tpc}
Measurements of the inclusive single particle cross section provided the first clear indications of the presence of saturation effects. However, more recently, the study of angular correlations between two particles has become the area of most interest. \emph{Two particle correlations} are in fact more sensitive than inclusive production to distinguish between model predictions. In a perturbative QCD description of the hadronic interactions, the leading contribution to the hard parton-parton scattering comes from the $2\rightarrow2$ process which produces a pair of particles in the final state back-to-back balanced in transverse momentum. In case of scattering off a saturated medium, instead, the jet associated with the parton from the probe (in high rapidity particle production, this is usually a fast $x$ valence quark) is balanced by many low-$x$ gluons in the target. In a saturated regime, the probe scatters coherently off the nuclear strong color field, composed of gluons characterized by a strong collective behaviour. The classical $2\rightarrow2$ QCD picture is then replaced by a more complex $2\rightarrow many$ process in which the correlation between the two leading particles is partially lost. As the density in the nucleus increases, it becomes more and more difficult to detect the recoil particle, leading eventually the scattering process to be described with a \emph{mono-jet} ($2\rightarrow1$) picture.

\begin{figure}\begin{center}
\includegraphics[width=0.60\textwidth]{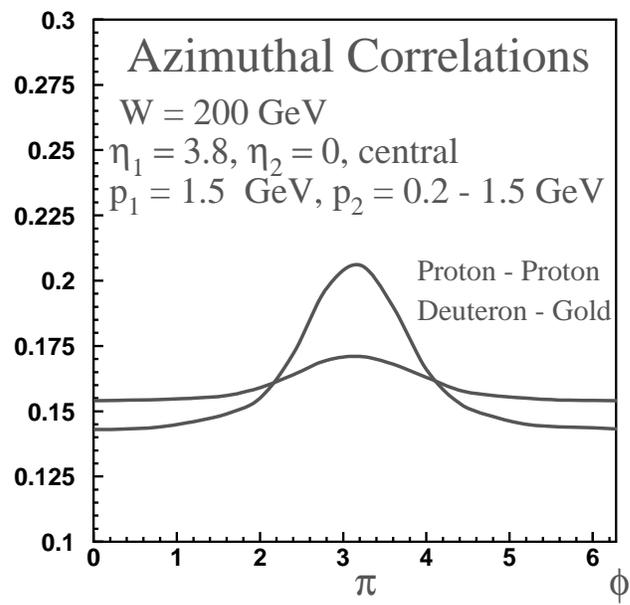}
\end{center}
\caption{Two-particle azimuthal correlations in p+p and d+Au interactions, as predicted in \cite{Kharzeev2005627}.}\label{proofs3}
\end{figure}

One can quantify the disappearance of the two particle correlation by looking at their azimuthal distribution \cite{Kharzeev2005627} (Figure \ref{proofs3}). Scattering between dilute systems will create preferentially back-to-back pairs of particles. This translates into a peak in the distribution of the difference of azimuthal angles at $180^{\circ}$. On the other hand, collisions with a saturated medium will cause this correlation to weaken, resulting into a broadening of the back-to-back peak, and eventually to its disappearance (mono-jet) when saturation sets in. One can distinguish the two pictures by comparing azimuthal correlations for p+p interactions, which (at least at RHIC energies) are described by perturbative QCD, with correlations in d+Au interactions. By selecting production of particles at high rapidity we select events were the low-$x$ component of the nuclear gluons is selected, while d+Au collisions remain free from large final state effects which affect nucleus-nucleus interactions.

This thesis will focus on the study of the \emph{two particle azimuthal correlations}. The goal will be to look for broadening or disappearance of the back-to-back peak in d+Au compared to p+p. Saturation effects will be studied as a function of rapidity and transverse momentum of the particles, and centrality of the collision. 
Interactions where the two particles are both forward probe in fact the lowest $x$ component in the nuclear gluon field, where the largest effect from saturation is expected. Lower transverse momentum particles are also more affected by saturation since the typical scale of the interaction $Q^{2}\equiv p_{T}^{2} $ is closer to the scale $Q_{S}^{2}$ where saturation sets in. Finally, since the most central and thicker part of the nucleus has the highest gluon density, central d+Au collisions are expected to reveal stronger signatures of saturation than peripheral d+Au collisions, which we expect to be more similar to p+p interactions.

\bigskip

%%% Chapter heading commands %%%

\chapter{Experiment: detector setup}
\label{chapter:Exp}

\index{Experiment: detector setup}

%%% Abstract %%%

%\begin{Abstract}
%\end{Abstract}

%%% Chapter sections %%%
\section{Relativistic Heavy Ion Collider}
The Relativistic Heavy Ion Collider (RHIC) is a multipurpose collider \cite{Hahn2003245} \cite{Ludlam2003428}  located at the Brookhaven National Laboratory (BNL) on Long Island, New York. It is a storage ring particle accelerator capable of accelerating protons, deuterium nuclei (deuterons) and heavy ions (such as Copper, Gold and Uranium ions) over a broad energy range. The main purposes of the RHIC physics program are:
\begin{itemize}
\item the study of the \emph{Quark-Gluon Plasma}, a hot and dense state of matter consisting of deconfined partons, that may have characterized the first instants of the Universe. It can be recreated at RHIC by colliding ultra-relativistic heavy ions with center-of-mass energy per nucleon pair as large as $\sqrt{s_{NN}}=200$ GeV
\item the study of the spin structure of the proton trying to solve the \emph{proton spin-puzzle} by colliding two beams of polarized protons up to $\sqrt{s}=500$ GeV
\end{itemize}
For a decade RHIC has been the most powerful heavy-ion collider in the world. Only recently, with the Large Hadron Collider (LHC) at the European Organization for Nuclear Research (CERN), it lost its primacy on this field. However RHIC will keep providing unique insights both in heavy ions physics, in an energy regime where discrimination between initial and final state effects is still clean, and in spin physics, in which RHIC will still be the highest energy accelerator for studying polarized protons collisions.   

\begin{figure}
\includegraphics[height=0.52\textwidth]{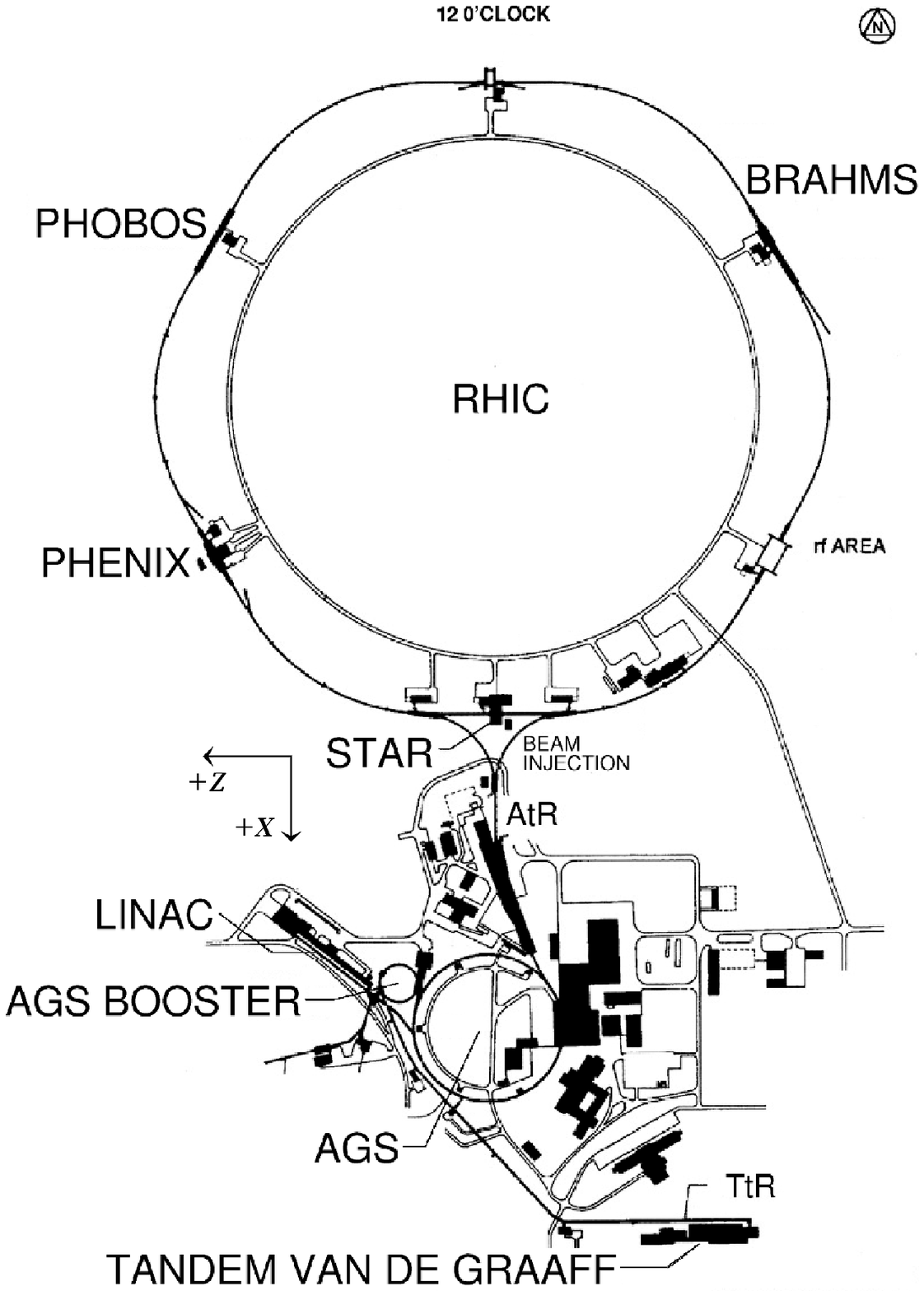} \hspace{0.04\textwidth}
\includegraphics[height=0.52\textwidth]{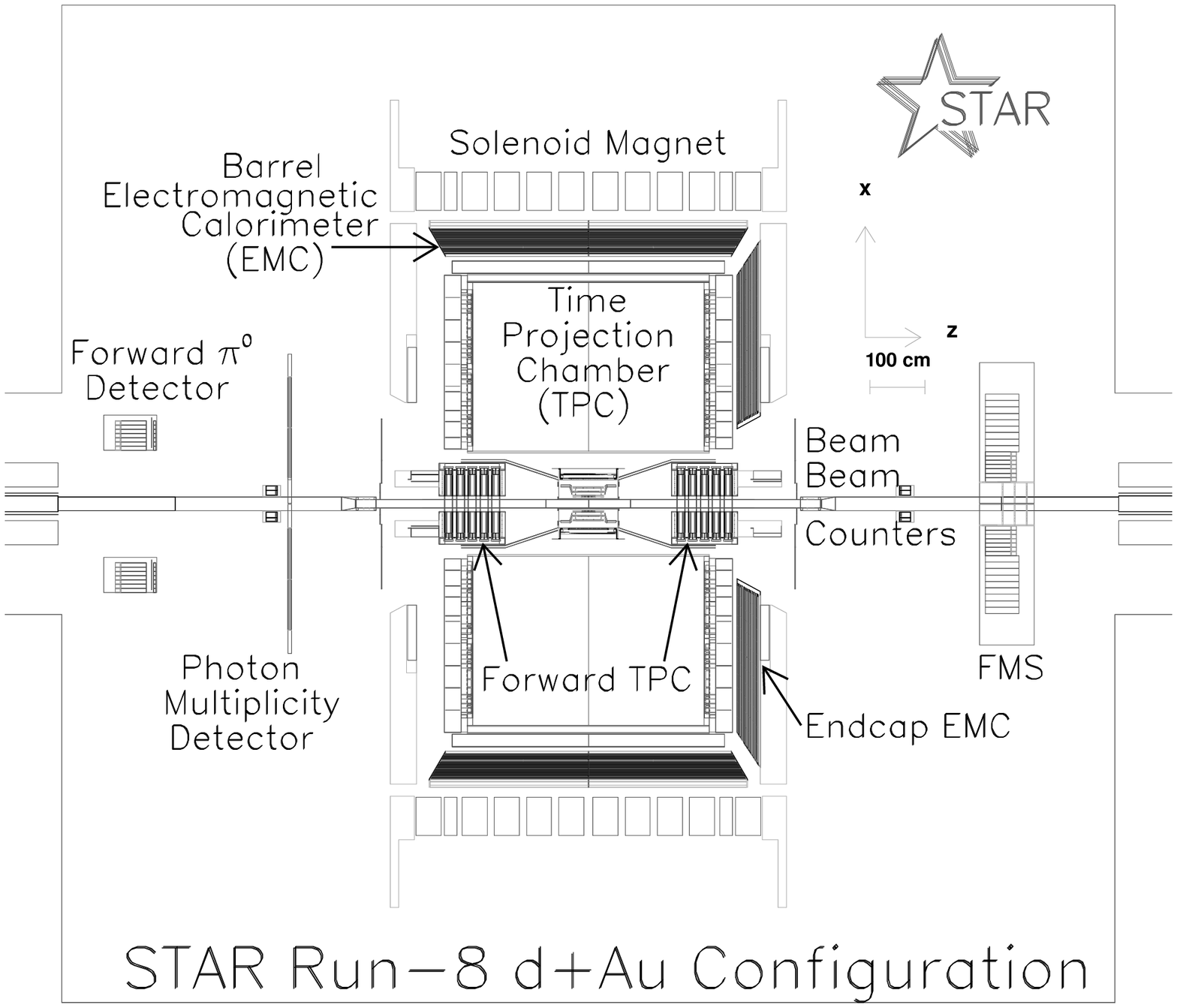}
\caption{Layout of the RHIC complex (left, figure from \cite{Hahn2003245}) and schematics of the STAR detector as used during the 2008 run (right).}\label{figrhic}\end{figure}

Picture \ref{figrhic} shows a schematics of the accelerator complex. Deuterium and heavy ions are extracted in the Pulsed Sputter Ion Source and pre-accelerated in the Tandem Van der Graaff accelerator. They pass through a series of stripping foils where they loose electrons and acquire a positive charge of +32e. The particle pulses are then injected into the Booster synchrotron where they are bunched and further accelerated up to 95 MeV/nucleon. Protons are instead pre-accelerated in the Linear Accelerator (LINAC) before being injected into the Booster. Bunches of ions are then further stripped as they reach the Alternating Gradient Synchrotron (AGS) with a charge of +79e (Gold) or +39e (Copper). In the AGS, particles are brought to an energy of 10 GeV/nucleon before being send to RHIC, where the finale stage of the acceleration takes place. Here particle beams circulate with opposite directions in two rings of 3.8 km length, where heavy ions are accelerated up to 100 GeV/nucleon and protons up to 250 GeV. Beams with a lifetime of about 10 hours cross each other at six points along the ring. The main RHIC detectors are located at four of these interaction points: STAR \cite{Ackermann2003624}, PHENIX \cite{Adcox2003469}, PHOBOS \cite{Back2002rb} and BRAHMS \cite{Adamczyk2003sq}, the latter two being de-commisioned after having fulfilled they purposes, and PP2PP \cite{Bltmann2004415}, dedicated to spin physics in pp interaction.

\section{STAR detector}
The Solenoidal Tracker At RHIC (STAR) is a detector constructed to investigate the behaviour of strongly interacting matter at high energy and density and to search for signatures of quark-gluon plasma (QGP) formation. This requires the possibility of simultaneous measurements of many different observables in order to clearly access the features of the complex dense matter so created. For this purpose, the design of the  STAR detector allows measurements of hadron production over a large solid angle, with several detectors, each specialized in detecting specific types of particles with high granularity.
The STAR tracking system is composed essentially by a large volume Time Projection Chamber (TPC) which covers the pseudo-rapidity range $|\eta|<1.0$ in the full azimuth. The TPC is also used at STAR as the main source of particle identification by measuring energy loss of ionizing particles.
Tracking at higher rapidity are achieved by two Forward TPC (FTPC) modules, located at $2.5<|\eta|<4.0$. The mid-rapidity TPC is surrounded, for trigger purposes, by the Central Trigger Barrel (CTB), a layer of scintillator tiles covering $|\eta|<1.0$ in the full azimuth. The outermost mid-rapidity detector is the Barrel ElectroMagnetic Calorimeter (BEMC). It covers the full acceptance of the TPC within the rapidity gap $|\eta|<1.0$ and it is designed to detect energy deposition from photons, electrons and electromagnetically decaying hadrons. The STAR barrel detectors are placed inside the 0.5 T magnetic field of a solenoidal magnet. The STAR configuration used for the 2008 d+Au and p+p run presented also a more forward calorimeter module, the Endcap ElectroMagnetic Calorimeter (EEMC), covering the rapidity range between $1.0<\eta<2.0$. Finally, the most forward calorimeter, newly installed at STAR, is the Forward Meson Spectrometer (FMS). It is a high granularity neutral meson spectrometer with large acceptance in pseudo-rapidity $2.5<\eta<4.0$ and in the full azimuth. 
For trigger purposes, two disks of scintillators (Beam-Beam Counter, BBC) are located at a distance of 3.7 meters from the interaction point. They provide the minimum bias trigger in p+p interactions. Trigger conditions in d+Au interactions are instead provided by two Zero Degree Calorimeter (ZDC) modules, located 18 meters far from the interaction point, into the RHIC tunnel.
In the following sections, detectors that are relevant for the present analysis are further described. Finally, we will describe in detail the main detector used for the measurements: the Forward Meson Spectrometer.

\subsection[Time Projection Chamber (TPC)]{Time Projection Chamber}

\begin{figure}[h!]
\begin{center}
\includegraphics[width=0.8\textwidth]{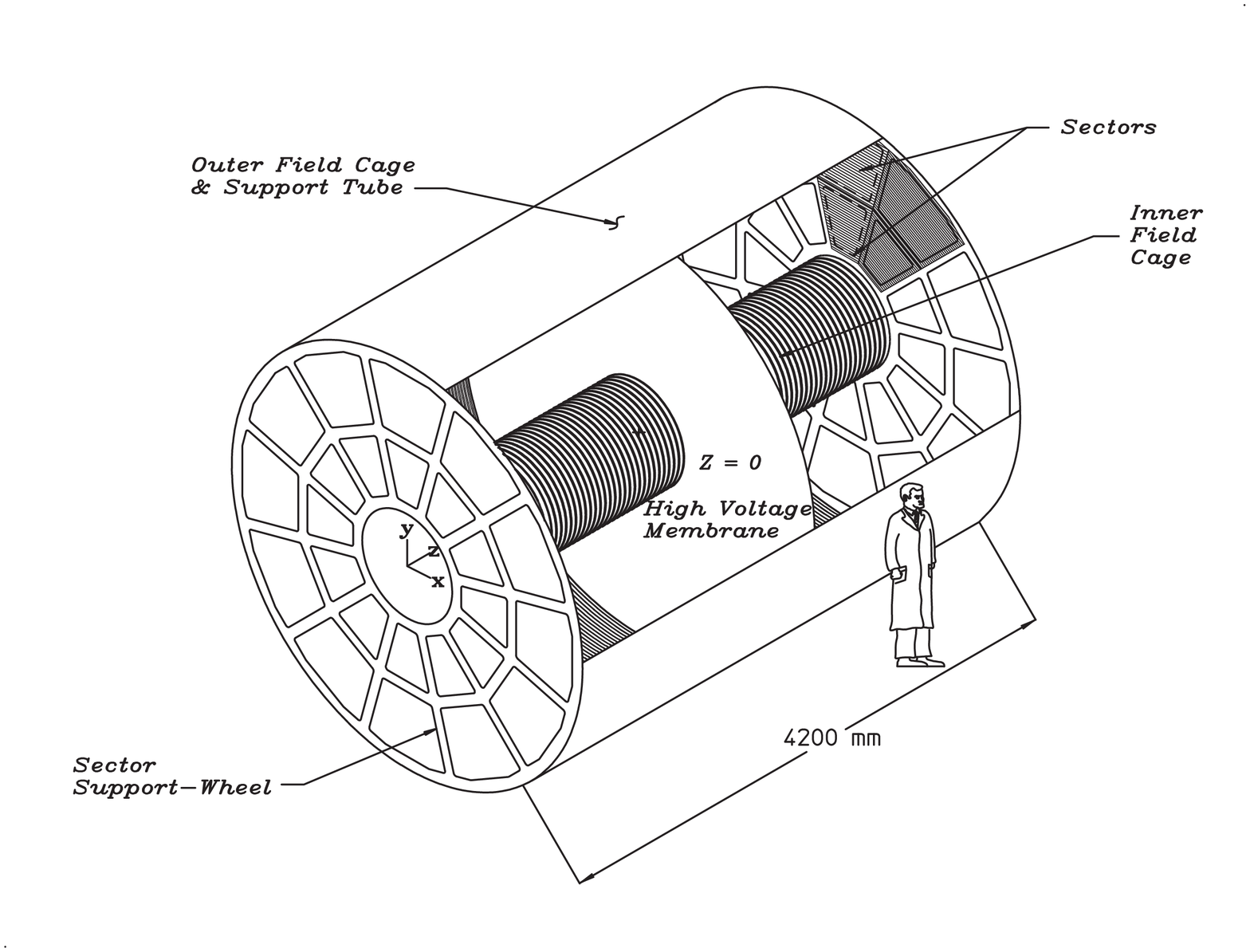}
\end{center}
\caption{Schematic view of the STAR TPC. Figure from \cite{Anderson2003659}.}\label{TPC}
\end{figure}

The Time Projection Chamber (TPC) \cite{Anderson2003659} is the main tracking and particle identification device at STAR. It records the tracks of the charged particles, measures their momenta and identifies them by measuring the ionization energy they lose while passing through the volume of gas. A schematic view of the TPC is shown in Figure \ref{TPC}.  The TPC cylindric body is 4.2 meter long and 4 meter in diameter. It is filled with a mixture of gas ($\mathrm{10\%}$ methane, $\mathrm{90\%}$ argon). The secondary electrons generated by the passage of ionizing particles through the gas drift towards the readout at the TPC end caps thanks to an uniform electric field of about 135 V/cm., generated between a central membrane and the end caps. The readout system is based on Multi-Wire Proportional Chambers (MWPC). The MWPC hits allow to reconstruct the transverse coordinates of the tracks, while the longitudinal position is calculated from the measured drift time. The TPC is designed to measure high multiplicities typical of heavy ions collisions. Its tracking efficiency depends on multiplicity, particle $p_T$ and particle type and it is in general of the order of $\mathrm{90\%}$. However tracking efficiency drops considerably at higher rapidity. For this reason the rapidity range of the TPC has been limited in the present analysis to $|\eta|<0.9$. In this analysis the TPC has been used to detect charged hadrons in the mid-rapidity range to correlate with the forward neutral mesons detected in the FMS.

\subsection[Barrel Electromagnetic Calorimeter (BEMC)]{Barrel Electromagnetic Calorimeter}

\begin{figure}[h!]
\includegraphics[width=0.5\textwidth]{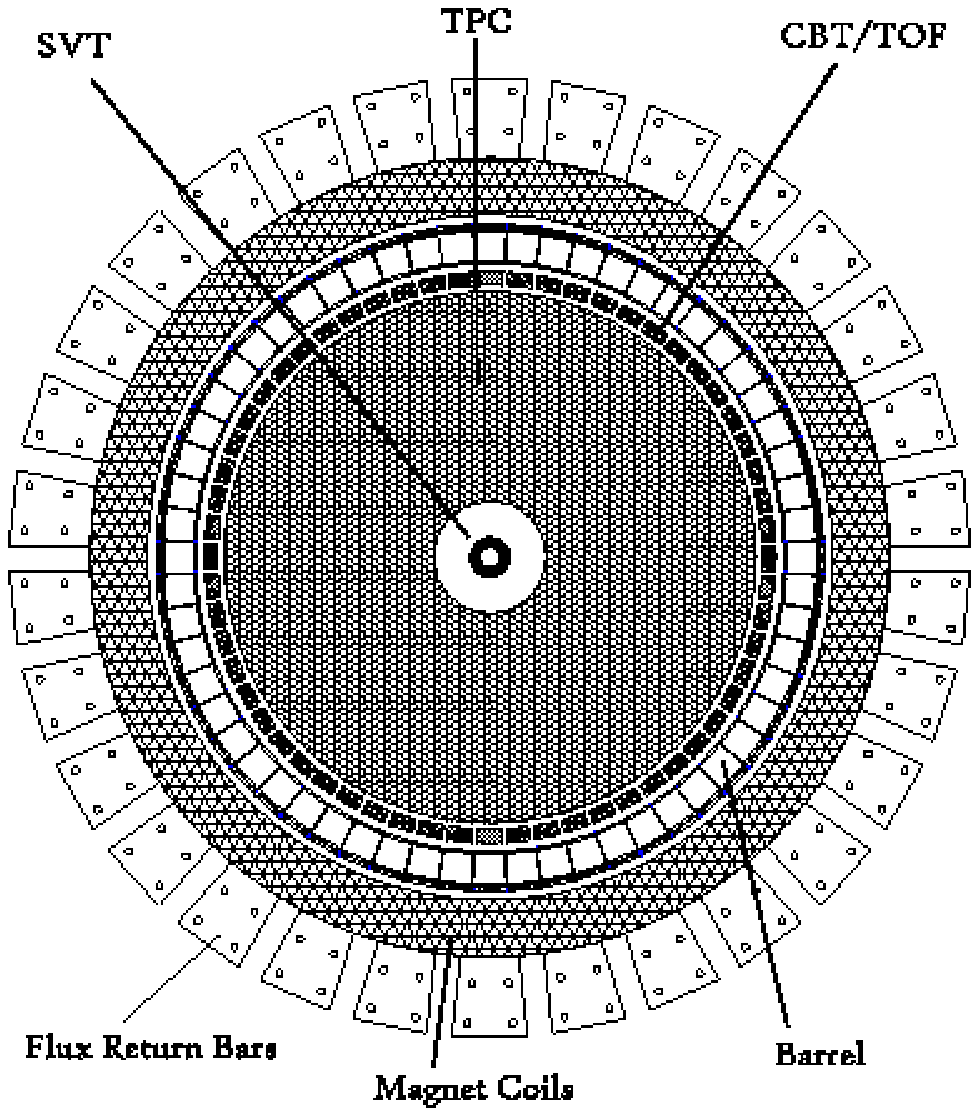}
\includegraphics[width=0.5\textwidth]{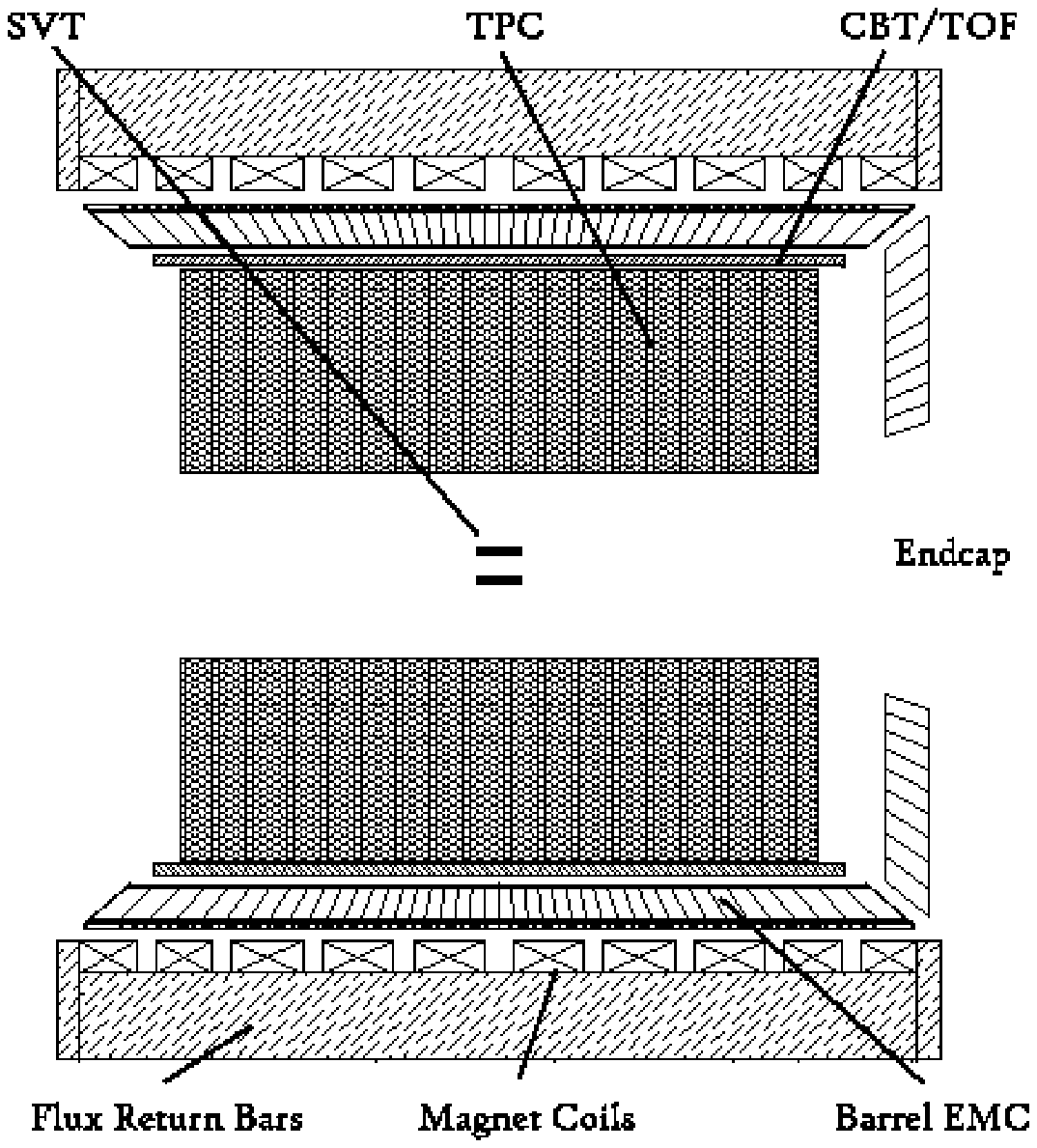}
\caption{Cross sectional view of the STAR detector showing the layout of the BEMC. Figure from \cite{Beddo2003725}.}\label{BEMC}
\end{figure}

\begin{figure}\begin{center}
\includegraphics[width=1\textwidth]{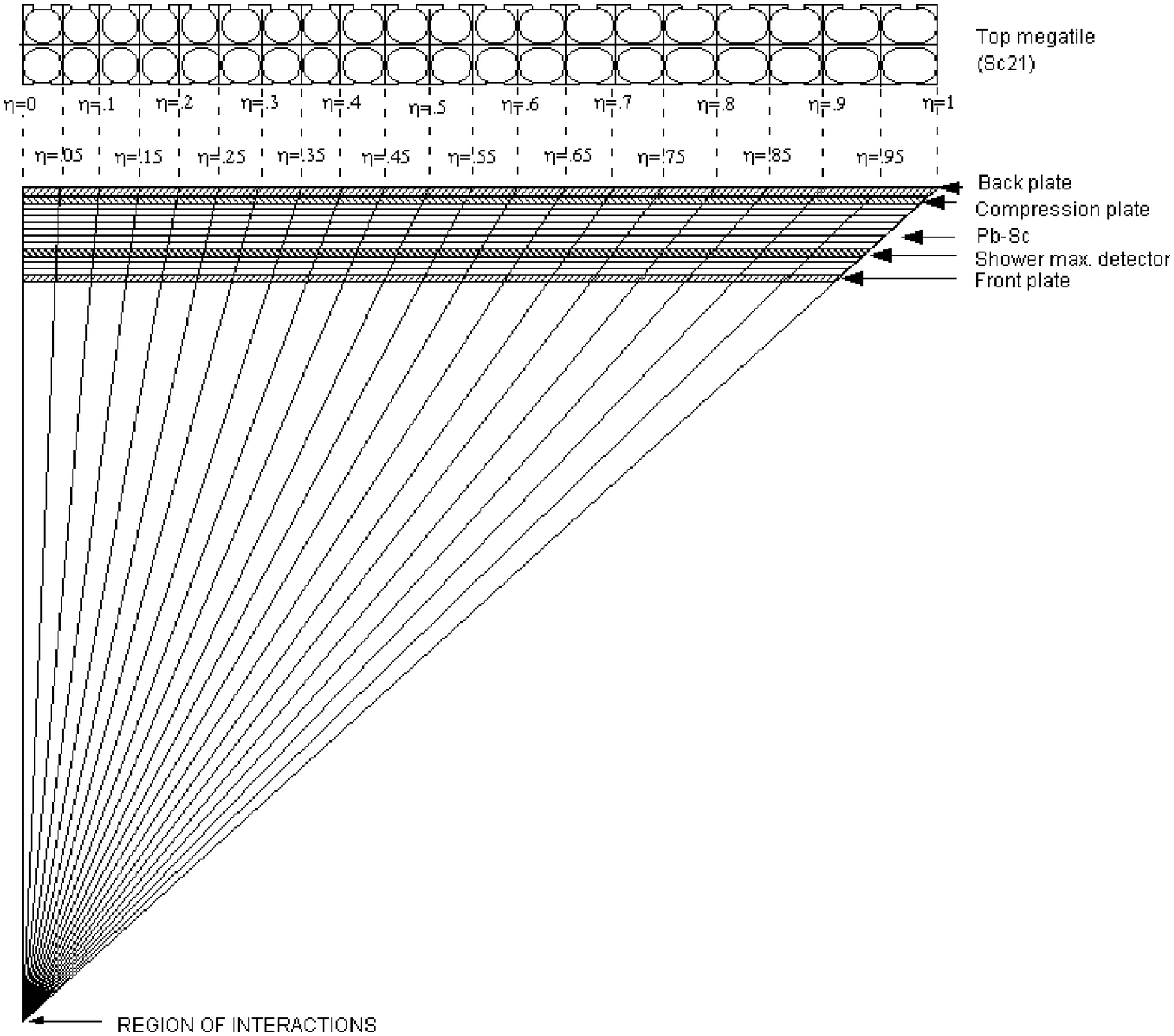}
\caption{Longitudinal segmentation of a BEMC calorimeter module as seen from side. Figure from \cite{Beddo2003725}.}\label{BEMC2}
\end{center}
\end{figure}

\begin{figure} \begin{center}
\includegraphics[width=1\textwidth]{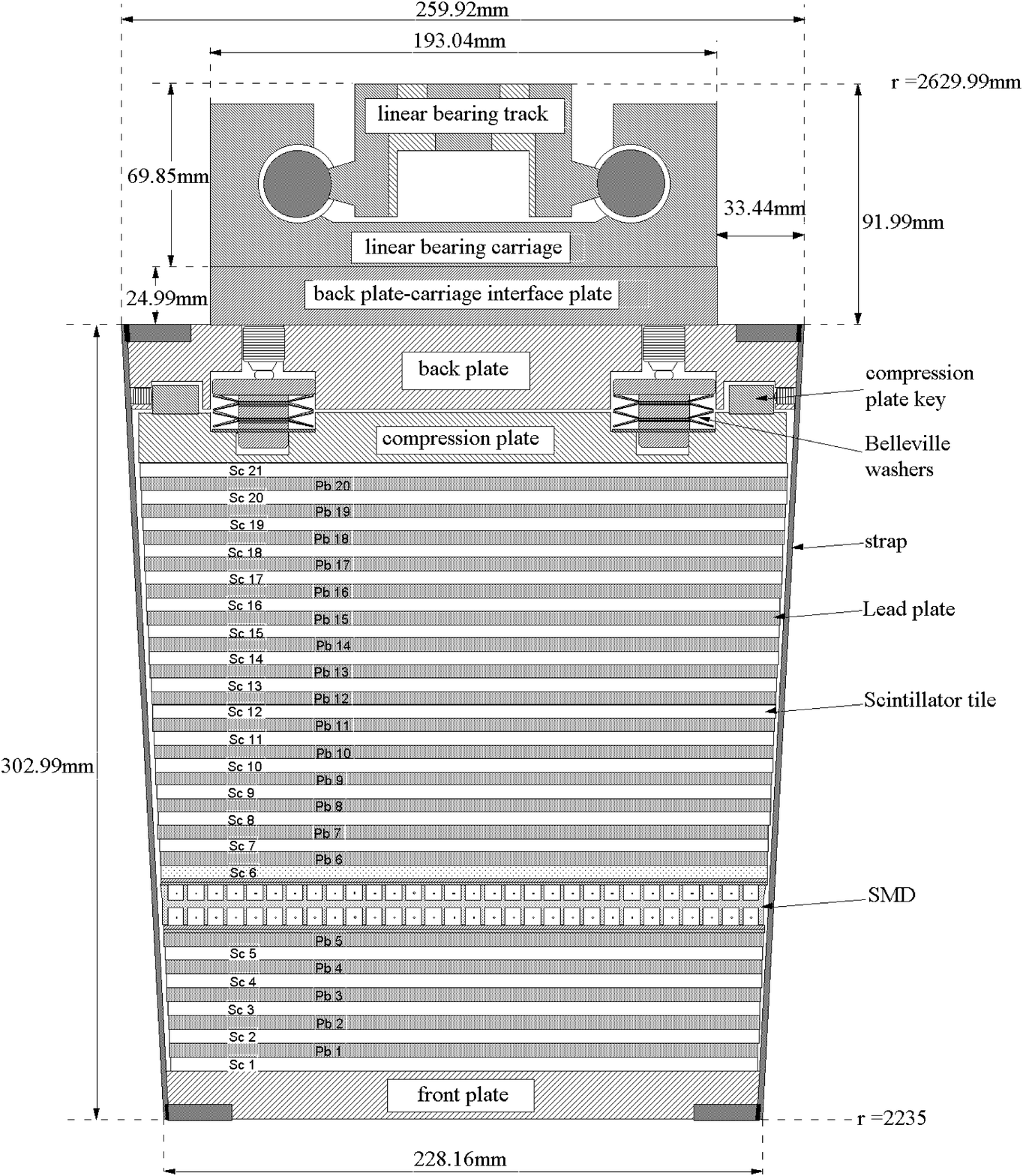}
\caption{Transverse segmentation of a BEMC calorimeter module as seen from the end view. Figure from \cite{Beddo2003725}.}\label{BEMC3}
\end{center}
\end{figure}

The Barrel Electromagnetic Calorimeter (BEMC) \cite{Beddo2003725} is a sampling calorimeter covering the rapidity region between $|\eta|<1.0$ in full azimuth, and it is placed around the STAR TPC, as shown in Figure \ref{BEMC}. Each of the two halves of the BEMC are 293 cm long and they extend between an inner radius of 223 cm and an outer radius of 263 cm. Each BEMC half is azimuthally segmented into 60 modules, each of which is composed of 40 projective towers of lead scintillator stacks, arranged so that there are 2 cells in $\phi$ and 20 in $\eta$. In total, the BEMC is then composed of 4800 tower, each of them covering 0.05 units in $\Delta\phi$ and 0.05 in $\Delta\eta$.  Each tower has a depth of 21 radiation length ($X_{0}$), corresponding to one interaction length for a hadron. This means that, while the electromagnetic shower is fully contained in the calorimeter (the maximum of the shower is at 5.6 $X_{0}$), the calorimeter is designed to let most of the hadronic signal go through before developing a shower. Each BEMC tower consists of two stacks of, respectively, 5 and 16 layers of scintillator, stacked one on top of the other and alternated by thin plates of lead. Between the two stacks, a Shower Maximum Detector (SMD) is positioned. A drawing of the calorimeter module is shown in Figures \ref{BEMC2} and \ref{BEMC3}.

The \textbf{plastic scintillator layers} are the main component of the calorimeter. They collect energy from photons, electrons and electromagnetically decaying hadrons. They are machined like mega-tile sheets with 40 optically isolated tiles. Signals from the single tiles are collected by wavelength shifting (WLS) fibers and transferred with optical connectors to decoder boxes. Here the signals from the 21 tiles composing the same tower are merged onto a single photomultiplier tube (PMT), powered by a Cockroft-Walton base. Layer by layer tests on the BEMC optical system, together with an analysis on cosmic rays and test beam data, measured the nominal energy resolution of the calorimeter as $\delta E/E=15\% \sqrt{E[\textrm{\,GeV}]}\oplus1.5\%$ \cite{bennet}.  

The \textbf{Shower Maximum Detector} (SMD) is used to provide fine spacial resolution in a calorimeter, which has segmentation (towers) significantly larger than an electromagnetic shower size. It is located between the inner and the outer scintillator stacks, at an approximative depth of $5.6$ radiation length at $\eta=0$ (up to $7.9$ at $\eta=1$). The SMD is a wire proportional counter detector with strip readout. The readout is done independently in $\eta$ and $\phi$, allowing the reconstruction of a two dimensional image of the shower. There are in total 36,000 strips in the full detector, and their coverage is $\Delta\eta\times\Delta\phi=0.0064\times 0.1$ for the $\eta$ strips and $0.1\times 0.0064$ for the $\phi$ strips. Tests at the AGS shown for the SMD an energy resolution of $\delta E/E=86\% \sqrt{E[\textrm{\,GeV}]}\oplus12\%$ in $\eta$ (and about $3-4\%$ worse in $\phi$), and a position resolution of $\sigma(r\phi)=5.6/\sqrt{E[\textrm{\,GeV}]}\oplus 2.4$ mm  and $\sigma(z)=5.8/\sqrt{E[\textrm{\,GeV}]}\oplus 3.2$ mm.

The first two scintillating layers in each tower are used as a \textbf{Pre-Shower Detector} (PSD). To achieve this, an independent readout system is included to collect a second sample of the energy from these first layers at the decoder box. The information from the PSD is stored together with the SMD signal. The PSD allows us to discriminate energy releases from hadrons (which are not supposed to shower in the first layers) from electromagnetic showers from electron and photons.  
 
The BEMC calibration consists of a process done at the beginning of the run and then further corrected before performing the analysis. Individual cell gains are corrected by aligning at the same value the single tower response to a Minimum Ionizing Particle (MIP), identified by the TPC. In addition, an overall gain correction, based on independent measurement of the electron energy release in the BEMC and its momentum in the TPC, is applied to all cells. 
Finally, a quality assurance (QA) procedure is routinely performed before the physics analysis. This is to check the status of each tower by reading a status table and to mask out possible badly responding channels. Since it has been found that tower calibration is less reliable at the edges of the calorimeter, signals detected in the towers of the two outermost rings at each side of the detector are removed from the analysis, effectively reducing the BEMC acceptance to the pseudo-rapidity range $|\eta|<0.9$.

\subsection{Beam-Beam Counter}

\begin{figure}[h!]
\begin{center}
\includegraphics[width=0.5\textwidth]{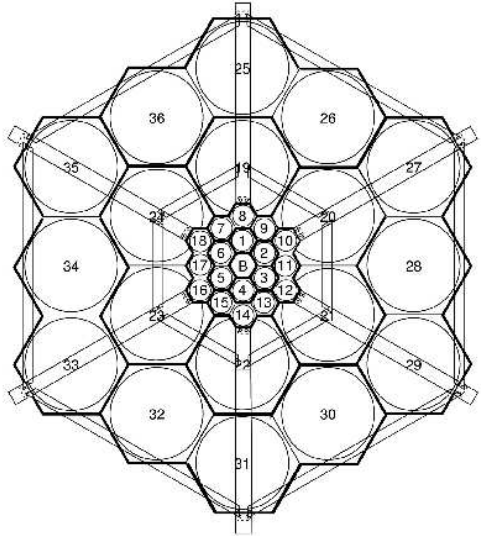}
\end{center}
\caption{Schematic view of the BBC detector. Figure from \cite{kiryluk2005}.}\label{BBC}
\end{figure}

The Beam-Beam Counter (BBC) \cite{kiryluk2005} detectors are two sets of hexagonal scintillator tiles located at a distance of $3.7$m from the interaction point, mounted around the beam pipe. Each of the two BBC modules is composed by two sets of scintillators, as shown in Figure \ref{BBC}. The innermost ring of smaller tiles has a radius between 9.6 and 48.0 cm and it covers the pseudo-rapidity range $3.5<|\eta|<5.0$. The outer ring is composed of larger tiles. It spans the pseudo-rapidity range of $2.0<|\eta|<3.5$ with a radius between 38 and 193 cm. The BBC is used in p+p collisions to provide a minimum bias trigger. Events are in fact selected when there is a coincidence of signals between one of the 18 small tiles on each of the BBC modules. The BBC modules are also used to measure the longitudinal position of the interaction vertex with an accuracy of 40 cm, by recording the time of flight difference between the two counters. Large values of the time of flight difference indicate the passage of beam halo, which is rejected at the trigger level. Finally, the BBC counting rate provides a measurement of the  absolute luminosity of the run. 
In this analysis the BBC was used as a trigger detector, as well as to provide a measurement of the event charge multiplicity to estimate the centrality of the collision. During the 2008 run, only the two inner rings of small BBC tiles were active.

\section[Forward Meson Spectrometer (FMS)]{Forward Meson Spectrometer}

In this section we focus on the Forward Meson Spectrometer (FMS). This is the crucial detector for the topic of this thesis and it will play a key role through all the steps of the analysis. The FMS is an electromagnetic calorimeter located 7.30 meter from the STAR interaction point in the west direction. It is a 2 m $\times$ 2 m matrix of lead-glass cells, intended to collect energy from photons, electrons and electromagnetically decaying hadrons. It is divided in two detachable halves (North and South), each of them composed of an inner (made with smaller size cells) and an outer part (made with larger cells). The whole FMS covers the forward pseudo-rapidity range $2.5<\eta <4.0$ in the full azimuth. The FMS extends the electromagnetic capability of STAR in the forward regions, making, together with the BEMC ($|\eta|<1.0$) and the EEMC ($1.0<\eta <2.0$), the coverage nearly hermetic in the wide $-1.0<\eta < 4.0$ range. This allows measurements of correlations of different species of forward and mid-rapidity particles, over this broad $\Delta\eta\times\Delta\varphi$ range. The variety of possible correlations includes signals from the STAR calorimeters (BEMC, EEMC, FMS) and time-projection chambers (TPC, FTPC). The FMS acts also as a fast-readout trigger detector. This is to optimize the sampled luminosity used for analysis involving the FMS. Minimum bias samples, in fact, provide very small occupancy in the forward phase space region, where the FMS operates. In order to have a statistically significant sample of data involving forward particles, it is therefore convenient to trigger the event in the forward region, using the FMS. 

The physics objectives STAR is expected to achieve, thanks to the addition of the FMS, can be summarized as follows:
\begin{itemize}
\item universality of the gluon distribution: the current knowledge of the gluon density distribution in heavy nuclei can be tested in the $x$ region between $0.001<x<0.1$;
\item gluon saturation: non-linear effects in the wave-function of the Gold nucleus can be accessed with a characterization of correlated pion cross sections as a function of $p_{T}$ and $\eta$;
\item spin puzzle: the origin of the large transverse spin asymmetry in $p_{\uparrow}+p\rightarrow \pi^{0}+X$ is expected to be resolved by measurements of forward $\pi^{0}$ production in transversely polarized proton interactions. 
\end{itemize}

The FMS is a relatively new detector. It was assembled and put in place during the second half of 2006 and it was ready for the first data taking during the FY08 run (2008). The quick transition from final prototype to full detector installation, was made possible by the availability of recycled lead-glass cells from decommissioned detectors, such as E831 at the Fermi National Accelerator Laboratory (FNAL) for the large cells, and E704 for the small cells made available by the Institute for High Energy Physics (IHEP) and the Thomas Jefferson National Accelerator Facility (TJNAF). Furthermore, the FMS used expertise achieved with previous prototype detectors, operating at forward rapidity regions at STAR since 2002. 

\subsection{Forward Calorimetry at STAR}
Forward calorimetry at STAR started in 2002, when a prototype forward $\pi^{0}$ detector (pFPD) was installed at 750 cm east of the interaction point. The pFPD was meant as a testbed for different detector solutions and had the goal of verifying the feasibility of neutral pion reconstruction in the forward regions at STAR. The pFPD consisted of two modules. On the north side was placed a Pb-scintillator sampling calorimeter with respective prototype scintillator strip `shower maximum detector' (pSMD). This was in fact a portion ($1/\mathrm{60^{th}}$) of an EEMC early design, at that time being installed at STAR. This EEMC prototype (pEEMC) proved the possibility of reconstructing neutral pions by detecting their decay photons with good resolution in energy and position. To rule out asymmetry effects, a $4\times4$ array of $3.8\times3.8\times45\textrm{\,cm}^3$ Pb-glass detectors was placed on the south side of the beam pipe. This detector was not able to identify $\pi^{0}$'s, due to its small size and poor resolution, but measured inclusive photon production. The pFPD was used to discover an energy asymmetry in the analyzing power \cite{Adams:2003fx}.

\begin{figure}
\includegraphics[width=0.46\textwidth]{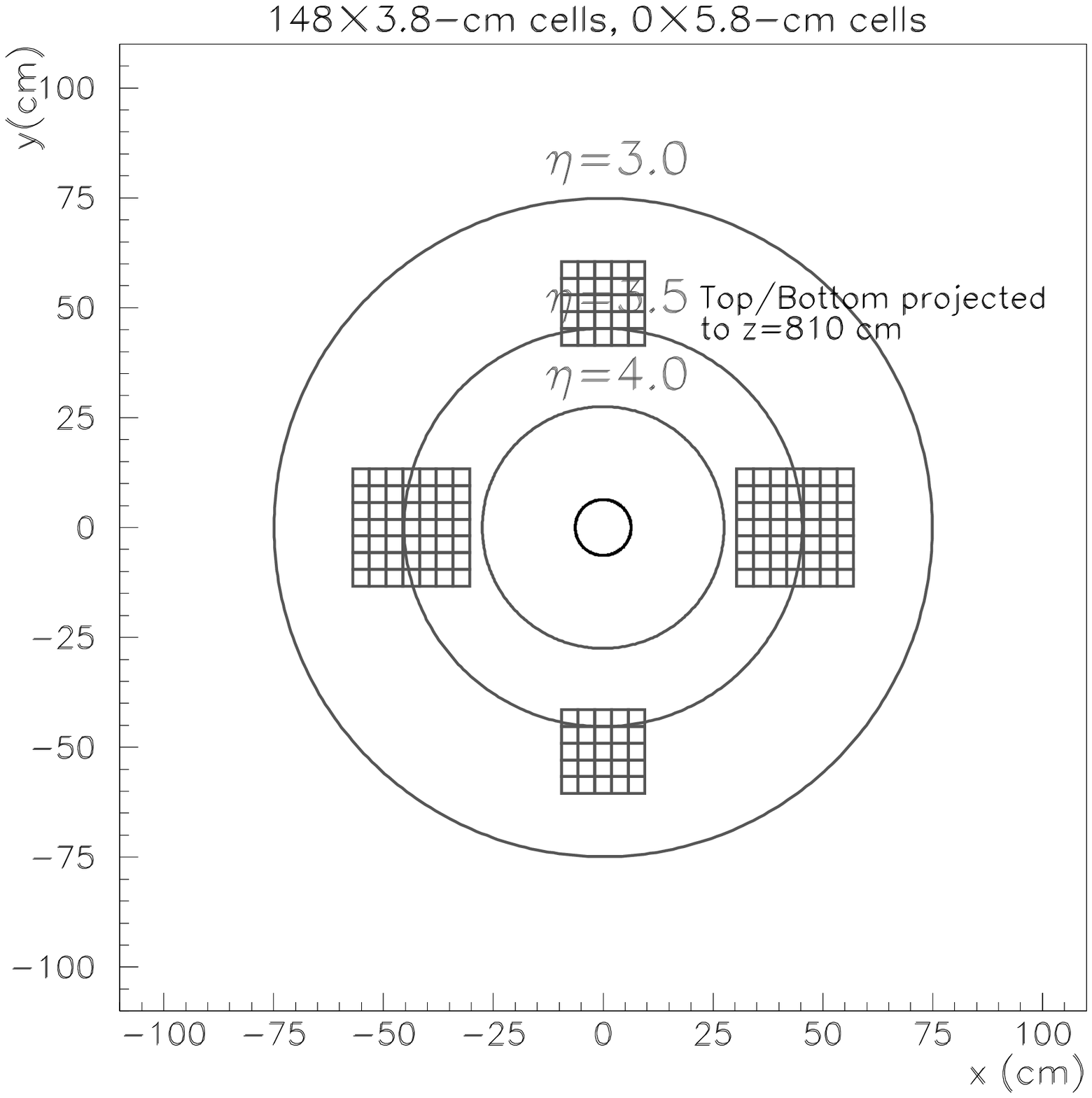}
\includegraphics[width=0.46\textwidth]{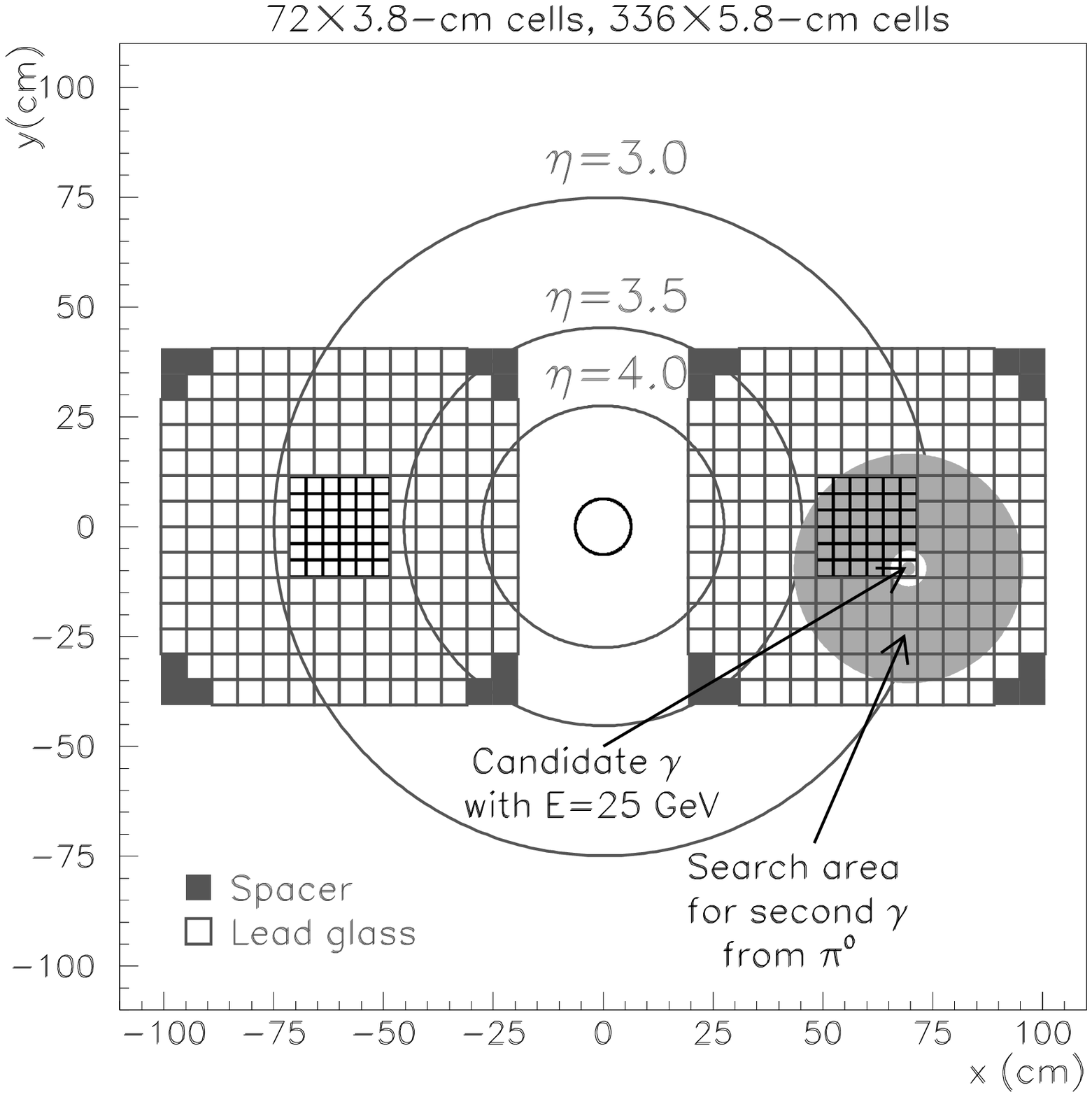}
\caption{Mechanical layouts of early forward calorimeters at STAR: FPD (left) and FPD++ (right).}\label{geoms}\end{figure}

\begin{figure}
\includegraphics[width=0.94\textwidth]{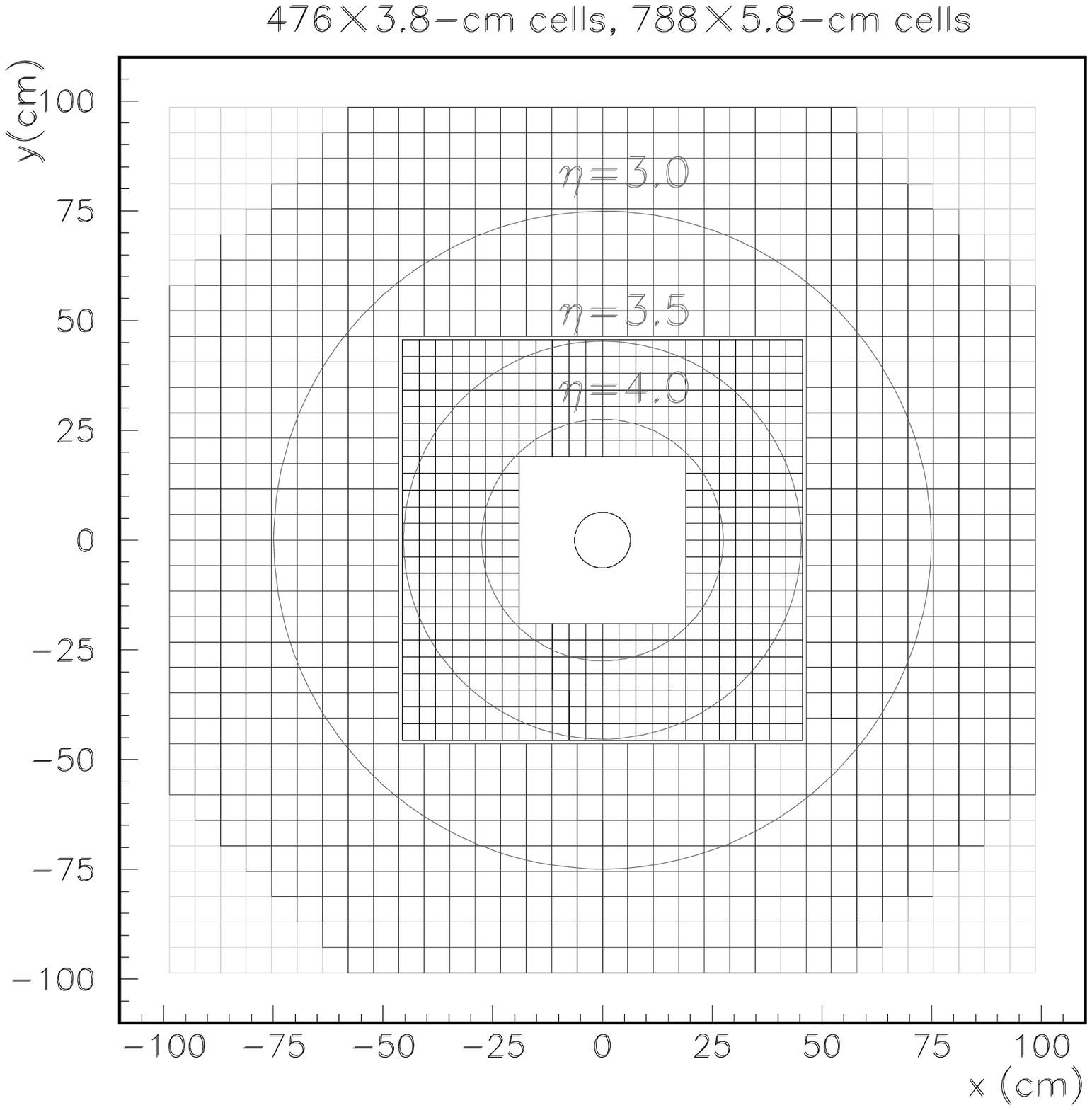}
\caption{Mechanical layouts of the Forward Meson Spectrometer (FMS) at STAR.}\label{geomFMS}\end{figure}

For the mixed p+p and d+Au run at $\sqrt{s_{NN}}=200\textrm{\,GeV}$, scheduled for 2003, the forward spectrometer was upgraded into a larger detector, thanks to the contribution of more Pb-glass cells from the decommissioned E704 detector. The resulting new Forward Pion Detector (FPD), positioned in place of its prototype, was composed of four modules of lead-glass detectors. Two main $7\times7$ arrays of cells were placed on the north and south side of the beam-pipe and supplied for SMD and Pre Shower Detectors (PSD). Two smaller $5\times5$ arrays were placed above and below the beam-pipe for systematics studies. In addition, a new pair of larger modules were incrementally installed on the west side of STAR, symmetrically to the east side. The position of the FPD modules along the longitudinal direction was meant to favor studies on the spin nature of the proton. The FPD proved to be able to reconstruct neutral pions. It collected data in un-polarized p+p interaction, proving the validity of Next to Leading Order (NLO) perturbative Quantum Chromodynamics (pQCD) in forward regions at RHIC. Exploratory measurements during the short 2003 d+Au run, together with p+p data acquired the year before, showed a strong suppression of azimuthal angular correlations between a forward $\pi^{0}$ and a mid-rapidity charge particle, detected with the TPC \cite{Adams:2006uz}. Such suppression, not seen in p+p, was consistent with the conjecture that the gluon density in the gold nucleus could have been saturated.

These striking  results in the 2003 run prompted a new series of upgrade plans for forward detectors at STAR. The west FPD side was incrementally improved and, in this position, for the 2006 p+p run a consistent upgrade has been possible thanks to a set of new and larger ($5.8\times5.8\times60\textrm{\,cm}^{3}$) Pb-glass detectors coming from the decommissioned E831 detector. The newly renamed FPD++ presented two halves in the west side of STAR, analogously to the FPD but composed by larger arrays of cells (Figure \ref{geoms}). Each half included an inner detector, a $6\times6$ matrix of small cells, and an outer ring of large cells to form a $14\times14$ stack with cut edges. Thanks to its larger acceptance, the FPD++ proved to be able to detect direct photons by gaining sensitivity in the reconstruction of photon pairs from neutral pion decays. The FPD++ was built as a temporary detector to allow engineering tests in preparation of the assembling of the larger FMS. Electronic and trigger schemes were tested during the 2006 run, together with systematic studies on mass and energy resolutions, reconstruction efficiency and topological analysis. This experience translated a year later into the commissioning of the FMS.   

\subsection{FMS: mechanical layout}

The Forward Meson Spectrometer (FMS) was built during the summer of 2006. Large and small cells from previous forward spectrometers, with an addition of new Pb-glass detectors, were prepared to be reassembled in the new geometrical configuration (Figure \ref{geomFMS}). This included two concentric square rings of cells: the inner detector, made of small cells and covering the pseudo-rapidity region $3.0<\eta<4.0$, and the outer detector, made of large calls and covering the pseudo-rapidity range $2.5<\eta<3.4$. Together they make up a $2\times2\mathrm{\,m^2}$ square matrix with cut corners. The inner ring counts of 476 smaller Pb-glass cells, while the outer ring counts 788 larger detectors. Each detector unit, of both inner and outer calorimeter, is composed by a Pb-glass cell that collects light from the electromagnetic shower, a PhotoMultiplier Tube (PMT) that collect and amplify the signal, and a High Voltage system to power the PMT. The Lead-glass cells collect the \v{C}erenkov radiation from an electromagnetic shower generated by a charge particle interacting with the Pb in the detector. They are respectively 18.00 and 18.75 radiation length long. Cells are optically glued to PMT units. Larger cells, provided by the FNAL E831 experiment, are coupled with a XP2202 PMT, powered by Zener-diode voltage divider. A set of smaller cells from IHEP presented FEU84 PMT's powered by Cockcroft-Walton HV bases, designed and built for this porpoise. A second set of small cells, provided by TJNAF to complete the inner calorimeter, were supplied by XP2972 PMT's.  

Each detector unit underwent a series of integrity tests and characterization checks before being placed in the FMS array. Pb-cells have been cleaned and wrapped in a thin (0.1mm) foil of reflective aluminized PolyEthylene Terephthalate (PET/Al) in order to contain as much as possible the shower radiation within the cell boundaries and to avoid external and cell-by-cell light contamination. Current-voltage characteristics (I-V curves) were analyzed to characterize the single PMT gain, and possibly to remove badly responding PMT's. PMT's characterized by a higher gain were placed in the innermost part of the detector. Detector response tests and positioning checks were completed \textit{in situ} for each detector unit by using a light-emitting diode (LED). In addition, a prototype LED light-flasher board covering a whole FMS quadrant was built prior to the 2008 run and used to test cell-by-cell  the presence of dead channels and possible mapping errors. The board was also used to monitor cell status "on the run" by allowing LED pulser events to enter the data stream. It is possible to easily identify such hits and remove them from the reconstruction algorithms afterwards.    

The readout system of the 1264 channels of the FMS is provided by so called QT boards, incrementally installed in replacement of older digitizer boards. Each 32-channel QT motherboard records ADC signals and TDC discriminators from 32 different detectors, whose signal is collected by four 8-channel QT8 daughterboards and then merged into the larger QT32 board. Before this step, the ADC signals are sent to a Field-Programmable Gate Array (FPGA) which compress the inputs and perform a first trigger selection. The signals are then passed to Data Storage and Manipulation (DSM) boards, where more refined trigger algorithms are performed. 

\subsection{FMS: performance}
As well as for its predecessor, calibration in the FMS is performed in two steps: on-line and off-line. Single cell-by-cell calibration is determined using $\pi^{0}$ reconstruction: each pair of clusters ($\pi^{0}$ candidate) reconstructed is associated with the lead-glass cell corresponding to the highest energy deposition; an invariant mass spectrum is then created for each ``high-tower'' and fitted with a gaussian function; the PMT gains of the single towers are then adjusted in order to move the centroid of the gaussian fit towards the nominal value of the $\pi^{0}$ mass. Once this is done for all cells, the procedure is iterated until reasonable convergence is reached. This procedure involves a different gain factor for each cell which is factorized into a ``basic''  gain factor, common to all the towers in the same module, and a correction factor typical of the single cell. Once these factors are calculated, they can be used for on-line calibration (both by applying effective gains and by modifying high-voltage values for single PMT's) for the later parts of the run. Further off-line corrections are included since the response of the FMS is found to be dependent on the energy of the detected particle (energy dependent correction) and on time and beam conditions (run dependent correction). The position of the $\pi^{0}$ peak in the invariant mass distribution is in fact moving as a function of the $\pi^{0}$ energy: after cell-by-cell calibration is applied, there still remains a energy dependence of the reconstruction which moves the peak to higher values of the mass as the energy of the detected particle grows. Dedicated Monte Carlo studies, simulating full $\mathrm{\check{C}erenkov}$ light, have shown that this dependence may be caused by missing energy in the reconstruction, due to deeper longitudinal shower profile, transverse leakage on the edges of the detector and ADC granularity. For this reason, an overall energy dependent correction is applied to all clusters and it is shown to work well in the range of interest $10<E<65$ GeV. In particular, calibration using the $\pi^{0}$ mass proved to well reproduce the values of the mass for heavier mesons ($\omega$, $J/\psi$). In addition to cell-by-cell and overall energy calibrations, a run dependent correction is necessary. In general the FMS has proved to be relatively stable during the production run. However, slight time dependence of the response of the detector are expected. As it has been done for the FPD++, a run dependent correction based again on the position of the centroid fit on the invariant mass spectrum compared to the nominal $\pi^{0}$ mass value has been applied. Single channel instability can be detected in the FMS thanks to the LED board, which also provides us with a useful calibration tool.

After these correction we expect a mass resolution around $\sigma_{m}\sim23$ MeV/$c^{2}$, based on the experience with the FPD++. Simulation studies have demonstrated an energy resolution smaller than $15\%/\sqrt{E/[\textrm{\,GeV}]}$ and a position resolution for $\pi^{0}$'s better than 0.5 cm. The efficiency of the FMS is expected, as well as for the FPD++, to be limited by geometrical acceptance only. Conservative estimates of 35$\%$ for reconstruction efficiency and (mainly) geometric acceptance are predicted.

%%% Chapter heading commands %%%

\chapter{Topology: event reconstruction}
\label{chapter:event_reconstruction}

\index{Event reconstruction}

%%% Abstract %%%

%\begin{Abstract}

%\end{Abstract}

%%% Chapter sections %%%

\index{Topology: event reconstruction}
\section{Data acquisition and trigger}
The STAR Data Acquisition (DAQ) system \cite{daq} is an electronic architecture that processes input from multiple STAR sub-detectors at different readout rates. Detectors in STAR are divided into fast and slow detectors, based on their readout rates. Fast detectors, such as ZDC, BBC, CTB, BEMC and FMS, are capable to provide trigger information and to cope with the RHIC beam crossing frequency of $\sim10$ MHz. On the contrary, the typical recorded event rate of $\sim1000$ Hz is limited by the slower detectors (TPC, FTPC, SMD) and in particular by the drift time of the slowest detector in STAR, the TPC. The main task of the STAR DAQ system is to read data from fast and slow detectors, to balance and reduce the data rates, and to store them in the High Performance Storage System (HPSS) facility. 

To facilitate this process, a pipelined trigger system has been designed. The STAR trigger system \cite{trg} consists of a series of four trigger levels, the first three based on fast informations and the last one including tracking from the slow detectors. For every bunch crossing, information on readout and status of each fast detector is sent to the Data Storage and Manipulation (DSM) boards which act as a fast decision tree. If the DSM decides that certain specific conditions are met, the Trigger Control Unit (TCU) issues a trigger which serves as a unique event identifier and determines whether the event is to be stored or not. This information is then passed to the detectors to start the digitization of the buffered signal. This part of the trigger process lasts for no more than 1.5 $\mathrm{\mu}$s after the collision and it is referred to as Level 0 trigger. During signal digitization further conditions can be applied to the fast stream of data (Level 1 and Level 2 triggers) before they are passed to the DAQ system. At this stage the typical size of a processed event can reach 200 MB in central Au-Au collisions. The largest part of is dominated by the output of slow tracking detectors, mainly the TPC. The main task  of the DAQ system is here to read data from the STAR detectors at a rate up to 20,000 MB/sec and to reduce the rate to 30 MB/s in order to be able to store the stream of data on tape (HPSS).  

The STAR trigger framework allows to select an event using different trigger confi\-gurations simultaneously. Each trigger configuration corresponds to a list of trigger definitions for the different detectors. An event that fulfills all these requirements is labelled with a identifier which reflects uniquely the features of the event. The STAR trigger system allows events to carry multiple trigger identifiers.

\begin{description}
\item[Minimum Bias Events.] 
This is the most loose condition to be applied to an event. It ensures the least possible bias towards the final state of the interaction. For this reason, Minimum Bias (MinBias) trigger is of a fundamental importance for measures of inclusive cross sections; however it does not provide a convenient selection of forward events which are crucial for studies of saturation effects. Therefore, MinBias selections will not be used in this thesis.  
In p+p collisions, a MinBias trigger is issued when a coincident signal in both BBC's modules is recorded. The $z$ position of the interaction vertex is calculated from the time difference between the East and West BBC signals. In d+Au collisions, a MinBias trigger requires the detection of at least one neutron in the East module of the ZDC, towards which the Gold beam is directed. The $z$ position of the interaction vertex is given, in this case, from time difference between the ZDC signal and the RHIC strobe which signals the bunch crossing. Since the trigger in d+Au collisions does not require a coincidence signal from both sides of the interaction point, it proves to be more susceptible to beam background events.

\item[Slow FMS Trigger Events.] 
In order to use a stream of data optimized for the selection of forward particle production, which is in the interest of this thesis, a trigger configuration using detection requirements on the FMS has been developed. FMS triggered events require a hit in one of the FMS cells (high-tower) above a certain threshold, which is defined differently for the inner (small cells) and the outer (large cells) modules of the FMS. Since the gains of the PMT's vary a lot, the energy threshold are also very different from cell to cell. They however correspond to a common number of ADC counts (400 ADC counts for the inner calorimeter and 200 ADC counts for the outer during the year 2008). These translate into a high-tower energy threshold using gain and correction factors from offline reconstruction. No MinBias conditions are required for such events, since the high values of the trigger threshold should prevent background fluctuations to trigger a fake event.   
The so called FMS triggered \emph{slow} events include information from all the sub-detectors in STAR. Since the rate of the data flow is constrained by the TPC, these events are labelled as ``slow''. This set of data has been used for the measurement of correlations between a forward $\pi^{0}$ (the trigger particle) and a mid-rapidity charged particle (the associated particle) using the TPC.     

\item[Fast FMS Trigger Events.]
The stream of data just discussed (FMS slow events) includes all the informations necessary to allow also measurements of correlations between two neutral pions (namely: FMS and BEMC tower information). However, as we have seen, the event rate of this data set is very small since it is driven by the rate of slow detectors. It is possible, for measurements that do not require the TPC, to use a different stream of data. The so called FMS triggered \emph{fast} events are characterized by the same trigger condition on the forward detector as for the slow ones. However the volume of data is stored on tape before the rate reduction performed by the DAQ system. This allows us to save more lighter data files which results into a large amount of events with informations from fast detectors only.    These include, besides the FMS which provides the trigger, the BEMC towers (but, notably, not the Barrel Shower Maximum Detector, BSMD) and both the EEMC towers and shower maximum (ESMD) detectors. This set of data has been used for the measurement of correlations between a trigger $\pi^{0}$ in the forward region (FMS) and either a mid-rapidity $\pi^{0}$ (BEMC) or another forward $\pi^{0}$ detected in the FMS. 
 
\end{description}

\begin{figure}
\begin{tabular}{l r}
\includegraphics[width=0.46\textwidth]{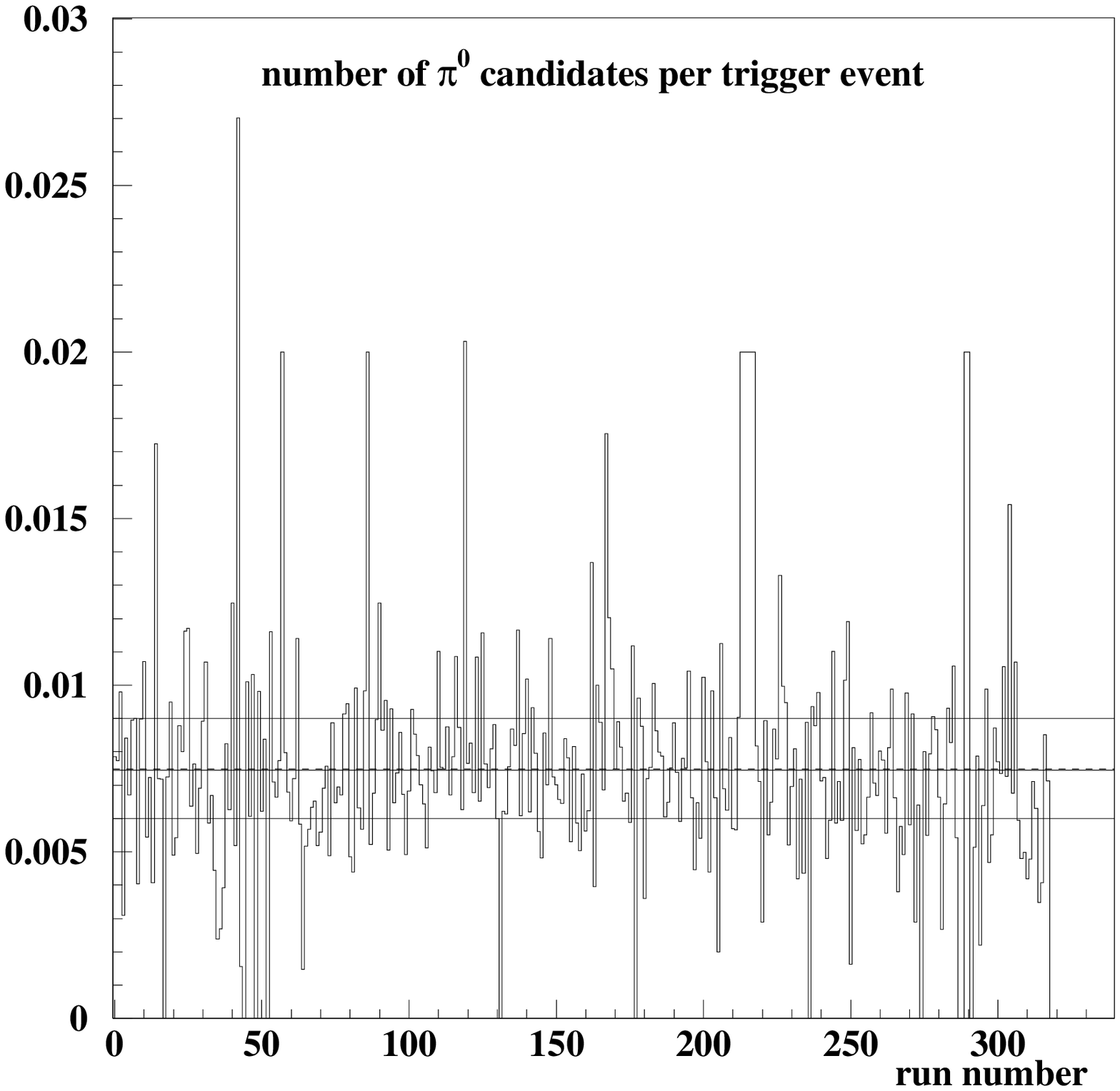} 
\includegraphics[width=0.46\textwidth]{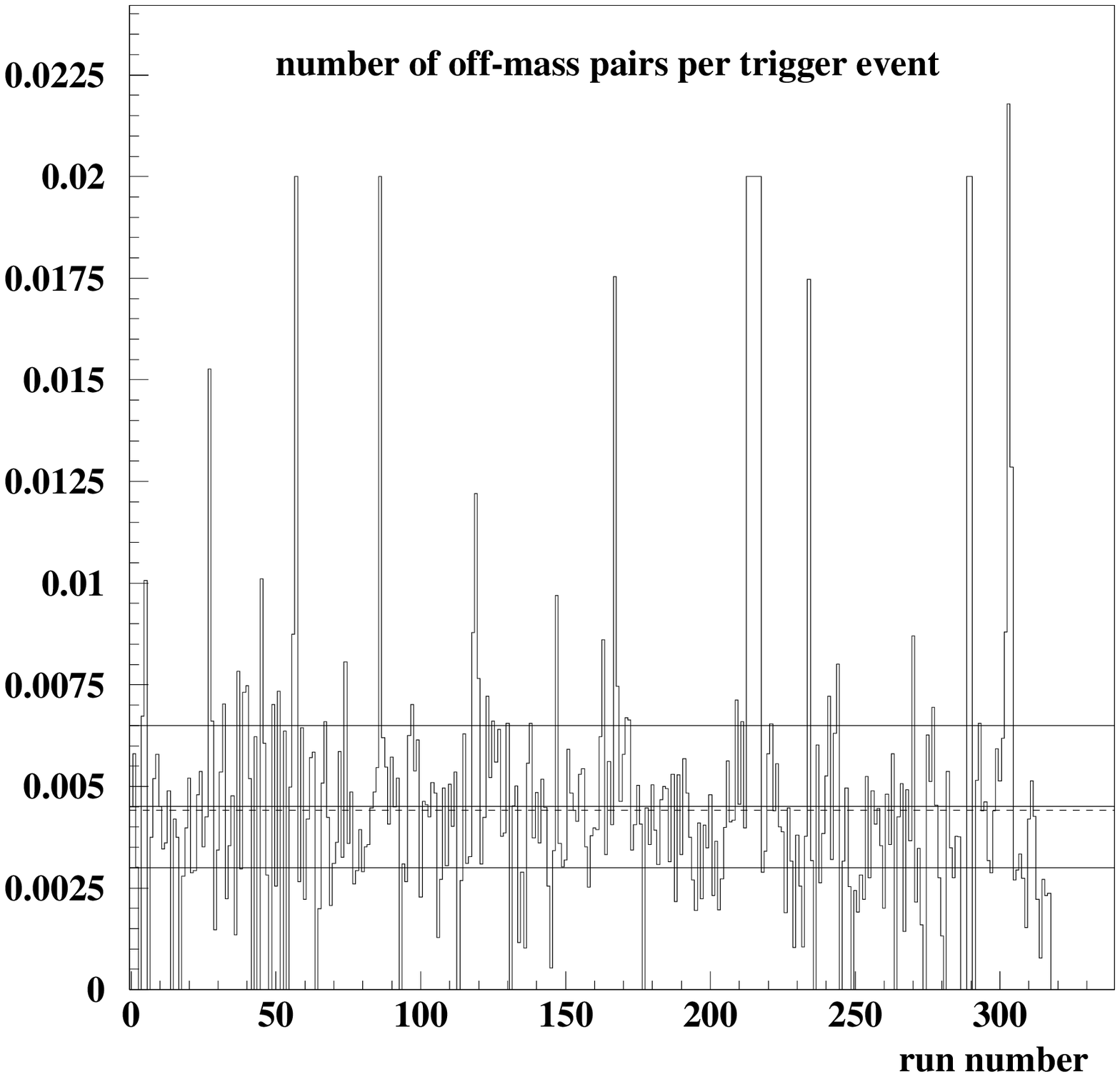} 
\end{tabular}
\caption{Run dependence of the number of BEMC $\pi^{0}$ candidates (left) and off-mass pairs (right) per trigger event. Averages over the whole period are indicated in dashed lines, while a selected number of ``good'' runs is indicated in solid lines, together with its multiplicity average. For runs with no FMS information available, the ratio has been set to $0.02$.}\label{5.rundep}
\end{figure}

The stability of the BEMC during the run period has been checked by counting the number of BEMC $\pi^{0}$ candidates per FMS trigger $\pi^{0}$ in function of the run number. As shown in Figure \ref{5.rundep}, the number of candidates is relatively stable along the run, with a limited number of runs with BEMC present only (where the ratio has been set to 0.02) and some runs with a very low number of candidates per run where, most likely, also the FMS was not operative. In addition to this, a reduction of the number of pairs can be seen as a general trend at the end of the run period. This is most likely attributable to worsening calibration of the BEMC during the very last period of data taking. In order to check if this affects the measurements, the number of BEMC pairs per trigger event have been calculated for a limited selection of good runs (indicated in Figure \ref{5.rundep} as the runs where the number of pair falls within the two outermost solid lines). The average number of pairs in this selection (central solid line) is not significantly different from the overall average (indicated with a dashed line). This is true for either the $\pi^{0}$ candidates and for the off-mass pairs used to estimate the background. As a precaution, however, such bad runs have been excluded from the measurement. 

\section{TPC: charged particle reconstruction}
The Time Projection Chamber (TPC) is used to reconstruct charged particles by collecting informations about their momentum and energy loss. When a charged particles passes through the TPC, it ionizes the gas the chamber is filled with. The electrons created drift in the electric field towards the readout at the end-caps of the detector. Each end-cap readout is composed of 12 sectors, each of them segmented into 45 pad rows. The transverse coordinates of the track are reconstructed from the hits on such readout pads. In order to improve the quality of the reconstruction, a minimum number of 25 hits per track is required.      
Moreover, each track used in this analysis is required to be generated from within 3 cm of the primary vertex and to have a pseudo-rapidity $|\eta|<0.9$ comparable with the BEMC acceptance.

\section{BEMC: neutral meson reconstruction}
The Barrel ElectroMagnetic Calorimeter (BEMC) has been used in this analysis for $\pi^{0}$ reconstruction in the mid-rapidity range $|\eta|<0.9$. This is achieved by detecting the two photons to which the $\pi^{0}$ decays: 
\begin{displaymath}
\pi\rightarrow\gamma\gamma \qquad , \qquad BR=98.8\%
\end{displaymath}
Each photon is reconstructed by clustering groups of adjacent towers with a non-zero energy deposition, using an algorithm which will be described in the next sections.
Since the $\pi^{0}$ lifetime is very short ($\tau=8.4\times10^{-17}\mathrm{\,s}$, corresponding to a decay length of $c\tau=0.025 \mathrm{\,\mu m}$), we can reasonably assume that the two photons originate from the primary vertex of the interaction. Therefore we can use the following expression
\begin{equation}\label{BEMCmass}
M_{\gamma\gamma}=\sqrt{E_{1}E_{2}(1-\cos{\psi})}
\end{equation}
to calculate the value of the invariant mass for each pair of photons detected in the BEMC. In Eq. \ref{BEMCmass}, $E_{1}$ and $E_{2}$ are the energies of the two photons and $\psi$ is the opening angle between them, in the laboratory reference system. Each pair of photons contributes with one entry to the invariant mass spectrum, in which the $\pi^{0}$ appears as a peak centered around its nominal mass value of 134.98 $\mathrm{\,MeV/c^{2}}$.
Together with the pairs of photons coming from $\pi^{0}$ decay (branching ratio equal to 0.988) the invariant mass spectrum is also populated with all other pairs of clusters detected in the BEMC. These combinations makes the measurement of the $\pi^{0}$ yield more difficult, and need to be studied and treated. As we will see in detail in the next sections, three main background contributions can be considered: $\eta$ decay, \emph{combinatorial background} and \emph{hadronic background}.

\subsection{BEMC cluster finding algorithms}

The goal of a cluster finding algorithm (cluster finder) is to identify a signal from a particle within a list of towers with some energy deposition. The basic idea is to collect groups of adjacent cells and combine them to create a two dimensional profile reflecting the position and the energy deposition of the electromagnetic shower. The resulting cluster of towers is composed of a peak, the tower with the highest energy deposition, plus usually some less energetic towers neighboring the peak. The STAR software framework provides a cluster finder algorithm which applies to BEMC tower and pre-shower signals, as well as for each of the two SMD layers. However, as mentioned before, the Shower Maximum Detectors are not used in this analysis for practical (SMD signals are not included in the fast stream of data) as well as for physical reasons (SMD is not optimized for low transverse momentum signals). For this reason the default cluster finder has been modified and optimized for working with tower signals only.

\begin{description}
\item[BEMC default cluster finder.]
The default STAR clustering algorithm works as follows. For each module, hits with an energy deposition above a certain threshold $E_{seed}$ are selected as peak candidates. Starting from the most energetic of these seeds, a cluster is created around it by adding hits which are adjacent\footnote{For tower clusters, two towers are to be considered \emph{adjacent} when they share a side and not just a corner. For one-dimentional ($\eta$ and $\phi$) SMD clusters the definition is trivial.} to the peak itself and present an energy deposition smaller than the peak, but above a threshold $E_{add}$. Neighbors are recursively added to the cluster until a pre-defined maximum cluster size $N_{max}$ is reached. At the end, the total cluster energy is compared to a third threshold $E_{min}$ with the goal of rejecting low energy clusters. After this, the clustering process is iterated around the next seed. Table \ref{tavolasoglia} shows the default thresholds for the four sub-detectors, while in Figure \ref{figuracluster} some examples of cluster assignment are illustrated. The two rightmost examples in this figure show how the algorithm can split a cluster into two adjacent ones, using the energy sorted ordering for association of the neighbors. However, statistical fluctuations in the energy of a hit can cause single photon signals to be erroneously split. When this happens, the cluster splitting becomes a source of background. 

\begin{table}\begin{center}
\begin{tabular}{|c|c|c|c|c|}
\hline detector & $E_{seed}$ [GeV] & $E_{add}$ [GeV] & $E_{min}$ [GeV] & $N_{max}$\\
\hline\hline towers & 0.35 & 0.035 & 0.02 & 4 \\
preshower & 0.35 & 0.035 & 0.02 & 4 \\
SMD$_{\eta}$ & 0.02 & 0.0005 & 0.1 & 5 \\ 
SMD$_{\phi}$ & 0.02 & 0.0005 & 0.1 & 5 \\
\hline 
\end{tabular}\caption{Default cluster finder thresholds.}\label{tavolasoglia}\end{center}
\end{table}

Once the cluster is finalized, the energy and the hit position of the incident photon can be estimated. The position of the cluster in $\eta$, $\phi$ coordinates is calculated as the energy weighted mean position of the hits composing the cluster. The geometrical center of each detector element is used as the hit position.
\begin{displaymath}
E=\sum_{i}E_{i}\qquad , \qquad \eta=\frac{\sum_{i}\eta_{i}\cdot E_{i}}{E} \qquad , \qquad \phi=\frac{\sum_{i}\phi_{i}\cdot E_{i}}{E}
\end{displaymath}

Once all the two-dimensional tower clusters and the two one-dimensional SMD clusters are found, they are combined into BEMC points by correspondence in position. A point is required to have at least a tower cluster. This provides the energy of the cluster and a rough estimate of the position. Additional information may come from SMD clusters associated to the tower cluster. If this is available, SMD provides a more precise measurement of the position of the shower, thanks to its finer resolution. Moreover, in cases where one tower cluster is associated with two SMD clusters, the SMD allows us to divide the energy of the towers and to reconstruct two showers by splitting the tower cluster.

\begin{figure}
\begin{center}
\includegraphics[width=1\textwidth,clip=]{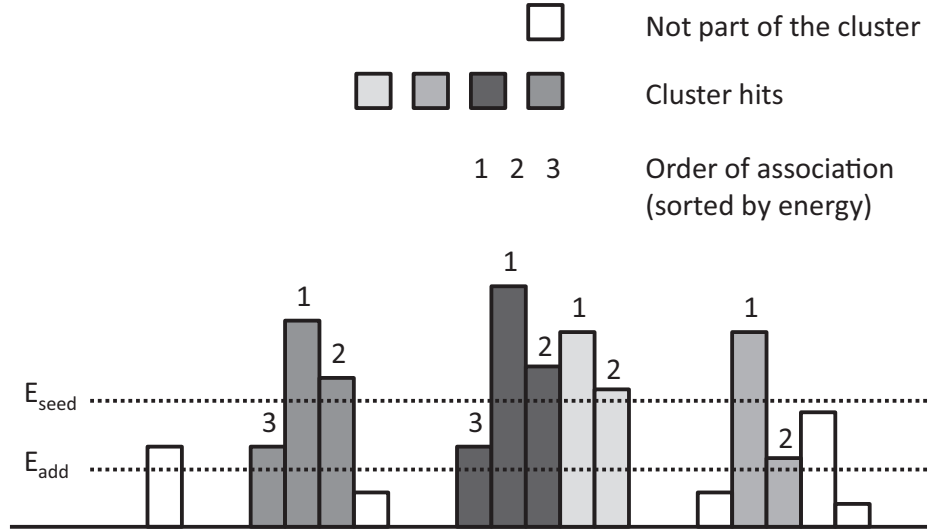}
\end{center}
\caption{Examples of cluster topology using the default BEMC cluster finder.}\label{figuracluster}
\end{figure}

\begin{figure}
\begin{center}
\includegraphics[width=0.8\textwidth]{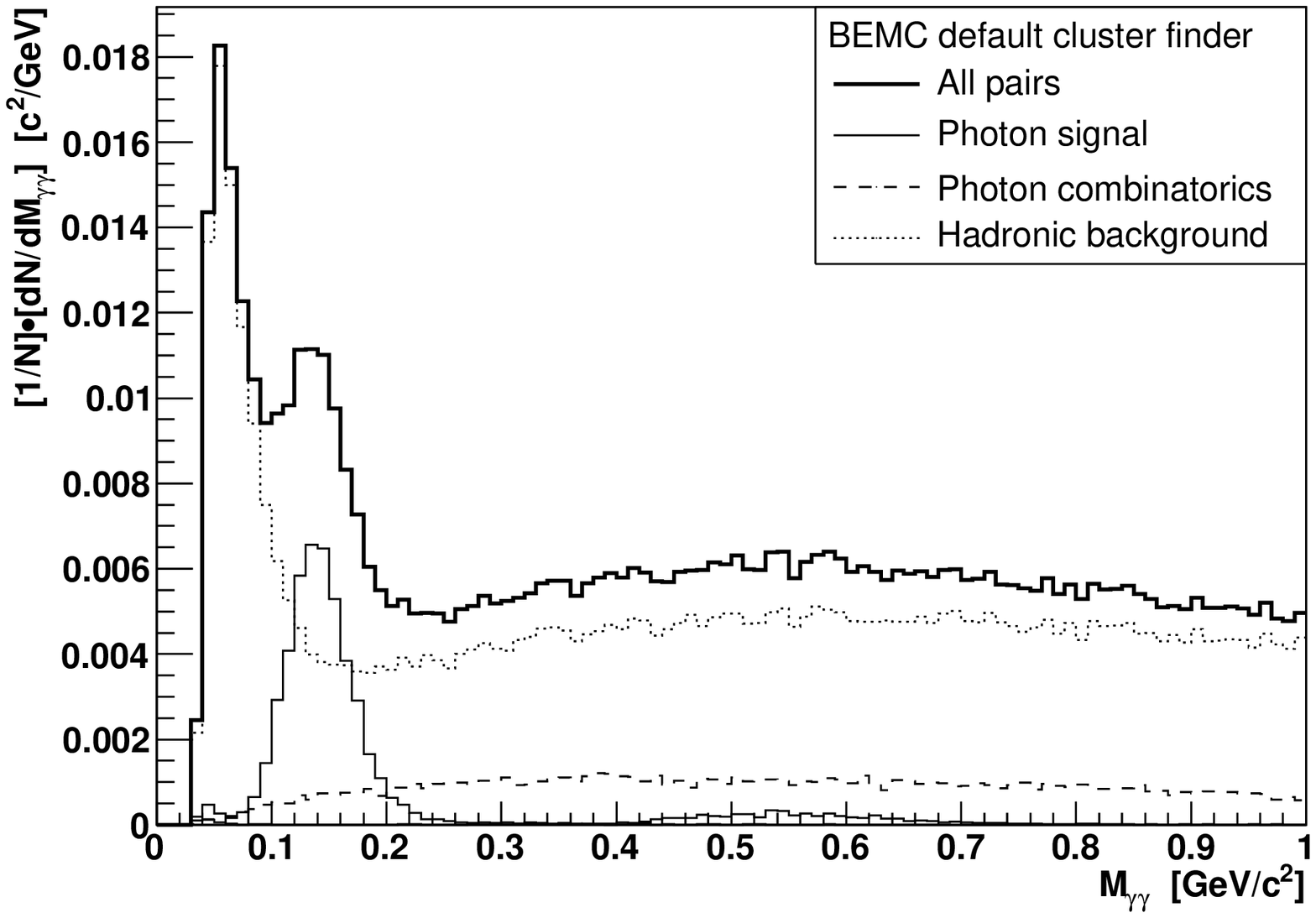}
\end{center}
\caption{Invariant mass distribution for pairs of clusters with $p_{T}>1.5\mathrm{\,GeV/c^{2}}$ in p+p collisions, simulated using PYTHIA and reconstructed using the default cluster finder.}\label{massadefault}
\end{figure}

\end{description}

Unfortunately, a series of difficulties occur when the default cluster finder algorithm is used in absence of SMD informations, as for this analysis. The crucial point is that the SMD provides a finer spatial resolution which allows the algorithm to split a group of towers into smaller clusters. This is important because of the large size of the BEMC tower, which may easily include the whole shower and additional background. Not using the SMD detectors translates into a poorer resolution, both in energy and in position. This leads photon signals to be merged or blurred into clusters with other signals, which generates a large background in the pair invariant mass spectrum used to determine neutral pions. Figure \ref{massadefault} shows the invariant mass spectrum for pairs of clusters in simulated (PYTHIA) p+p events, using the default BEMC cluster finder. Different line styles represent the different contributions to the distribution. This is possible by performing an association analysis on Monte Carlo events, where information about the nature of the simulated particles are available. Each reconstructed cluster is associated with the simulated track pointing to it. When more tracks hit the region where the cluster lies, the one which provides the largest fraction of energy to the cluster is associated to it. The overall histogram (the bold black line) presents two clear peaks and a broad background. The photon pair signal is composed by a peak centered around $M_{\pi^{0}}=0.135 \mathrm{\,GeV/c^{2}}$, representing the neutral pion contribution, and another smaller peak representing the $\eta$ meson contribution. The signal is composed by pairs of clusters associated with photons decaying from the same pion. The broad background is composed of two components: combinations of photons belonging to different mothers (\emph{combinatorial background}, the dashed line) and hadron-photon pairs (\emph{hadronic background}, the dotted line). At very low values of the invariant mass we can see the peak of what is called the \emph{low-mass background}. This is a background component of the invariant mass spectrum generated by erroneously split clusters, whose effect is to enhance the yield of pairs with minimal angular separation (and, hence, minimal invariant mass). Most of this background appears to be originated by erroneous splitting of hadronic showers. 

In order to reduce this background, it is convenient to select smaller clusters with a more peaked energy profile, in which the leading signal is better represented uncertain situations are avoided. The default thresholds of Figure \ref{tavolasoglia} (optimized for a cluster-finder which includes SMD information) have been tuned toward higher values ($E_{seed}=0.5\mathrm{\,GeV}$, $E_{add}=0.1\mathrm{\,GeV}$). This significantly reduces the background in the invariant mass spectrum, especially at low values of the transverse momentum $p_{T}$ of the pion candidate. At higher $p_{T}$ values ($\sim 3$ GeV) it becomes again more difficult to disentangle signals close to each other, and the gain in resolution is not so significant. However, for small clusters (i.e. cluster composed by 1 or 2 cells) the background reduction is still effective. These studies lead to a different algorithm.

\begin{description}

\item[BEMC modified cluster finder]
The developed cluster finder for the BEMC towers is been designed for working without the SMD detectors. The idea is to reduce the impact of different signals merged into the same cluster by reducing the size of the cluster. This is achieved by performing the same cluster finding algorithm described before, but associating only the peak tower to the cluster. In this way, the reference for energy and position of the reconstructed photons is given only by the hits above the seed threshold $E_{seed}$.  Operatively, this translates into setting the threshold $E_{add}$ for neighboring hits to a very high value (for example: $E_{add}\geqslant E_{seed}$). Moreover, in order to avoid neighboring hits with large energy to be considered as new clusters seeds, they need to be removed from the list of hits. This step has a double side effect. On one hand, it helps removing the so called \emph{low-mass background}. On the other hand, the removal of neighbor towers translates into an effective cut of two tower widths on the minimal distance between the pairs of clusters. The effect of the new algorithm on the clusters of Figure \ref{figuracluster} is shown in Figure \ref{fmodified}. 

\begin{figure}
\begin{center}
\includegraphics[width=1\textwidth,clip=]{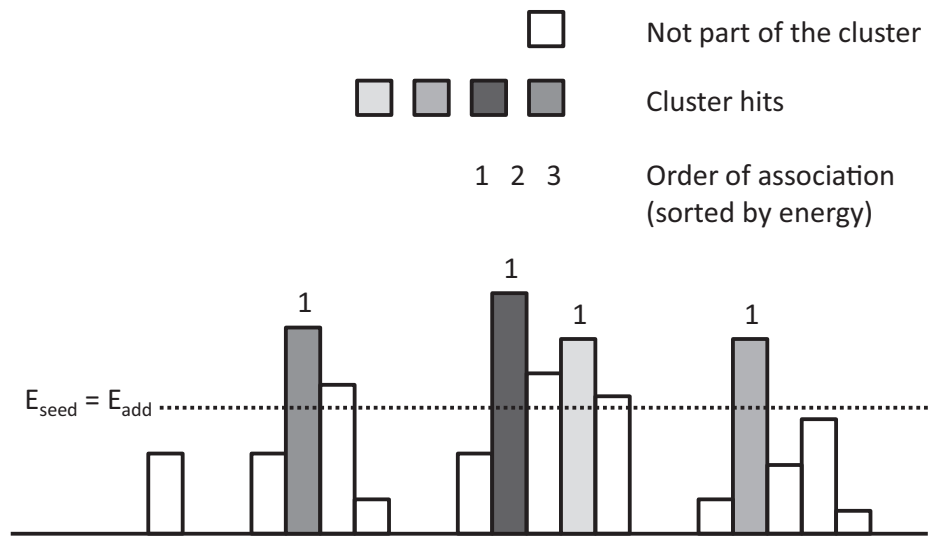}
\end{center}
\caption{Examples of cluster topology using the modified BEMC cluster finder with $E_{add}=E_{seed}$.}\label{fmodified}
\end{figure}

In order to make the selection even more effective and to remove the last traces of low-mass background, the algorithm that checks the neighboring hits has been modified in order to identify and exclude hits touching the peak tower only with a corner (diagonal neighbors). Finally, the requirement for a cluster to be confined within one module has been removed too. These modifications lead the algorithm to remove from the list of seed candidates a ring of 8 towers around each peak. As already said, this implies a cut on the minimal distance between clusters, but at the same time it reduces dramatically the possibility of wrong cluster merging which would happen due to the absence of the SMD. This benefits especially large clusters from hadronic shower which may cover a higher number of cells and could be easily mistaken for multiple signals by the default finder. The effect of the modified algorithm on the invariant mass spectrum is shown in Figure \ref{massamodified}. The background reduction along the whole spectrum is clear, when compared to the default algorithm. Moreover, there is no low-mass background due to the cut on the minimal angle.

\begin{figure}
\begin{center}
\includegraphics[width=0.8\textwidth]{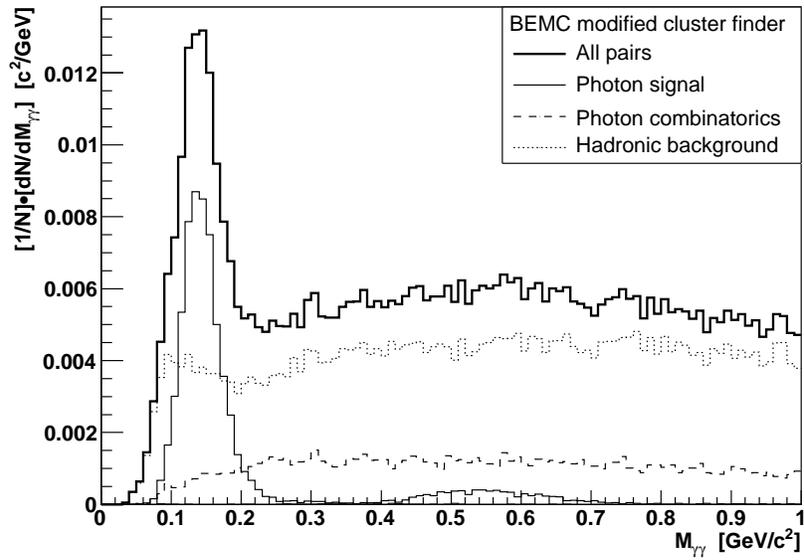}
\end{center}
\caption{Invariant mass distribution for pairs of clusters with $p_{T}>1.5\mathrm{\,GeV/c^{2}}$ in p+p collisions, simulated using PYTHIA and reconstructed using the modified cluster finder.}\label{massamodified}
\end{figure}

\end{description}

\subsection{Background treatment}\label{3.backgroundtreatment}
Once the invariant mass spectrum for pairs of cluster has been produced, one needs to estimate  the contribution of pairs of photons into which the $\pi^{0}$ decayed. To do so, one selects a mass window centered around the nominal value of the pion mass. However, as it is clear from Figure \ref{massamodified}, also in this area the contribution from the background is relevant and the possibility of considering a background pair as a pion signal is not negligible. In order to reduce this effect, one needs to select the proper mass window which maximizes the ratio between pion signal and background. Moreover, one can try to reduce the background itself by filtering out the particles that create it. A thorough understanding of the different components of the invariant mass spectrum is therefore necessary.
\begin{figure}
\begin{center}
\includegraphics[width=0.7\textwidth]{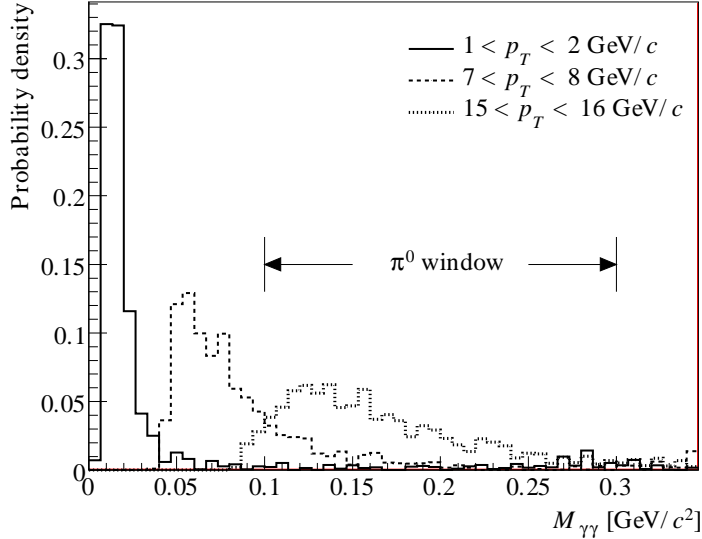}
\end{center}
\caption{Low mass background contribution in simulated single photon events, for different $p_{T}$ selections. Taken from \cite{Oleksandr}.}\label{lowmass}
\end{figure}

\textbf{Low-mass background.} As we discussed before, a easily identifiable component of the spectrum is represented by the \emph{low-mass background}. This is caused by wrong cluster splitting that creates a fake pair of photon candidates, characterized by a small separation and thus a small invariant mass. We have seen that, by applying an isolation cut, this component can be nearly eliminated. However, some care needs to be taken while dealing with the low-mass background. At a fixed opening angle $\psi$ between the two photon candidates, in fact, the invariant mass $M_{\gamma\gamma}=\sqrt{E_{1}E_{2}(1-\cos{\psi})}$ grows with the energy of the two photons or, analogously, with the transverse momentum of the parent. This means that the peak of the low-mass background moves towards higher values of $M_{\gamma\gamma}$ as the $p_{T}$ of the $\pi^{0}$ increases. This is illustrated in Figure \ref{lowmass}, taken from \cite{Oleksandr}, in which the amount of low-mass background is simulated for single photon events at different $p_{T}$. Although it is clear that the low-mass peak is moving towards the $\pi^{0}$ mass region, a significant contamination of the signal appears to happen only for neutral pions with a $p_{T}$ relatively high: $p_{T}\gtrsim 5$ GeV/c. This is significantly higher than the average value of the transverse momentum of the pions used for this analysis: $\langle p_{T} \rangle\lesssim 2$ GeV/c. For this reason, in the kinematical region of interest for this study, the pion peak is expected to be clearly separated from low-mass background, whose largest component at low $M_{\gamma\gamma}$ values  is removed by the effective opening-angle cut. 

\textbf{Combinatorial background.} Once the low-mass background is been cleared out, one can study the different components of the invariant mass spectrum by performing the association analysis. One of the main components in Figure \ref{massamodified} are pairs of clusters associated with photons which do not share a common mother. This is know as \emph{combinatorial background} since it reflects our blindness in pairing the correct photon candidates when trying to reconstruct the mother pion. It is not possible to remove this background component, since it is generated by the signal itself. However it is possible to reduce it by rejecting pairs of clusters with very different energy releases, by looking at the energy asymmetry between the two clusters: 
\begin{displaymath}
Z_{\gamma\gamma}=\frac{|E_{1}-E_{2}|}{E_{1}+E_{2}}.
\end{displaymath}
Neutral pions are in fact expected to decay into photons with an uniform asymmetry distribution. However, in a typical event, the number of soft pions is much larger than that of hard ones (the $p_{T}$ spectrum is falling exponentially). When pion candidates are created by combining inclusively pairs of clusters, the probability of picking an asymmetric pair of photons is enhanced by the higher number of low energy clusters. In order to reduce this effect, we apply a cut in the energy asymmetry distribution and reject pairs with $Z_{\gamma\gamma}>0.7$.     

\textbf{Hadronic background.} The second and largest background component in the $M_{\gamma\gamma}$ spectrum in Figure \ref{massamodified} is composed of pairs of clusters where (at least) one of the two hits is not associated with a photon. Most of the entries come from hadron-photon pairs, with the biggest hadronic contribution by charged pions, protons and neutrons. Hadronic showers are in fact expected to leave some energy deposition in the calorimeter as well. They usually have a significantly smaller energy deposition and a broader spatial distribution, both longitudinal and transverse, than electromagnetic showers. Because of this clear topological difference, it is usually not difficult to isolate the signal from electromagnetic showers from the background provided by hadron showers. A number of techniques for shower discrimination have been considered for this analysis. However, most of them rely on the use of slow detectors whose informations are not available in the fast stream of data. These include the use of the TPC as a charge particle veto (a BEMC cluster is rejected when there is a TPC track pointing towards it) and the use of the Barrel Shower Maximum Detector for discriminating electromagnetic showers where their development is maximum (around 5.6 radiation lengths, where the SMD is placed; on the contrary, hadrons develop a broader shower with a maximum at the end of the calorimeter). Similarly, the Barrel Pre-Shower Detector could be used for shower discrimination. This is made of the first two layers of scintillating materials of the BEMC towers, and it is therefore a fast detector. However its read out has been delivered to the slow stream of data, making its use not convenient for this analysis.

\begin{figure}
\begin{center}
\includegraphics[width=1\textwidth]{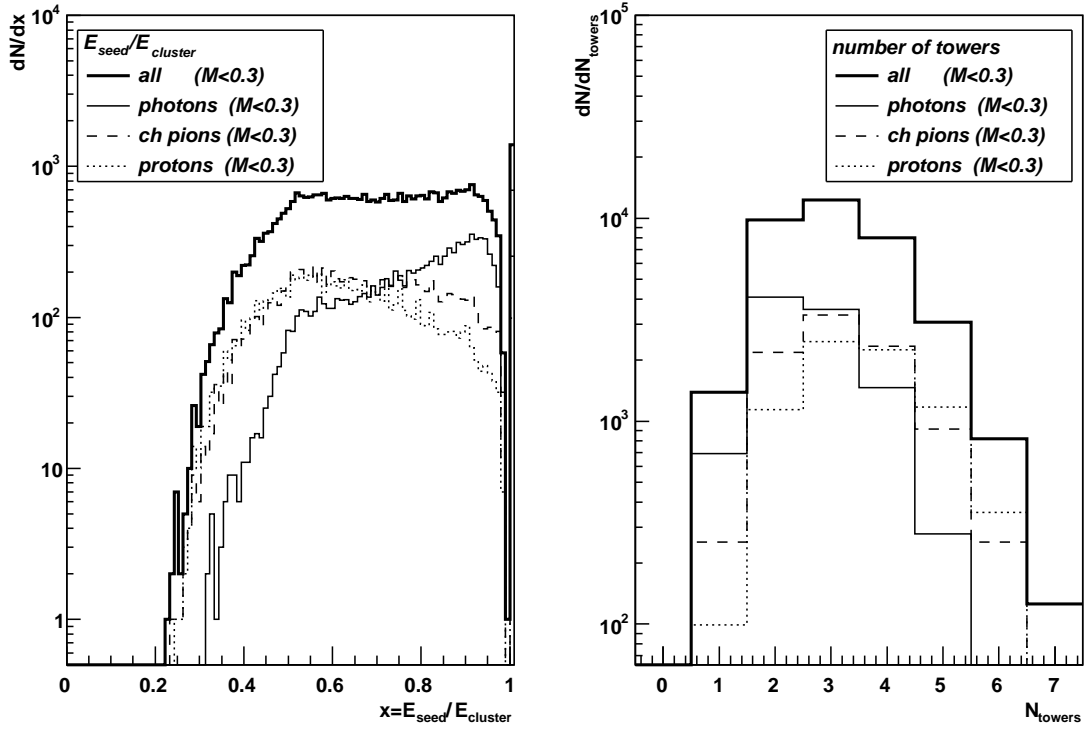}
\end{center}
\caption{Distributions of $x=E_{seed}/E_{cluster}$ (left) and number of towers (right) of the largest cluster for all pairs of clusters with $M_{\gamma\gamma}<0.30\mathrm{\,GeV/c^{2}}$ and $p_{T}>1.5\mathrm{\,GeV/c}$ in p+p simulation. The total distributions (bold) are displayed together with the contributions of the $\pi^{0}$ signal (solid) and of the background pairs where the largest cluster is either a charged pion (dashed) or a proton (dotted lines).}\label{qualitytower}
\end{figure}

Another attempt involved the use of the Central Trigger Barrel (CTB). The CTB consists in 240 scintillator slats arranged around the TPC, covering the same $\eta$-$\phi$ region than TPC and BEMC. It is a fast detector designed to respond to charged particle signals for trigger purposes. For this reason it can be used as a veto for charged hadrons by rejecting clusters in correspondence to a CTB hit. Unfortunately, during run-8, the granularity of the CTB was heavily reduced because of shortage of read-out modules. In particular, the read-outs of the 4 modules segmenting the CTB along the $\eta$ direction were merged. Because of the consequent poor resolution of the CTB, the veto capabilities are heavily affected and random rejections of BEMC clusters are not excluded. As a result, the CTB veto reduces the background level only marginally, while the signal is also affected because of the poor CTB resolution.

\begin{figure}
\begin{center}
\includegraphics[width=0.8\textwidth]{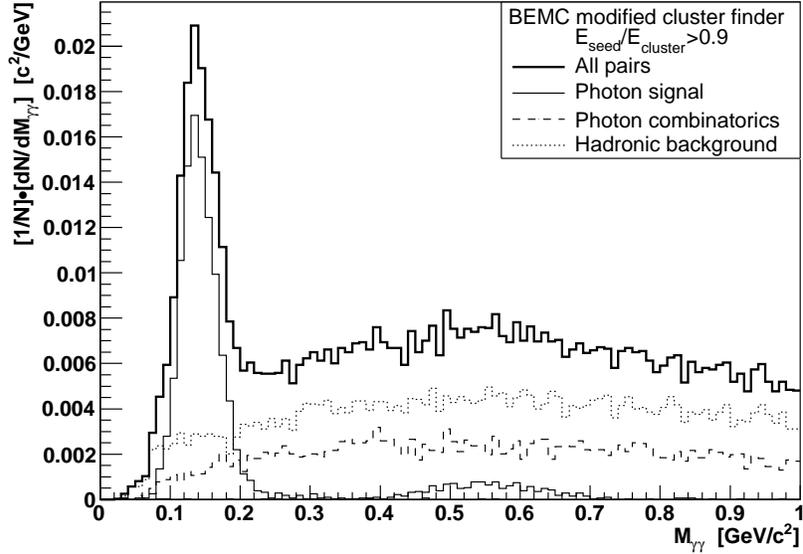}
\end{center}
\caption{Invariant mass distribution for pairs of clusters with $p_{T}>1.5\mathrm{\,GeV/c}$ in p+p collisions, as reconstructed using the modified cluster finder with additional requirement on the peak tower over cluster energy ratio $E_{seed}/E_{cluster}>0.9$.}\label{massaquality}
\end{figure}

Since using information from other detectors to reduce the background component proved to be not convenient or not effective enough, the only possibility left is to go back to the cluster finder. It is in fact possible to relax the energy thresholds in the algorithm and try to characterize the shower shape in the transverse direction. Electromagnetic showers are expected to have a more compact shape, mostly contained into one BEMC tower. On the contrary, hadronic showers present a much larger transverse profile. Once the $E_{add}$ threshold in the modified cluster finder has been lowered, neighboring cells are allowed to contribute to the cluster. In this way it is possible to characterize the cluster with additional information about the transverse widening, such as number of cells and amount of the energy of the seed tower with respect to the total cluster energy. A pion candidate is then associated with the largest cluster of the pair.  Figure \ref{qualitytower} shows the distributions of these parameters for different species of particles. It is clear from this how photon generated showers present a smaller number of BEMC hits and a narrower transverse distribution (the peak tower gets most of the energy of the shower). Based on this, one can improve photon identification by applying a quality cut on the clusters. Figure \ref{massaquality} shows how the invariant mass spectrum changes with a cut on the energy ratio between the seed tower and the total cluster. The seed-over-cluster energy ratio that optimizes the ratio between signal and background, and rejects the smallest number of good photon clusters, is found to be $E_{seed}/E_{c}>90\%$. 

By comparing the background reduction we can achieve by using the four different methods (default cluster finder, modified cluster finder with isolation cut, modified cluster finder with CTB veto, modified cluster finder with quality cut), we conclude that the last provides the best results. Results are summarized on table \ref{summaryf}. Note that signal and background in the study using the CTB (on data) have been separately scaled in order to make the two components compatible with simulation (which, as we will see later, differs drastically from data).

\begin{table}\begin{center}
\begin{tabular}{|c|c|c|c|c|c|}
\hline algorithm & additional cut & $ E_{seed}$ [GeV] & $E_{add}$ [GeV] & $N_{max}$ & S/B \\
\hline\hline default &  & 0.350 & 0.035 & 4 & 0.55 \\
modified & & 0.500 & 0.500 & 9 & 1.00 \\
modified & CTB veto & 0.500 & 0.500 & 9 & 1.33 \\
modified & quality cut & 0.500 & 0.070 & 9 & 2.07 \\
\hline 
\end{tabular}\caption{Summary of the BEMC finding algorithms. Signal over background ratios for simulated p+p interaction with $p_{T}>1.5\mathrm{\,GeV/c}$ are indicated. The ratio for the modified finder with additional CTB cut has been extrapolated from data.}\label{summaryf}\end{center}
\end{table}

\section{FMS: neutral meson reconstruction}
In the FMS pions are similarly identified by selecting pairs of photon candidates in the FMS which present an invariant mass  $M_{\gamma\gamma}=\sqrt{E_{1}E_{2}(1-\cos{\psi})}$ close enough to the nominal value of the $\pi^{0}$ mass. Also the algorithm used to group cells into clusters, representing photon signals, is mostly similar to what we have previously discussed. However, the geometry of the FMS and its position in the far forward region at STAR slightly changes the approach. The typical separation distance $d_{\gamma\gamma}$ between two photons originating from a $\pi^{0}$ decay, measured at the surface of the two detectors, is roughly comparable: $\langle d_{\gamma\gamma}\rangle\approx 10 \mathrm{\,cm}$. In fact, while the pions detected in the FMS have, on average, larger energy (and therefore smaller opening angle), the distance of the FMS from the interaction point is much larger than the BEMC, so that the two effects compensate. The FMS granularity is, however, higher than in the BEMC (in cases where no SMD is used), and this makes it much easier to distinguish the two photons and it allows a more detailed reconstruction. A BEMC tower covers an area in $\Delta\phi\times\Delta\eta$ of 0.05$\times$0.05, which translates to an average transverse size of 10$\times$10 cm$^2$ (smaller at mid-rapidity and increasing at the edges of the detector). As comparison, a large FMS lead-glass cell measures 5.8$\times$5.8 cm$^2$, while a small one measures 3.8$\times$3.8 cm$^2$. This means that a typical electromagnetic shower is usually fully confined within a BEMC tower, while it stretches along few FMS towers, allowing a better measurement of its shape. 

The parameterization of the transverse shower profile is done using a well tested method \cite{Lednev1995292}, already used for the FPD/FPD++ and ported into the FMS reconstruction code. 
The method applies to the reconstruction of photons in homogeneous electromagnetic calorimeters (like lead-glass), when the particles are within a reasonably good approximation hitting the detector perpendicular to its front surface \cite{Binon198686}. This is the case for the FMS because of its large distance from the vertex. On the transverse plane, the fine granularity of the FMS array allows a quite accurate description of the shower profile. The first step is to draw a two-dimensional cumulative distribution $F(x,y)$ (normalized to 1) to fit the transverse energy density distribution
\begin{equation}\nonumber
F(x,y)=\frac{1}{2\pi}\sum_{i=1}^{3}a_{i}\left(\arctan{\frac{xy}{b_{i}\sqrt{b_{i}^{2}+x^{2}+y^{2}}}}\right)
\end{equation} 
with the parameters 
\begin{displaymath} 
\begin{array}{l l l} 
a_{1}=0.80 & a_{2}=0.30 & a_{3}=-0.10  \\
b_{1}=8.0 \mathrm{\,mm} & b_{2}=2.0 \mathrm{\,mm} & b_{3}=76 \mathrm{\,mm} \\
\end{array}
\end{displaymath} 
The energy deposition in each cell is calculated from the value of the cumulative function at the cell corners:
\begin{eqnarray}
G(x,y)&=&F\left(x+\frac{d}{2},y+\frac{d}{2}\right)-F\left(x+\frac{d}{2},y-\frac{d}{2}\right)+\\\nonumber
&\phantom{=}-&F\left(x-\frac{d}{2},y+\frac{d}{2}\right)+F\left(x-\frac{d}{2},y-\frac{d}{2}\right) \nonumber
\end{eqnarray}
where $d$ is the cell size, and $(x,y)$ are evaluated at the center of the cell. It should be noted that the cumulative function that describes the transverse energy profile of the shower is independent of the cell size. Moreover, the Moliere radius $R_{M}$, which gives the transverse dimension of the electromagnetic shower in a cell of a given material, is similar between large and small cells. For these reasons, the same cumulative function can be applied to both large and small FMS towers.

In addition, an asymmetry in the x-y shower profile has been added, while porting the algorithm from the FPD++ to the FMS version, to meet the asymmetrical configuration in the small cell module geometry, due to the addition of plastic spacing layers between the rows of cells. The approximation of the hit position with the center of the cell leads to a small discrepancy in the reconstructed position of photons hitting the edge of the cell, given the fact that the energy distribution is fitted.

\subsection{FMS finding algorithms}

The shower shape profile just described is used in the FMS photon finding algorithm to fit the energy deposition in a cluster of cells. From the result of the fit, one determines an estimate of the energy deposition and the $(x,y)$ hit position of the photon. This procedure is simple for relatively low energy pions ($E_{\pi}<30$ GeV), where the decay photons hit the detector in two relatively distant positions. In this case, the two signals appear as two separated peaks that can be fitted individually. On the contrary, when the energy of the pion is higher ($E_{\pi}\simeq50$ GeV), the hit position of the two photons can be close enough to cause the two clusters to merge. In this case, one needs to separate the two contributions before being able to fit them. Obviously this procedure can prove to be difficult when a clear two-peak structure cannot be found. For this reason, a moment analysis on the structure of the cluster is implemented in the clustering algorithm to help distinguishing single photon clusters from two photon clusters.

\begin{figure}
\begin{center}
\includegraphics[width=0.8\textwidth]{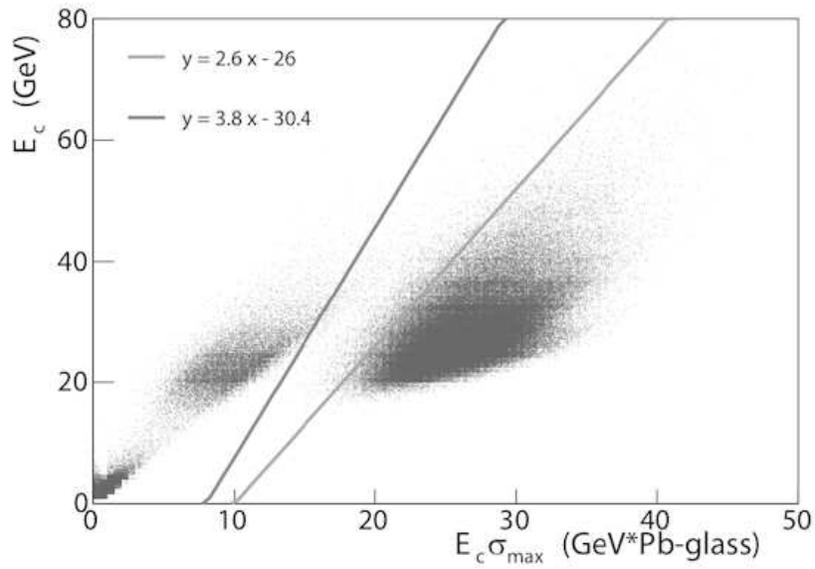}
\end{center}
\caption{Cluster distribution in the $E_{C}^{2}\sigma_{max}^{2}-E_{C}^{2}$ plane for simulated single pion events, as reconstructed in the FPD++ small cells by the default cluster finding algorithm. The two lines divide the plane into three regions in which single or double (or both) signal fit is tried. Figure from \cite{2004PhDT........87W}. }\label{yiquncategory}
\end{figure}

\begin{description}
\item[FMS default cluster finder]
In order to deal with events which may contain two merged photon signals, a more sophisticated algorithm has been developed \cite{2004PhDT........87W} than the one used for the BEMC. The algorithm uses the finer FMS granularity to find potential \emph{valleys} between two towers. A cluster is first created, similarly to the BEMC, by identifying a \emph{peak} tower, that is a tower with non zero energy deposition surrounded by towers with lower energy. Neighboring towers are incrementally added to it if they share a side (not just a corner) with the peak tower to form the cluster. This procedure is then iterated to include in the clusters additional towers with lower and lower energy deposition. The final (transverse) energy density shape of the cluster does not present, in this way, any substructure. However, when two peaks are relatively close to each other (say: one tower distant to each other), the cells in between the two look like a valley. In general, this valley tower is added to the cluster of which peak tower is closest to the valley itself. In case of multiple equidistant peaks, the valley is associated to the partial cluster with highest energy. In this way, groups of adjacent cells which present a multiple peak structure are split in pieces, each of them representing a single photon signal. For this default cluster finder, the definition of a cluster then reads: a cluster is a group of cells with non zero energy deposition, topologically connected in the column-row two-dimentional lattice space, whose transverse energy distribution does not present points of local minimum (valleys). An energy threshold of $E_{th}=2$ GeV is applied to all so defined clusters to separate real photon signals from statistical fluctuations or MIP energy deposits. 

There are however cases where two photons hit the calorimeter in two nearby positions and create a single cluster with no evident two-peaks structure. When this happens, no valley is found, the cluster is not split and a single photon fit would be erroneously applied. To avoid this, a moment analysis is performed to each cluster and used to characterize the transverse spreading of the energy distribution. Single photon clusters are most likely to have a small transverse size while  clusters containing two photon signals are expected to be larger in size. One can characterize the collection of towers in a cluster by computing its first and second moments which provide, respectively, the position of the center of gravity and information about the orientation of the cluster. 

\begin{eqnarray}\nonumber
&E_{C}&=\sum_{i}E_{i} \\\nonumber
&x_{0}&=\frac{\sum_{i}x_{i}E_{i}}{E_{C}}\qquad y_{0}=\frac{\sum_{i}y_{i}E_{i}}{E_{C}}\\\nonumber
&\sigma^{2}_{xx}&=\frac{\sum_{i}E_{i}(x_{i}-x_{0})^{2}}{E_{C}}\qquad  \sigma^{2}_{yy}=\frac{\sum_{i}E_{i}(y_{i}-y_{0})^{2}}{E_{C}} \\\nonumber
&\sigma^{2}_{xy}&=\frac{\sum_{i}E_{i}(x_{i}-x_{0})(y_{i}-y_{0})}{E_{C}}
\end{eqnarray}     
One can evaluate the maximum spread along the long axis of the cluster by diagonalizing the $2\times2$ matrix of the $\mathrm{2^{nd}}$-moments, from which one obtains:
\begin{equation}\nonumber
\sigma_{max}^{2}=\frac{\sigma^{2}_{xx}+\sigma^{2}_{yy}+\sqrt{(\sigma^{2}_{xx}-\sigma^{2}_{yy})^{2}+4(\sigma_{xy}^{2})^{2}}}{2}
\end{equation}

It can be shown \cite{2004PhDT........87W} that the quantity $E_{C}^{2}\sigma_{max}^{2}$ for a cluster formed by two photon contributions is related to the invariant mass squared $M_{\gamma\gamma}^{2}$ of the pair of photons. On the other hand, if the cluster is generated by a single photon, $\sigma_{max}^{2}$ follows a different dependence on $E_{C}$. This means that we can catalog clusters by looking at the region they populate in the $E_{C}^{2}\sigma_{max}^{2}-E_{C}^{2}$ plane. In Figure \ref{yiquncategory}, obtained for the FPD++ small cells, we can clearly distinguish two major areas, one with a higher $\sigma_{max}^{2}$ than the other, relative to the cluster energy $E_{C}$. The plane is therefore divided into three regions that correspond to different classes of clusters: one-photon clusters, two-photon clusters and ambiguous clusters. On the first class (relatively narrow clusters) only a single photon fit is applied to determine energy and hitting position of the particle, using the shower shape function described before. On the second class (broader clusters) a two-photon fit is applied, using a double one-photon fit function and assuming the two signals are generated from photons originated from the same pion decay. This allow us to correlate the fitting parameters and have a better description of the shower shape. In the third class of ambiguous clusters both fits are tried and the best one is chosen based on the $\chi^{2}$ value of the two fits. The same method has been ported from the FPD++ and adapted to the FMS. The parameters of the lines dividing the space in three areas have been optimized for the FMS response. 
\end{description}

\begin{figure}
\begin{center}
\includegraphics[width=1\textwidth]{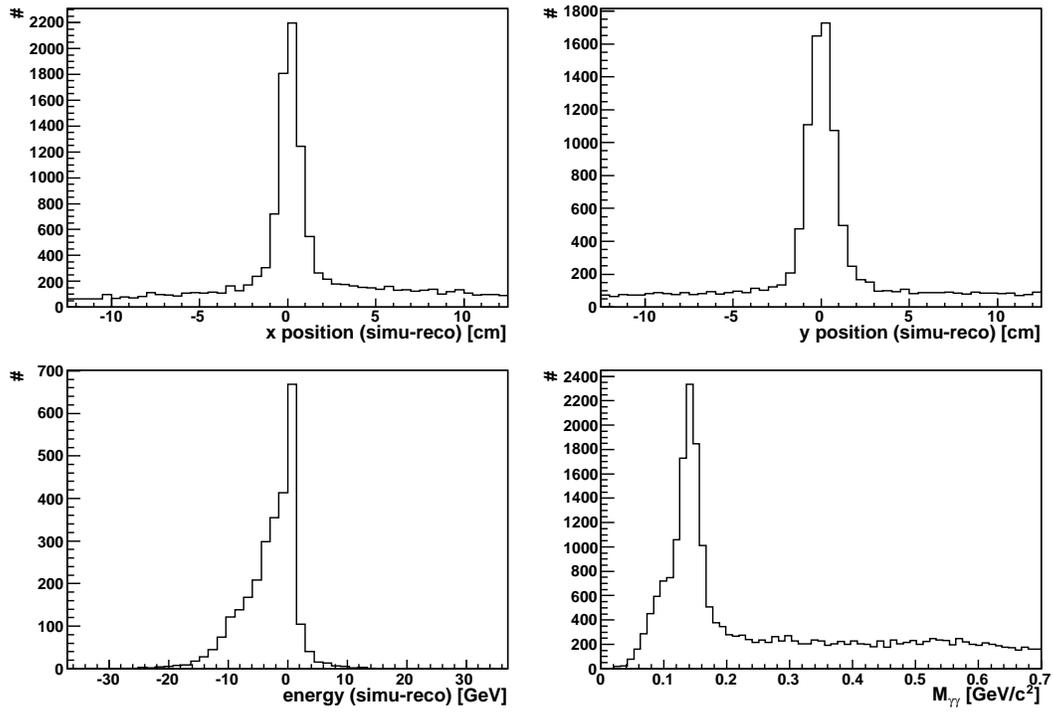}
\end{center}
\caption{FMS default cluster finder. Comparison of position and energy between simulated and reconstructed neutral pions in a full PYTHIA simulation. On the bottom right panel: invariant mass for all pair of reconstructed clusters}\label{FMSfinder}
\end{figure}

The FPD++ original cluster finder proved to be very reliable in estimating energy and position of the photon signals in the FMS (Figure \ref{FMSfinder}), even when the two merge together. However, uncertainties remain in the splitting procedure which precedes the moment analysis. The sharp cut applied to the two peak structure of the double photon cluster, in fact, does not allow a perfect description of the tail of the energy distribution. This is more important when the energy of the photons increases and the contamination in the valley becomes larger. In order to study this effect and its relevance on the finding process, a modified cluster finder has been developed.

\begin{description}
\item[FMS modified cluster finder]
To ensure a better representation of the two-photon shower profile in the contamination region between the two peaks, a simplified version of the cluster finder has been tested. In this version, the splitting procedure has been removed from the FMS cluster finder. A cluster is then defined as a group of non-zero adjacent towers. No matter if the cluster presents a two-peak structure, clear sign of a double signal, the split is not performed. On these clusters, formed as for the other version of the algorithm, the moment analysis is applied and the three categories are created (this time with different parameters). The three-options fitting procedure described before, is applied to the three classes of clusters. Large clusters with a double peak structure populate the region of the plane where the two-photon fit is applied. 
\end{description} 

A comparison between the two algorithms has been performed. The reconstruction efficiency and the resolution in position and energy of the reconstructed photons have been studied for single pion events simulated with GEANT \cite{Agostinelli2003250}, and for full PYTHIA \cite{Sjostrand:2001yu} events (with detector response from GEANT). In general, both algorithms perform very well in the reconstruction of a pair of photons. While the position resolution is comparable between the two algorithms (in both cases it is consistent with half cell width), the default cluster finder has a slightly higher accuracy in reconstructing the energy of single photons and pions (the spread around the simulated value is: $\sigma_{E}=0.79\mathrm{\,GeV}$). This appears to be true for both large and small cells. In the energy range of interest for this analysis, in addition, the default algorithm proves to better measure the separation distance of the photon pair at the detector surface ($\sigma_{d}=3.10\mathrm{\,cm}$) and the energy asymmetry between the two. As a consequence, also the mass resolution for reconstructed photon pairs is better for the default algorithm (the spread around the simulated value is: $\sigma_{M}=0.05\mathrm{\,GeV/c^{2}}$). The parameters used to associate each cluster to a topology class (single photon, double photon or ambiguous clusters) are separately optimized for the two finding algorithms. However the separation between the different regions in the  $E_{C}^{2}\sigma_{max}^{2}-E_{C}^{2}$ plane is clearer for the default algorithm in both single pion and full PYTHIA events. For all these reasons, the default algorithm has been used through all this analysis.

\subsection{Dead cell prescription}
An additional study has been performed in order to understand the impact of a dead cell in the cluster finding algorithm. A hole in the FMS matrix can happen because of multiple reasons: PMT failing  or not responding as expected, bad electrical connections, broken optical couplings. For this reason, a map of ``good'' channels is continuously updated to mask out ``bad'' channels from the reconstruction chain. This has however an impact on the cluster algorithm, since holes in the matrix can cause loss of clusters or erroneous cluster splitting, and can interfere in the reconstruction of energy and position of clusters near (or involving) the dead cell. 

\begin{figure}
\begin{center}
\includegraphics[width=1\textwidth,clip=]{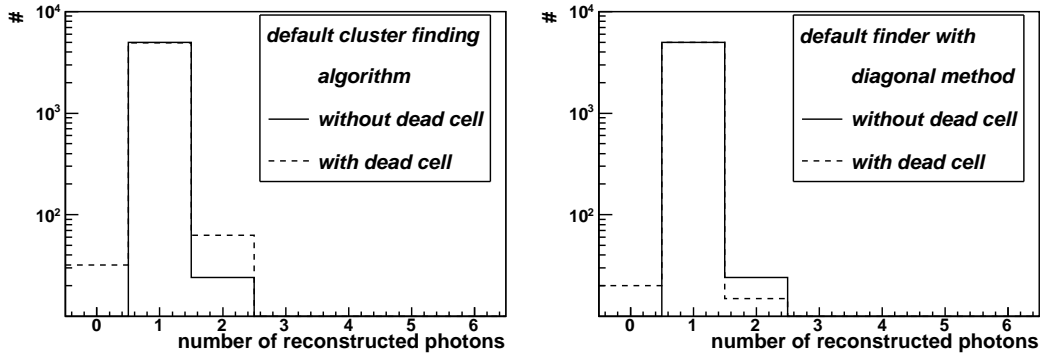}
\end{center}
\caption{Number of photons detected (in solid lines) within a sample of 5000 single photon events, simulated to hit a particular position in the FMS. In dashed lines, the effect of the removal of this cell from the matrix (to emulate a dead cell) on the reconstruction, as performed by the default finding algorithm (left) and with the additional ``diagonal prescription'' described in the text (right).}\label{deadcell1}
\end{figure}

To address this problem, the gain of a specific cell in the FMS array has been set to zero to emulate the dead-cell effect. Single photon events were simulated using GEANT and the reconstruction of such signals has been tested as a function of the distance between the center of the dead cell and the expected impact position of the photon on the FMS (estimated by projecting the photon track to the FMS surface). The results of this analysis show that a dead cell causes the following effects in the vicinity of the dead cell: loss of clusters when the main portion of the deposited energy would be in the dead cell, cluster splitting when the dead cell creates a fake valley in the group of cell that forms the cluster (Figure \ref{deadcell1}), degradation of the energy and position resolution for clusters affected by the dead cell. However, thanks to the high granularity of the FMS, this degradation only affects a limited region around the dead cell, for which a particular prescription in the cluster finder can be used to limit the effect.

When the dead cell corresponds to the peak of the energy deposition the energy of the photon is mainly lost, and so is the cluster. When the dead cell is instead slightly shifted from the center of the energy density distribution, the main effect is fake splitting of the cluster. To avoided this, a additional check on the towers adjacent to the peak cell can be performed before deciding if to split the cluster or not, in cases where the cluster contains the dead cell. This can be done either by performing a full eight-neighbors check all around the peak tower, or by limiting the additional check to the two towers adjacent to the dead cell and diagonal with respect to the tower to be checked. The first approach is similar to apply the modified cluster finder algorithm described before, which proved to be less reliable in normal cases. For this reason the second, less intrusive, approach is preferred. In general, by doing this additional check, a fake identification of a valley can be avoid when the two ``diagonal'' neighbors are not valleys themselves. In addition to the diagonal check prescription, one needs to remove the dead cell from the fit.

\begin{figure}
\begin{center}
\includegraphics[width=1\textwidth]{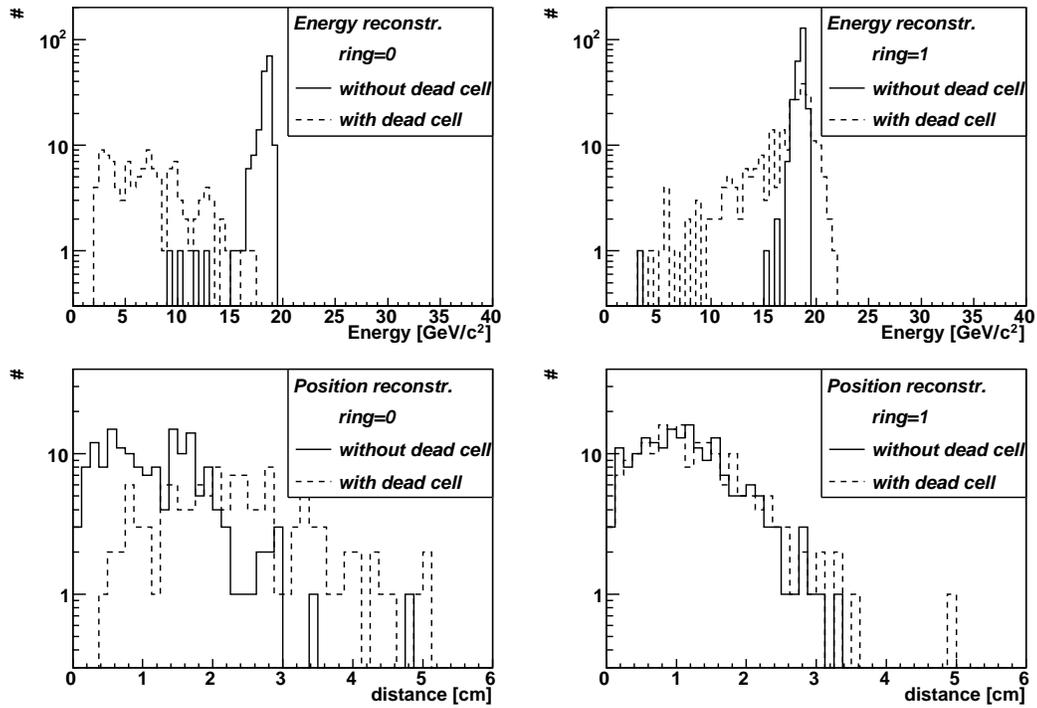}
\end{center}
\caption{Reconstructed energy (top) and distance from the ``real'' hit position (bottom) for a sample of 5000 single photon events (solid lines). The effect of a dead cell on the reconstruction algorithm, using the ``diagonal'' prescription described in the text, is indicated (dashed lines) for cases where the simulated photon hits the dead cell (ring 0, left) or the first ring of cells around it (ring 1, right).}\label{deadcell2}
\end{figure}

This ``diagonal'' check highly reduces the fake splitting of the clusters and minimizes the loss of clusters due to the dead cell. Once the algorithm has been tested, the impact of the dead cell on the reconstruction was studied as a function of the distance to the expected impact position. Both circular and square bins have been used to check the amount of clusters affected by the dead cell, both in energy and position reconstruction. The results show that the region of poor reconstruction efficiency is limited to a small area around the dead cell. As shown in Figure \ref{deadcell2}, removing the single dead cell from the cluster algorithm allows us to reconstruct fairly well the energy and the position of the clusters affected. Photons that hit the dead cell are poorly reconstructed both in energy and in position, as expected. However, photons hitting the calorimeter within a distance of a cell width from the dead cell are already well reconstructed, especially in their position, when compared with the same case without the presence of the dead cell. In conclusion, the diagonal check has been integrated in the finding algorithm and used only to check towers adjacent to a dead cell. In all other cases, the usual algorithm is used. 

%%% Chapter heading commands %%%

\chapter{Analysis: azimuthal correlations}\label{chapter_results}

%%% Abstract %%%

%\begin{Abstract}

%\end{Abstract}

%%% Chapter sections %%%

The topic of this thesis is the measurement of the coincidence probability between two particles as a means to search for saturation effects. As discussed in Chapter \ref{chapter:theoretical_background}, non linear contributions need to be included to the hadronic wave-function as the density of the gluons increases. In other words: when the longitudinal momentum fraction $x$ of the probed gluon in the nucleus is low enough, saturation effects should become visible. In order to access very low-$x$ values and determine if the boundaries of the saturation region are accessible at RHIC, one can select events which present a leading particle in the forward region.

In the perturbative QCD picture, the scattering between two hadrons is described at leading order by the $\mathrm{2\rightarrow 2}$ elementary interaction of two partons, which creates two back-to-back jets (in the rest frame), balanced in $p_{T}$. If we indicate with $\eta_{A}$ and $\eta_{B}$ the pseudo-rapidities of the two outgoing jets, the longitudinal momentum fractions of the two colliding partons are:
\begin{eqnarray}\nonumber
x_{1} &=&\frac{p_{T}}{\sqrt{s}}\left(e^{+\eta_{A}}+e^{+\eta_{B}}\right)\nonumber \\
x_{2} &=&\frac{p_{T}}{\sqrt{s}}\left(e^{-\eta_{A}}+e^{-\eta_{B}}\right)\nonumber
\end{eqnarray}  
When one of the two particles is detected in the large rapidity region, the scattering selected is most likely between a large longitudinal momentum (valence) quark and a low-$x$ gluon. The Forward Meson Spectrometer (FMS) is placed on the west side of the STAR hall and it faces the deuteron beam in d+Au collisions. When an event is triggered by the FMS, the low-$x$ component of the nuclear gluon is probed. Once the trigger particle is detected (i.e. when $\eta_{trg}$ is fixed), one can scan $x$ by varying the pseudo-rapidity of the second particle.
\begin{figure}
\includegraphics[width=0.5\textwidth]{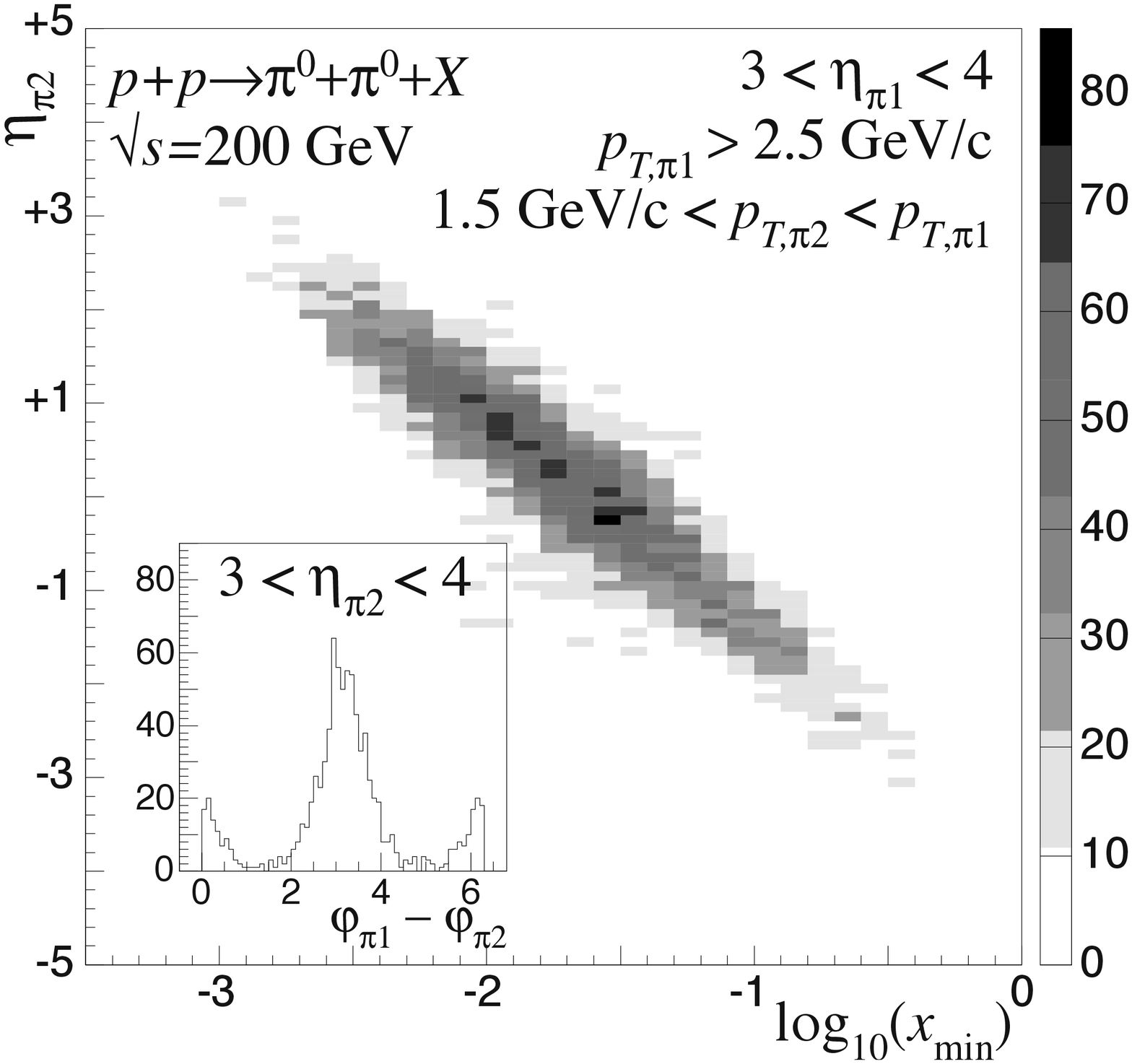} 
\includegraphics[width=0.49\textwidth]{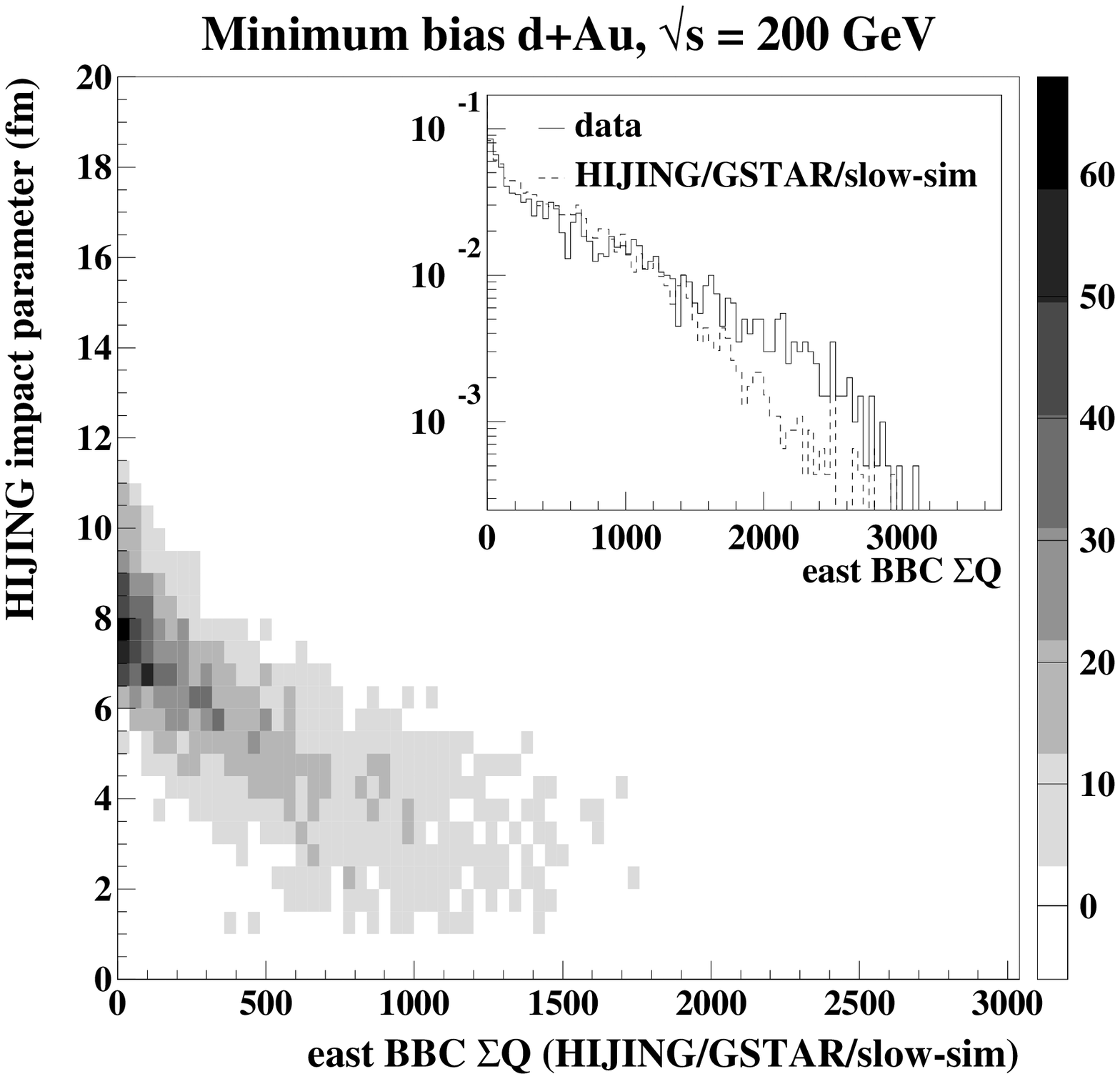} 
\caption{Left: PYTHIA simulation of di-pion production at large $\eta$ p+p collisions at $\sqrt{s}=200$ GeV. The $\eta$ of the associated particle is strongly correlated to the $x$ value of the soft parton probed in the partonic scattering. Figure from \cite{Bland:2005uu}. Right: HIJING impact parameter versus charge sum as recorded by the STAR BBC for simulated 
minimum bias d+Au events. Comparison of charge distribution with data is in the inset.}\label{4.compa}
\end{figure}
Figure \ref{4.compa} shows a study done using a PYTHIA \cite{Sjostrand:2001yu} simulation of two neutral pions. The pseudo-rapidity of the leading particle is chosen in the forward region ($3<\eta_{trg}<4$) while the associated particle is reconstructed over a broad range of rapidity. The plot shows that the pseudo-rapidity $\eta_{asc}$ of the associated particle (indicated as $\eta_{\pi2}$ in Figure \ref{4.compa}) is indeed strongly correlated to the momentum fraction $x$ of the probed parton. In particular, the higher the rapidity of the associated particle, the lower the $x$ probed. This tells us that the kinematical configuration where the largest effect from saturation is expected is when both particles are reconstructed in the forward region (which at STAR means: using the FMS). 

The STAR collaboration is pursuing a systematic plan of measurements of azimuthal correlations with the goal of defining the boundaries of the saturation region. The drawing on the right-hand side of Figure \ref{dis} illustrates how the density of partons in a hadron is expected to change as a function of the variables $Q^{x}$ and $x$ or, correspondingly, of the observables $p_{T}$ and $\eta_{asc}$. Measurements of azimuthal correlations between two mid-rapidity jets in d+Au collisions at STAR \cite{collaboration-2010} has shown no sign of non-linear initial state effects, indicating that, for this kinematic range, the best description for the hadronic structure is that of a dilute system of partons (bottom-right portion of the drawing). The saturation region can be however approached, as discussed before, by selecting events with a forward trigger particle and increasing the pseudo-rapidity $\eta_{asc}$ of the associated particles. This corresponds to moving vertically from the bottom to the top in Figure \ref{4.compa}. The $\eta$-scan is studied in STAR by comparing azimuthal correlations $\Delta\varphi$ involving a forward neutral pion, reconstructed using the FMS ($2.5<\eta_{trg}<4.0$) together with:
\begin{itemize}
\item a mid-rapidity particle in the range $-1.0<\eta_{asc}<1.0$. This can be either a neutral pion reconstructed with the BEMC or a charged track from the TPC;
\item a intermediate-rapidity neutral pion in the range $1.0<\eta_{asc}<2.0$, reconstructed using the EEMC. This analysis is not part of this thesis work;
\item a second large-$\eta$ neutral pion $2.5<\eta_{asc}<4.0$, reconstructed in the FMS.
\end{itemize}
In addition, as we can see from the sketch in Figure \ref{dis}, the saturation region can also be approached by lowering the typical momentum transferred of the interaction. This corresponds to moving horizontally in the scheme towards $\Lambda_{QCD}$ and it is achieved by lowering the $p_{T}$ cut requirement for both trigger and associated particle. In this analysis, the $p_{T}$ dependence of azimuthal correlations has been studied by using two different $p_{T}$ selections, as it will be  specified later.

\begin{figure}
\begin{center}
\includegraphics[width=0.60\textwidth, angle=90]{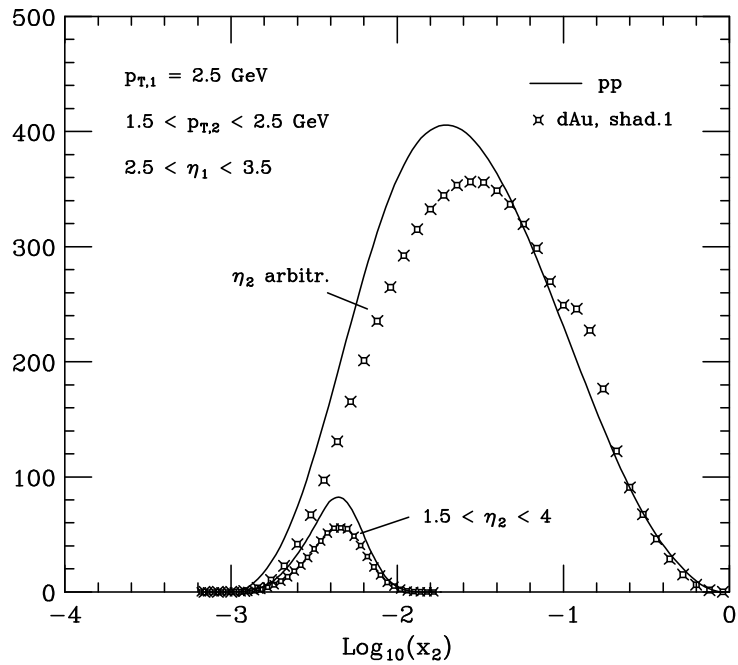} 
\caption{Leading order distribution of  $\log_{10}{x_{asc}}$ of the di-pion cross section for p+p and d+Au at $\sqrt{s}=200\mathrm{\,GeV}$. Figure from \cite{Guzey2004173}. }\label{4.guzey}
\end{center}
\end{figure}

The starting point for our analysis is provided by a perturbative QCD inspired calculation \cite{Guzey2004173}. In this calculation, a trigger $\pi^{0}$ with pseudo-rapidity in the range $2.5<\eta^{(trg)}<3.5$ (comparable with the FMS acceptance) and $p^{(trg)}_{T}>2.5\mathrm{\,GeV}$ is associated with a second $\pi^{0}$ with no restriction in rapidity and with $p_{T}^{(trg)}>p_{T}^{asc}>1.5\mathrm{\,GeV}$. Here the upper limit for the associated pion momentum is set by the $p_{T}$ of the trigger pion, in order to assure the event  to be indeed triggered in the forward region. Such kinematical selection leads to the distribution of $x$ for the probed gluon shown in Figure \ref{4.guzey}. It shows how the FMS is capable of probing the gluon momentum distribution down to $x\approx 10^{-3}$, well into the range where saturation is expected to set in. Moreover, it also shows how the lowest $x$ range is accessed when both particles are forward.

\begin{figure}
\includegraphics[width=0.47\textwidth]{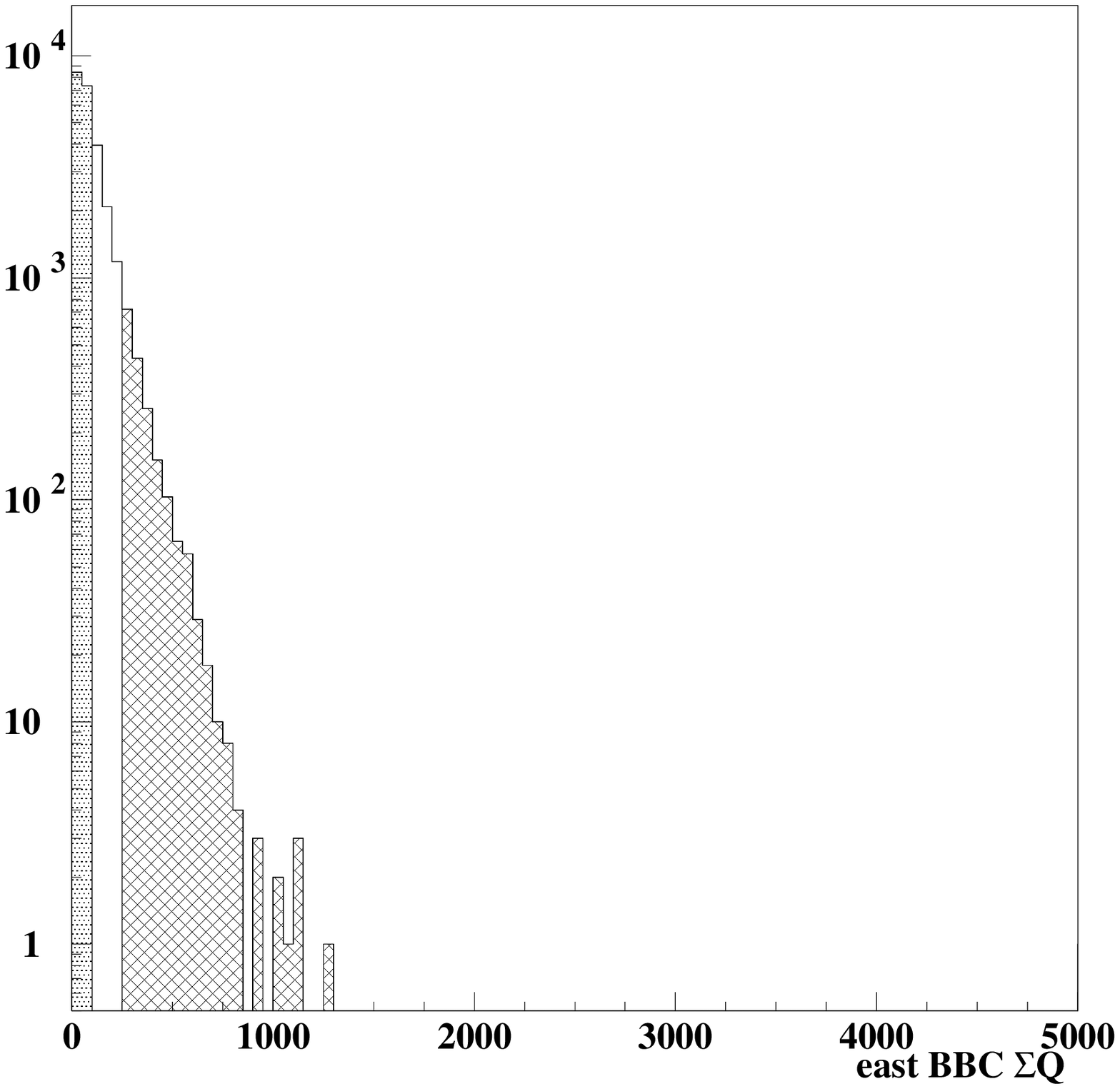} \hspace{0.05\textwidth}
\includegraphics[width=0.47\textwidth]{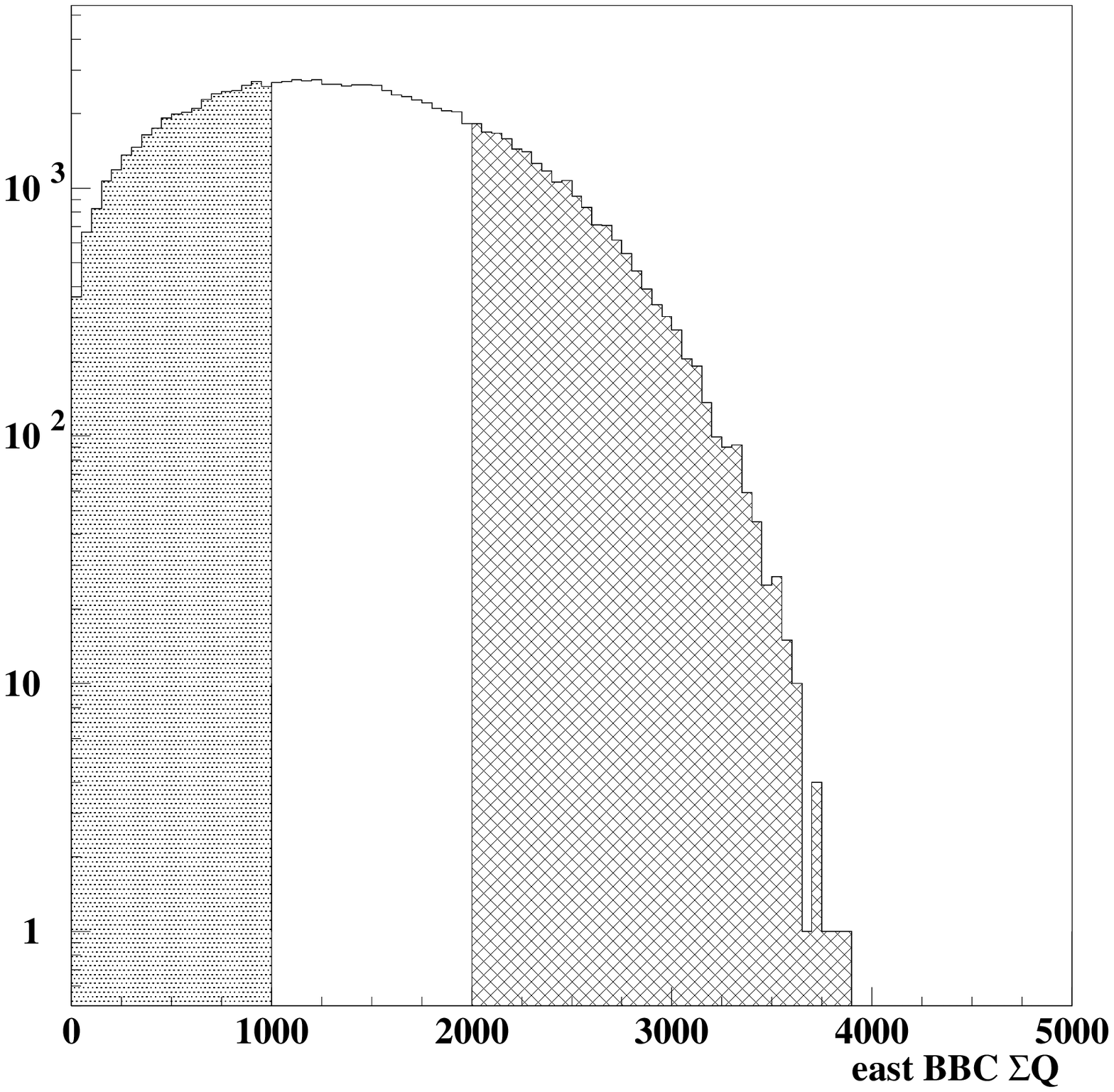} 
\caption{East BBC charge multiplicity distribution for p+p (left) and d+Au (right) interactions. Multiplicity classes are indicated.}\label{4.bbc}
\end{figure}

Finally, saturation depends on the density of the probed medium. Therefore, its effects are expected to be enhanced when the denser part of the gold nucleus is probed. For this reason, azimuthal correlations have been studied as a function of the multiplicity of the event. High multiplicity d+Au collisions can be in fact associated with central collisions, characterized by a low impact parameter $b$. In order to disentangle peripheral from central collisions, we can use the information from the STAR Beam-Beam Counters (BBC). In particular, the sum of charges $\sum{Q_{BBC}}$ recorded in the east BBC module, facing the Gold beam, provides a indirect measurement of the impact parameter $b$ of the collision. Dedicated studies using a state of the art event generator for heavy ions collisions, HIJING 1.383 \cite{PhysRevD.44.3501}, shows that the impact parameter is directly correlated to the multiplicity of the event, as quantified by the east BBC charge sum. The right-hand plot of Figure \ref{4.compa} shows such correlations between impact parameter and multiplicity with, in the inset, a comparison between triggered data and minimum bias simulation of $\sum{Q_{BBC}}$. The plots in Figure \ref{4.bbc} show the east BBC charge sums for p+p and d+Au data. The distributions have been divided in three multiplicity regions and azimuthal correlations have been computed for each region.

\section{FMS-TPC correlations}

The first part of the analysis concerns azimuthal correlations between a forward $\pi^{0}$, reconstructed in the FMS, and a mid-rapidity charged track from the TPC. The first approach is to consider coincidence probability between leading particles. A forward $\pi^{0}$ is built by pairing all possible combinations of two clusters in the FMS, selecting those which falls within the FMS acceptance, precautionary taken to be $2.8<\eta^{(trg)}<3.8$. Pion candidates are then selected within those pairs (in the FMS acceptance) that present a transverse momentum $p_{T}^{(trg)}>2.5\mathrm{\,GeV}$, an energy sharing asymmetry $Z_{\gamma\gamma}<0.7$ and an invariant mass within the range $(0.07<M_{\gamma\gamma}<0.30)\mathrm{\,GeV/c^{2}}$. In order to select the best reconstructed (and most reliable) pion, the candidate with the largest $p_{T}$ is selected as \emph{leading} $\pi^{0}$. The mid-rapidity charged particle is selected from those tracks in the rapidity range $|\eta^{TPC}|<0.9$ that present at least 25 hits in the TPC readout system. As for the neutral pions, the charged track with the largest $p_{T}$ is selected as the \emph{leading associated particle}. The additional requirement $p_{T}^{(FMS)}>p_{T}^{(TPC)}$ is added for ensuring the event to be triggered in the forward region\footnote{We are in fact interested in studying events in which the scattering between a valence quark in the deuteron and a soft gluon in the nucleus generate, in first approximation, two jets of which (at least) one of them is forward. By applying this condition, events with two high-$p_{T}$ jets at mid-rapidity and some secondary pion in the FMS are not included in our study.}. For each event, the difference between the azimuthal direction of these two particles is computed and normalized with the number of trigger (FMS) pions. In this way, the coincidence probability per trigger event is measured. In order to avoid further vertex efficiency corrections, a selection on the quality of the vertex reconstruction is applied to both charged tracks and neutral pions, so that events with no signal are thrown out. In particular, this cut requires at least one reliable vertex to be found by the TPC vertex algorithm. Finally, a BBC reconstructed vertex is required for the event to be processed.

\begin{figure}
\begin{center}
\includegraphics[width=0.60\textwidth]{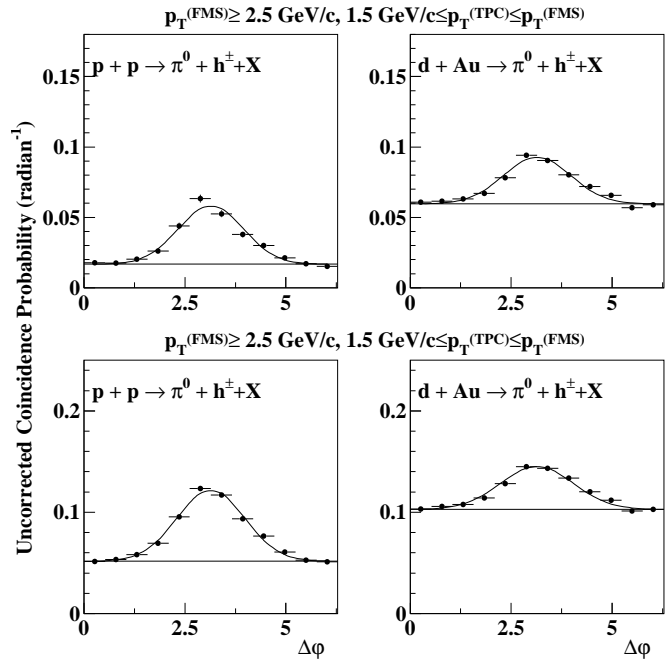} 
\caption{Uncorrected coincidence probability versus azimuthal angle difference between a leading forward $\pi^{0}$ and a leading mid-rapidity charged track. Comparison are made between p+p (left) and d+Au (right) collisions at $\sqrt{s}=200\mathrm{\,GeV}$ for two sets of $p_{T}$ cuts.}\label{4.leading}
\end{center}
\end{figure}

Figure \ref{4.leading} shows the azimuthal correlations $\Delta\varphi$, normalized per trigger event, for this selection of cuts. Data from p+p and d+Au interactions are compared. The top row shows $\Delta\varphi$ for the $p_{T}$ selection specified before. In the bottom row, instead, a looser $p_{T}$ cut is applied: $p_{T}^{(FMS)}>2.0\mathrm{\,GeV/c}$, $p_{T}^{(FMS)}>p_{T}^{TPC}>1.0\mathrm{\,GeV/c}$. This is done to study the $p_{T}$ dependence, as mentioned before. Data are fitted with a constant function plus a periodic Gaussian\footnote{In order to reproduce the periodicity of the distribution in $\Delta\varphi$, it is required the fitting function to have null derivative in the interval extremes $\Delta\varphi=0,2\pi$. To do so, the fitting Gaussian for the interval $\left[0,2\pi\right]$ has been used alongside two additional Gaussian functions with same standard deviation $\sigma$ as the original and centered in $\Delta\varphi=-\pi$ and $\Delta\varphi=3\pi$ respectively.} centered at $\pi$. All correlations present a constant  background representing the uncorrelated underlying event. This contribution appears to be stronger in d+Au events than in p+p, as well as at lower $p_{T}$. All correlations present a peak as well, representing the correlated back-to-back contribution.

\begin{figure}
\begin{center}
\includegraphics[width=0.6\textwidth]{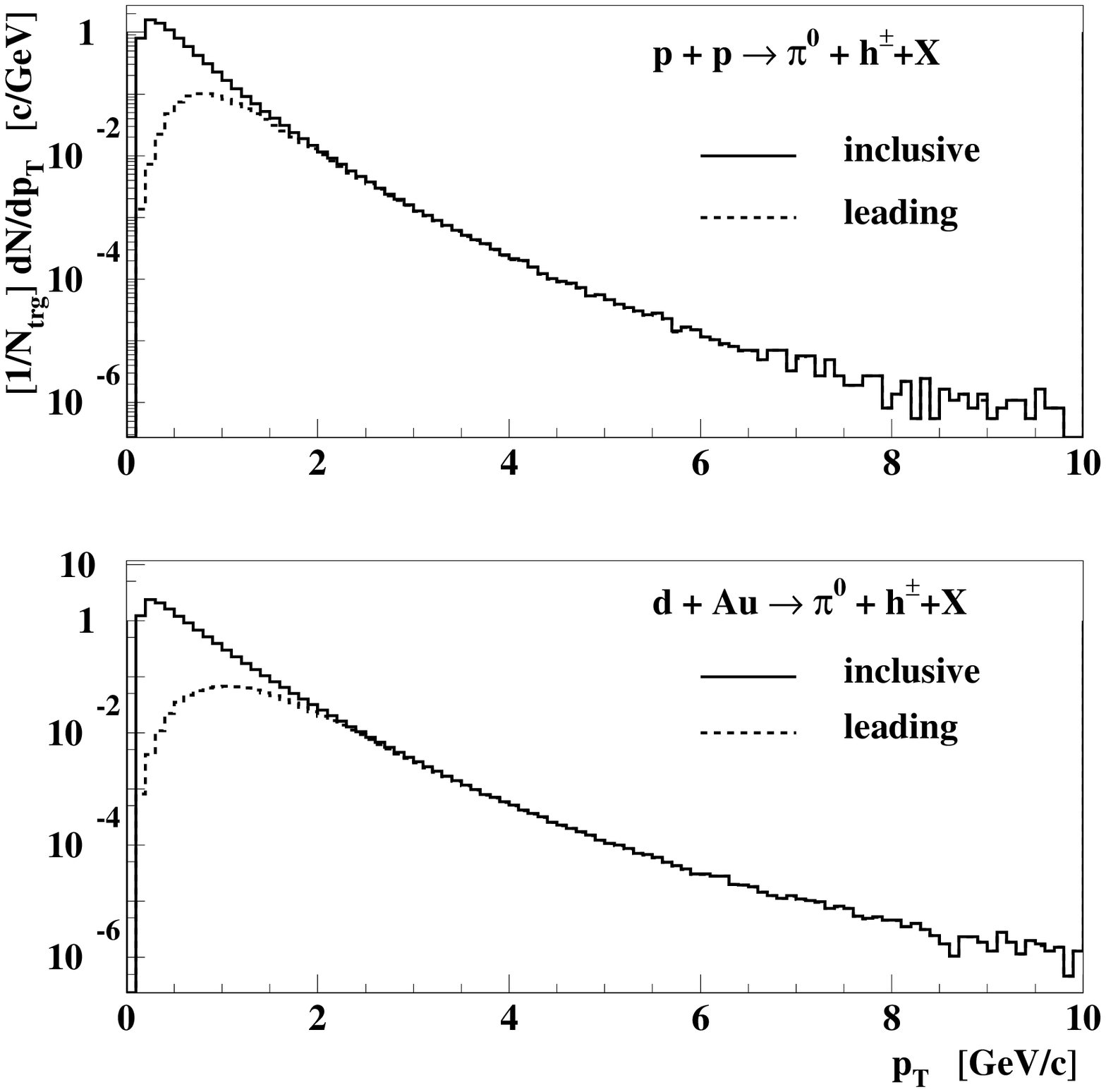} 
\end{center}
\caption{$p_{T}$ spectra for mid-rapidity charged particles per triggered event in p+p and d+Au interactions at $\sqrt{s}=200\mathrm{\,GeV}$.}\label{4.spectra}
\end{figure} 

\begin{figure}
\begin{center}
\includegraphics[width=0.60\textwidth]{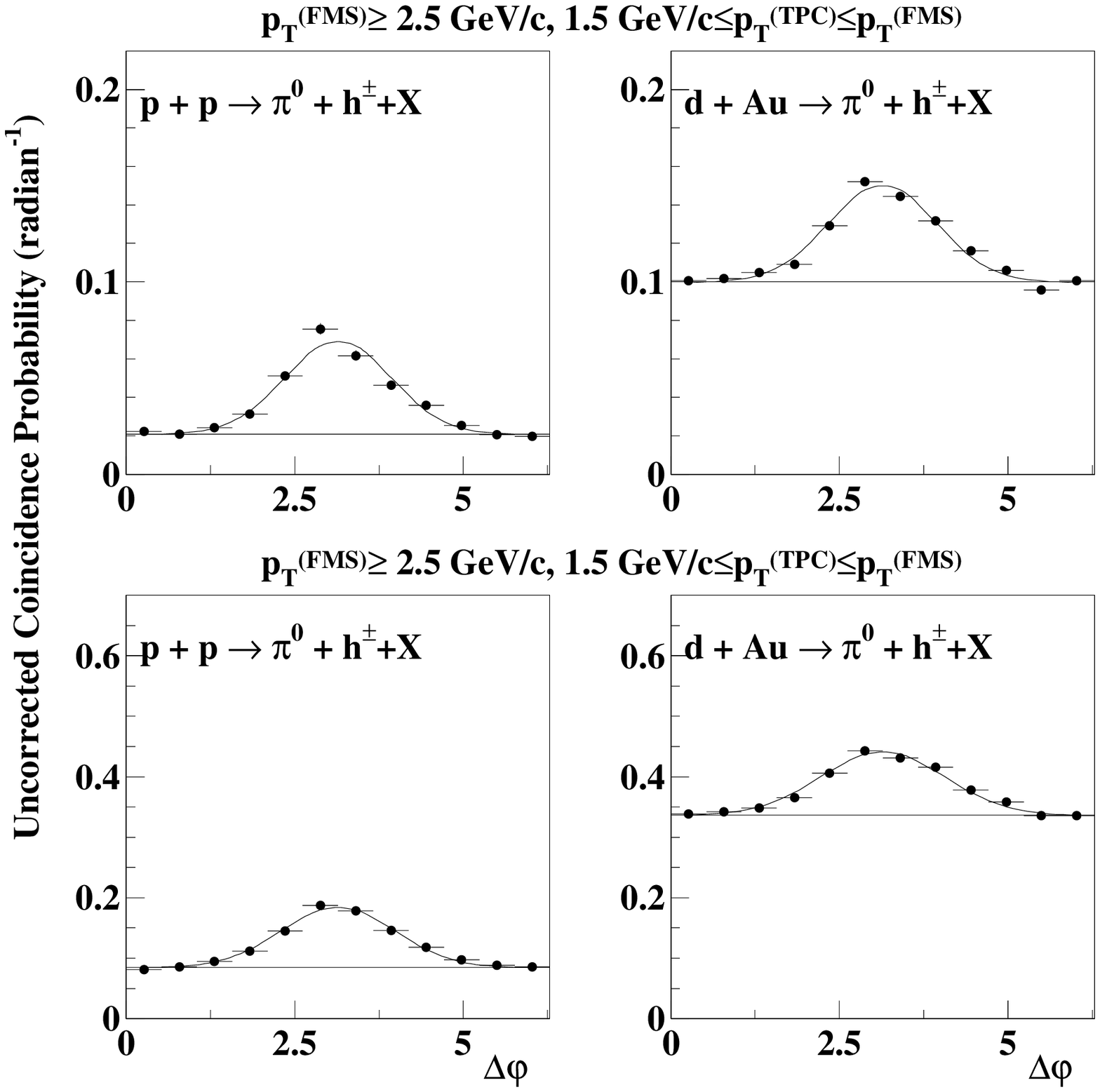} 
\end{center}
\caption{Uncorrected inclusive coincidence probability versus azimuthal angle difference between a forward $\pi^{0}$ and a mid-rapidity charged track in p+p (left) and d+Au (right) collisions at $\sqrt{s}=200\mathrm{\,GeV}$ for two sets of $p_{T}$.}\label{4.spectradue}
\end{figure} 
 
We have so far considered correlations between leading particles, defined as those particles that carry the largest transverse momentum $p_{T}$. This should provide a cleaner measurement of the di-jet correlation, since hard particles represent more accurately the direction of the jet they belong to. Figure \ref{4.spectra} shows that, at relatively high $p_{T}$, the only particle surviving the cuts is indeed the leading one (inclusive and leading spectra coincide). At lower $p_{T}$, however, the number of sub-leading particles starts to become important. As the $p_{T}$ decreases, it becomes more and more possible that the correlation is carried by another particle than the leading one. By performing the analysis on leading particles only, we may find a weaker correlation just because the soft, correlated particles are ``hidden'' by the harder, uncorrelated one. This is true especially in d+Au interactions, where the multiplicity is higher and the number of sub-leading particles per event is larger. To avoid this, azimuthal correlations can be measured inclusively, including in the plot all combinations between a forward pion and a mid-rapidity charge tracks. Inclusive correlations allow, in addition, an easier estimate of the reconstruction efficiency, as well as a natural comparison with quantities theoretically calculated. Figure \ref{4.spectradue} reproduces the inclusive version of the azimuthal correlation plots of Figure \ref{4.leading}. Inclusive correlations are consistent with what is seen in the leading analysis. All of the features seen for the correlations between leading particles are qualitatively reproduced in the inclusive analysis. 

\subsection{Efficiency correction and systematics}

In this section we will discuss the necessary corrections to apply to the reconstruction of TPC tracks. Our first goal is to compare $\pi^{0}$-$h^{\pm}$ (FMS-TPC) and $\pi^{0}$-$\pi^{0}$ (FMS-BEMC) forward-midrapidity correlations. Since all correlations are normalized per trigger pions, inefficiencies related to the FMS pion reconstruction will cancel out in this comparison; therefore, FMS corrections will not be discussed at this point. 

\begin{figure}\begin{tabular}{c c}
\includegraphics[width=0.5\textwidth]{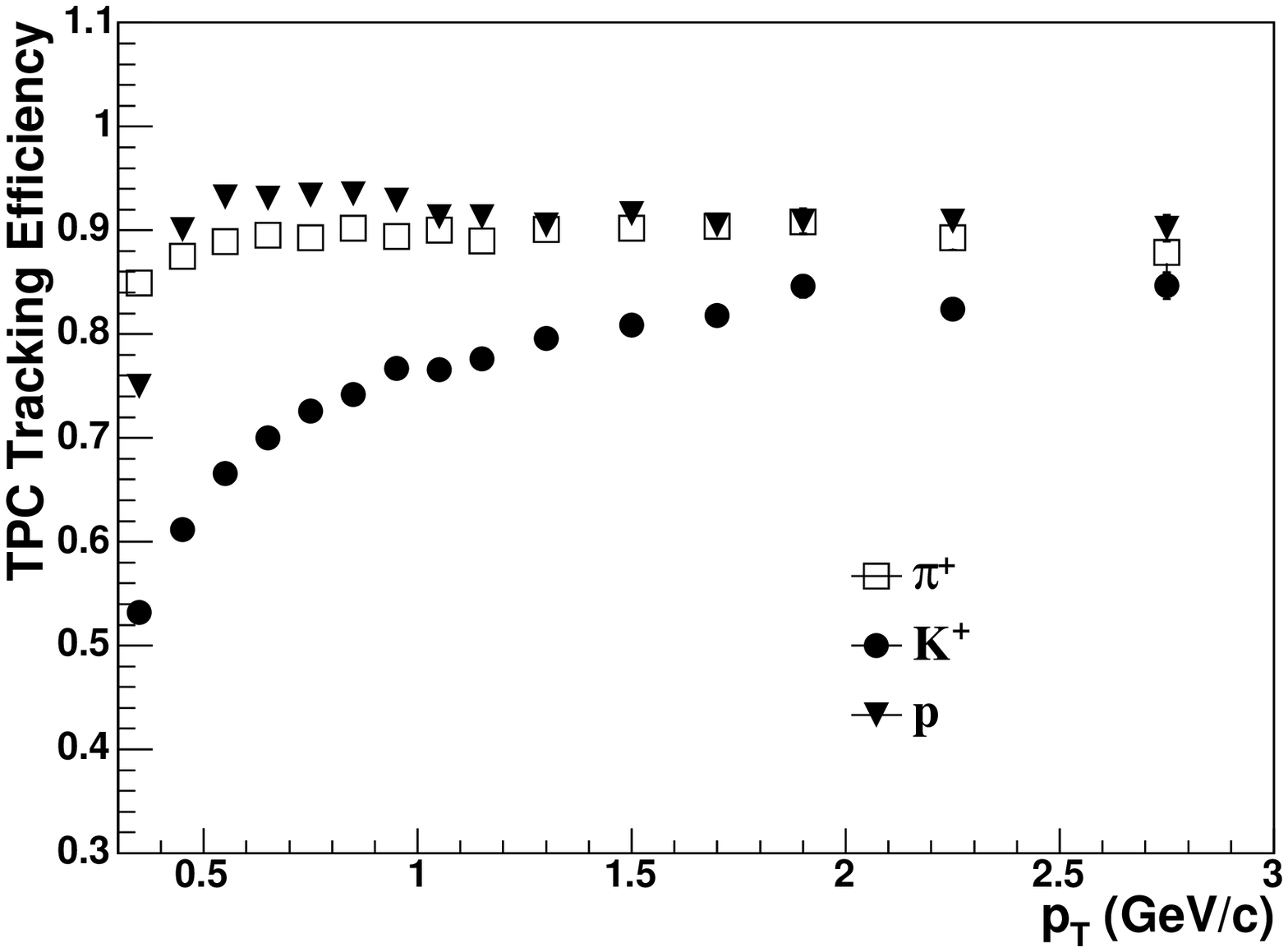} 
\includegraphics[width=0.5\textwidth]{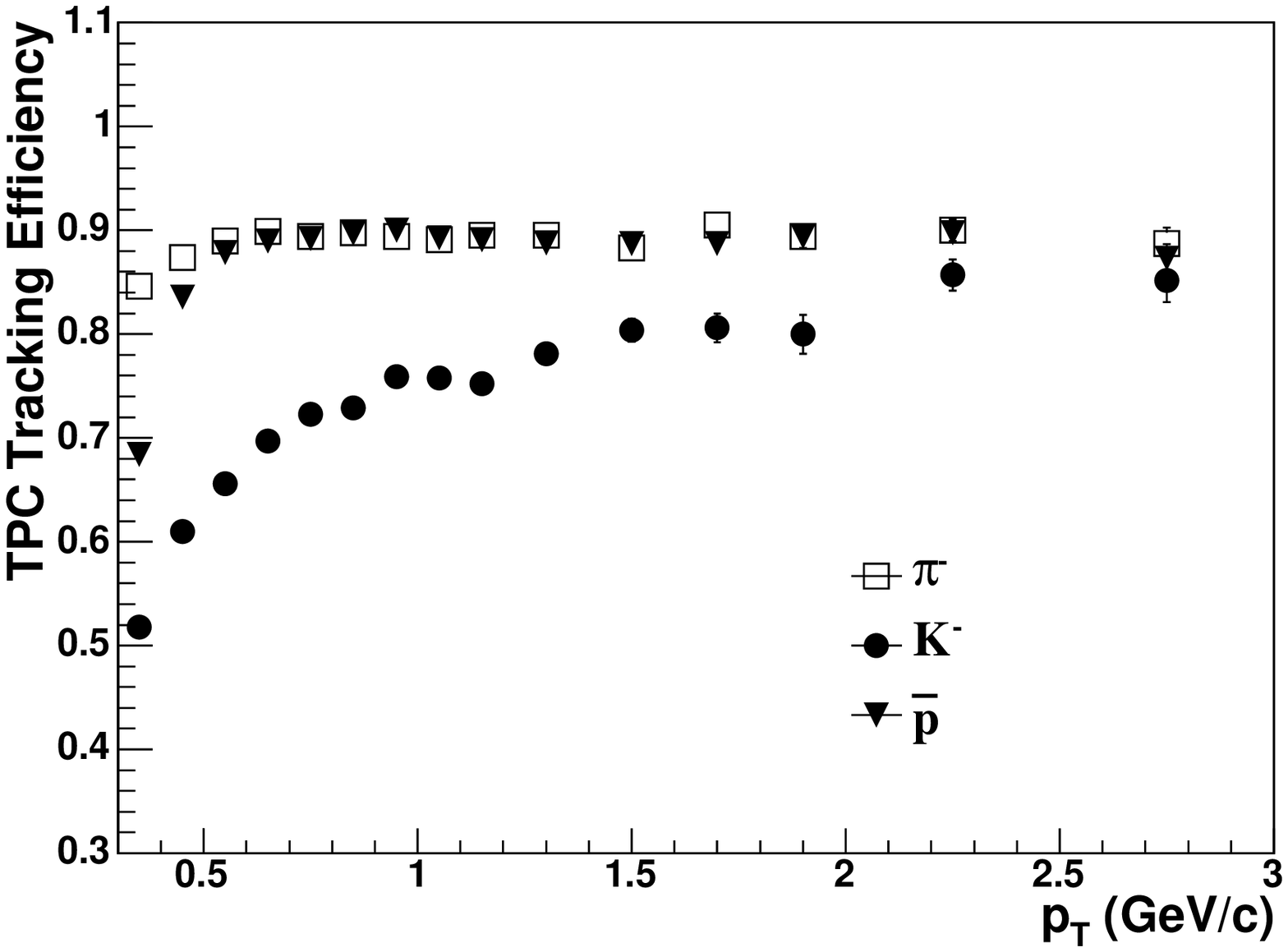} 
\end{tabular}
\caption{TPC reconstruction efficiency of $\pi^{\pm}$, $K^{\pm}$, $p$ and $\bar{p}$ as a function of $p_{T}$. Positive (negative) charged particles are on the left (right). Figure from \cite{Ruan:2005hy}.}\label{5.tpceff}
\end{figure}

Reconstruction of charged tracks in the TPC proves to be relatively clean. The tracking efficiency depends on the acceptance of the detector and the detection efficiency. The TPC acceptance is $\mathrm{96\%}$ for tracks traveling perpendicular to the beamline \cite{Anderson2003659}. The $\mathrm{4\%}$ inefficiency is caused by spaces between the TPC sectors. The total reconstruction efficiency, including the TPC acceptance, is estimated from the number of associated tracks in reconstructed Monte-Carlo events. The efficiency has been studied for different species of charged hadrons \cite{Ruan:2005hy} and it proves to be fairly constant in $p_{T}$ for particles with $p_{T}>0.3\mathrm{\,GeV/c}$ (Figure \ref{5.tpceff}) and not dependent on the pseudo-rapidity. The TPC efficiency is therefore taken to be a constant $\epsilon_{TPC}=90\%$ for both p+p and d+Au interactions. 

\begin{figure}
\begin{tabular}{c c}
\includegraphics[width=0.45\textwidth]{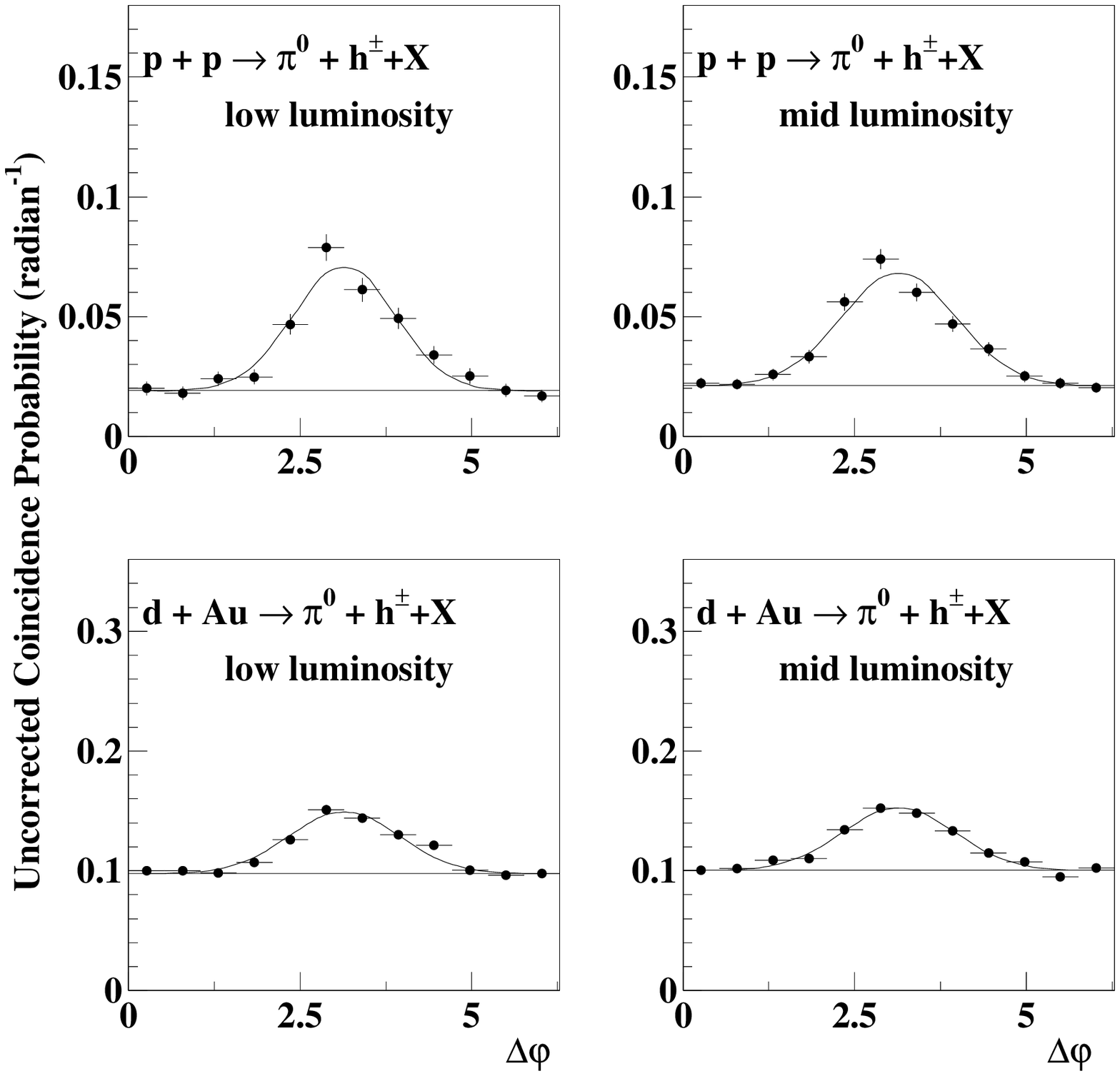} 
\includegraphics[width=0.45\textwidth]{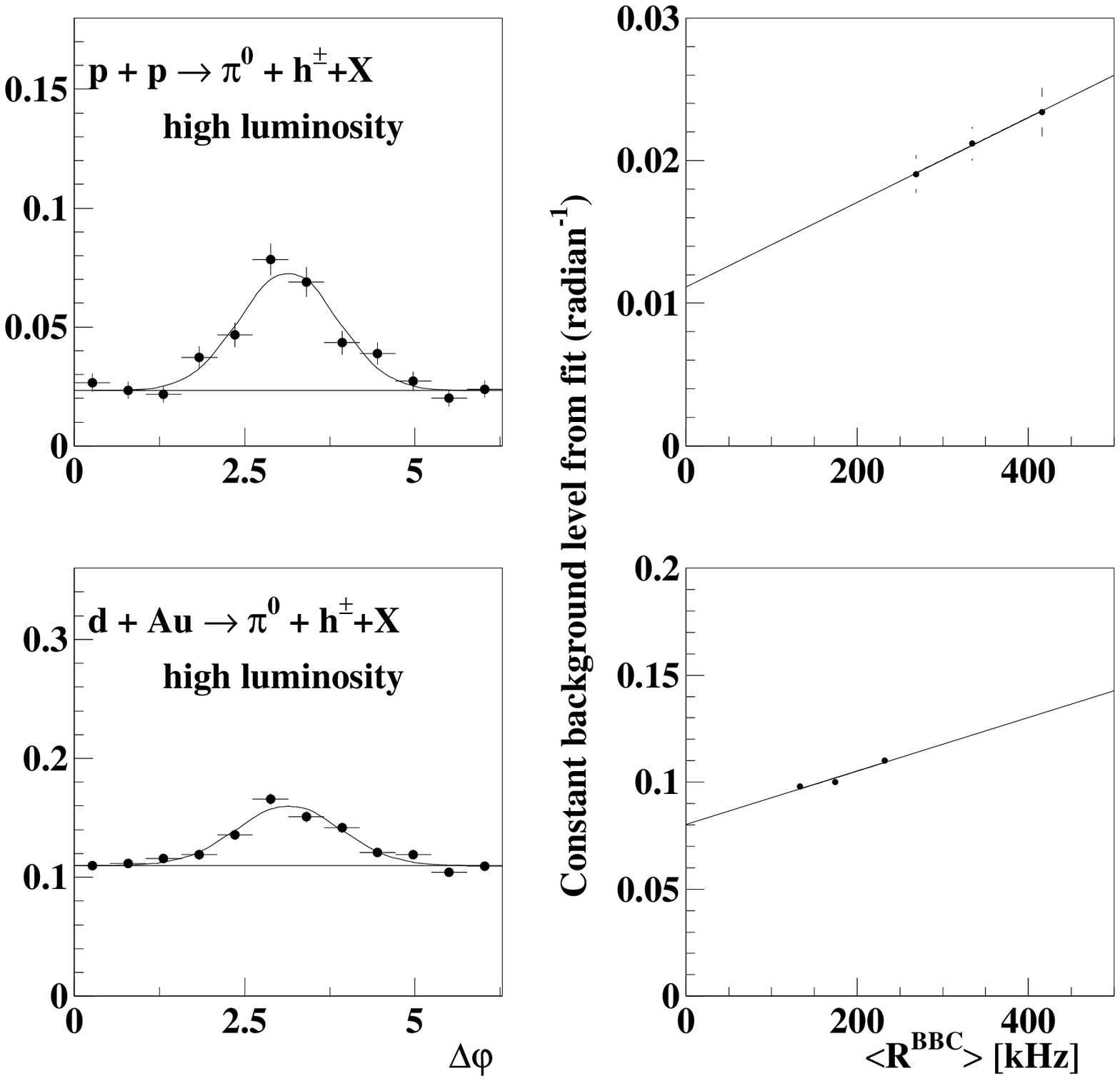} 
\end{tabular}
\caption{Uncorrected inclusive azimuthal correlations for three luminosity classes (lower to the left, higher to the right) for higher $p_{T}$ selection of p+p (top) and d+Au (bottom row) collisions. The right panels show the uncorrelated background component, as extrapolated from the fit, versus the luminosity.}\label{5.lumi}
\end{figure}

Although the TPC allows a relatively clean and efficient reconstruction of charged tracks, its slow readout time, much slower than the bunch crossing time, may cause many of the tracks to be actually from piled up collisions. When this happens, multiple vertexes are found in the event. In cases where the piled up vertex is generated from earlier or later bunch crossings, the track topology shows a discontinuity at the TPC central membrane. Such cases are recognized by the vertex finding algorithm \cite{Reed:2010zz} and the vertex is rejected. When the piled up collisions happen within the same bunch crossing, the vertexes have the same quality and it is more difficult to disentangle them. For this reason, it is necessary to apply a correction for piled up events to the azimuthal correlations. Since the average number of pile up events is proportional to the instantaneous luminosity, it is useful to study the variation of the $\Delta\varphi$ distribution with the luminosity in order to estimate its contribution. Production runs have been divided in three luminosity classes, based on BBC rates, labeled as low ($\langle\mathcal{R}^{BBC}_{pp}\rangle=269\mathrm{\,kHz}$ and $\langle\mathcal{R}_{dAu}^{BBC}\rangle=128\mathrm{\,kHz}$), mid ($\langle\mathcal{R}^{BBC}_{pp}\rangle=334\mathrm{\,kHz}$ and $\langle\mathcal{R}^{BBC}_{dAu}\rangle=173\mathrm{\,kHz}$) and high luminosity ($\langle\mathcal{R}^{BBC}_{pp}\rangle=416\mathrm{\,kHz}$ and $\langle\mathcal{R}^{BBC}_{dAu}\rangle=234\mathrm{\,kHz}$). Azimuthal correlation have been measured for these three run selections and results (for one set of $p_{T}$ cuts) are show in Figure \ref{5.lumi}. Table \ref{5.lumitable} shows instead the values of the correlated signal yield $A$ and standard deviation $\sigma$ and the uncorrelated constant background $b$, as obtained from the fit on data. While signal yields and widths do not change with luminosity, the uncorrelated background increases with increasing luminosity. This effect is larger in p+p, where the luminosity is larger, than in d+Au interactions. To correct for this uncorrelated pile up contribution, it is useful to extrapolate the background level to zero luminosity (right-hand plot in Figure \ref{5.lumi}). The contribution that exceeds this value corresponds to pile up to the coincidence probability and has to be subtracted from the azimuthal correlations. Once this constant (uncorrelated) background is subtracted from the coincidence probability, the distribution can be corrected for the TPC efficiency. The luminosity dependence is the only contribution to the systematic uncertainties here addressed. 

\begin{table}[h!]\begin{center}
\begin{tabular}{|c|c|c|c|c|c|}
\hline int. & $ p_{T} $ cuts & $\langle\mathcal{R}^{BBC}\rangle$ & A & $\sigma$ & b  \\ \hline\hline 
\multirow{6}{*}{p+p} & \multirow{3}{*}{higher} & low & $0.097\pm0.008$ & $0.736\pm0.061$ & $0.019\pm0.001$   \\ 
& & mid & $0.098\pm0.007$ & $0.829\pm0.055$ & $0.021\pm0.001$  \\ 
& & high & $0.089\pm0.010$ & $0.710\pm0.082$ & $0.023\pm0.002$  \\ \cline{2-6}
& \multirow{3}{*}{lower} & low & $0.211\pm0.012$ & $0.856\pm0.043$ & $0.081\pm0.002$ \\
& & mid & $0.213\pm0.010$ & $0.827\pm0.034$ & $0.087\pm0.002$  \\ 
& & high & $0.230\pm0.017$ & $0.866\pm0.061$ & $0.092\pm0.003$ \\ \hline
\multirow{6}{*}{d+Au} & \multirow{3}{*}{higher} & low & $0.104\pm0.009$ & $0.798\pm0.056$ & $0.098\pm0.002$   \\ 
& & mid & $0.107\pm0.009$ & $0.811\pm0.058$ & $0.100\pm0.002$  \\ 
& & high & $0.096\pm0.010$ & $0.759\pm0.073$ & $0.110\pm0.002$ \\ \cline{2-6}
& \multirow{3}{*}{lower} & low & $0.251\pm0.013$ & $0.850\pm0.036$ & $0.334\pm0.002$ \\
& & mid & $0.262\pm0.014$ & $0.960\pm0.042$ & $0.347\pm0.003$  \\ 
& & high & $0.243\pm0.018$ & $0.947\pm0.056$ & $0.377\pm0.003$ \\ \hline 
\end{tabular}\caption{Summary of the values of signal yield ($\mathrm{A}$), signal width ($\mathrm{\sigma}$) and uncorrelated background ($\mathrm{b}$) from fit on Figure \ref{5.lumi}.}\label{5.lumitable}\end{center}
\end{table}

\section{FMS-BEMC correlations}

\begin{figure}
\begin{tabular}{c c}
\includegraphics[width=0.45\textwidth]{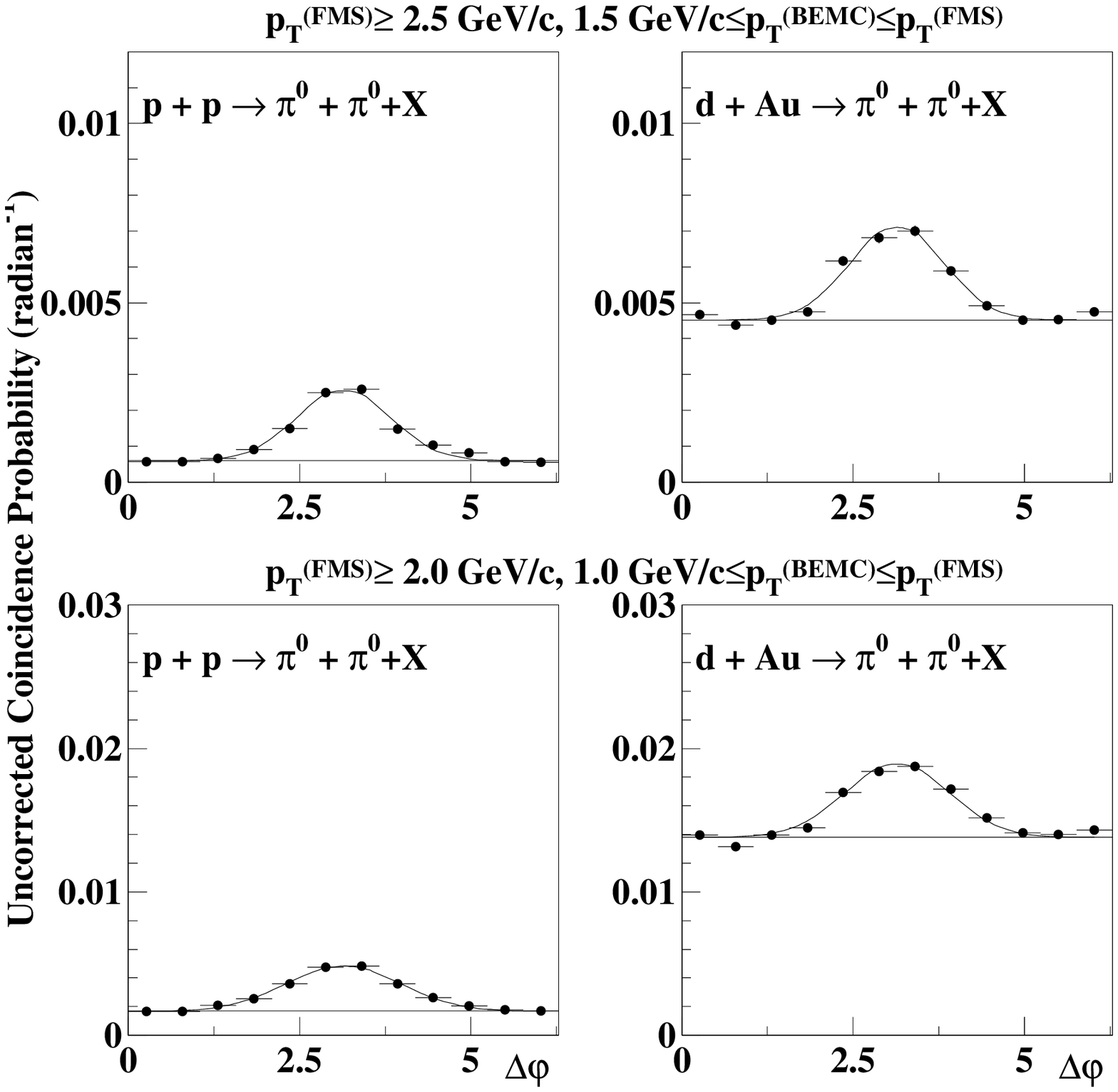} 
\includegraphics[width=0.45\textwidth]{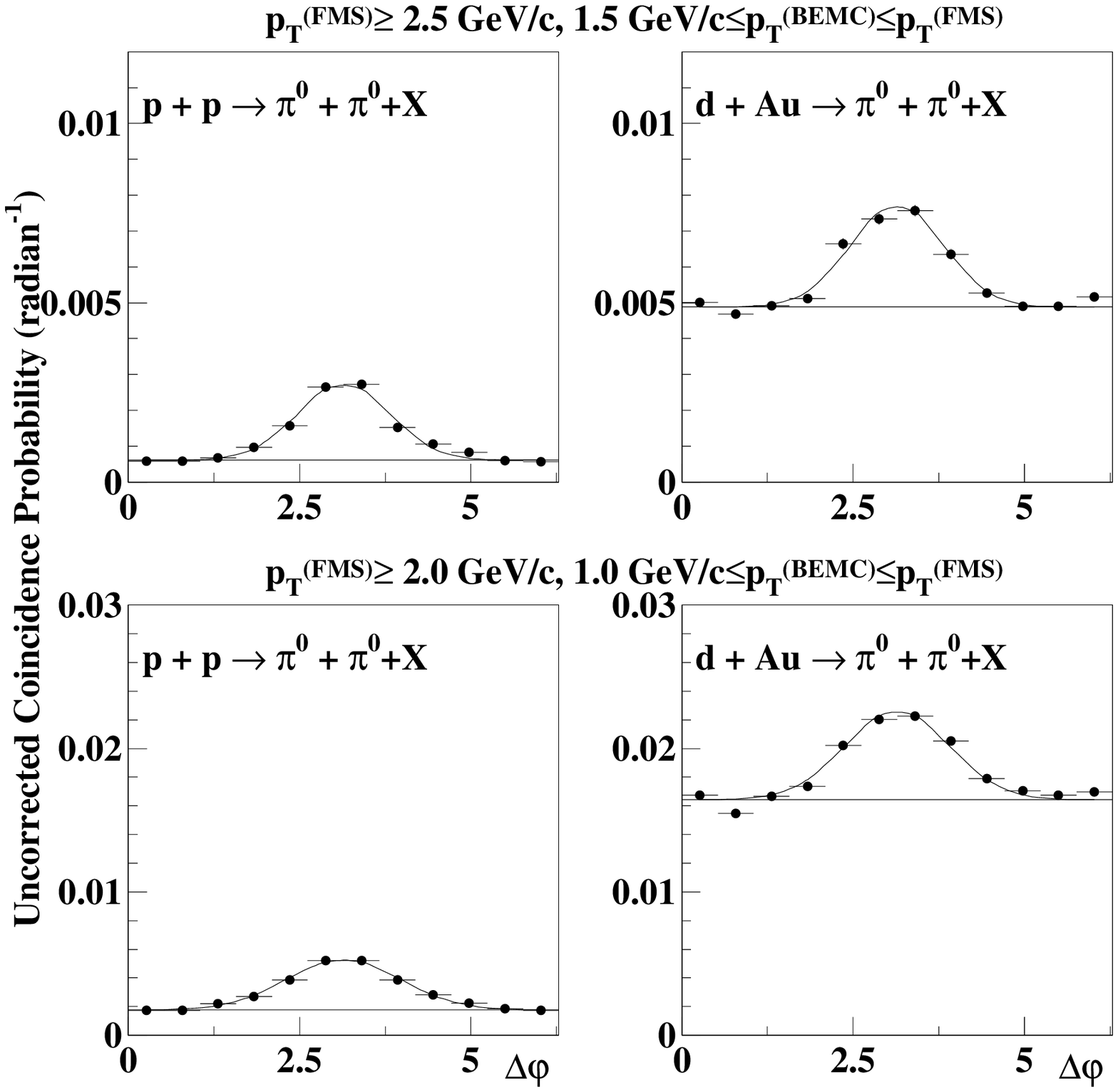} 
\end{tabular}
\caption{Uncorrected coincidence probability versus azimuthal angle difference between a forward $\pi^{0}$ and a mid-rapidity $\pi^{0}$. Correlations are computed between leading particles (left panel) and inclusively (right panel). Each panel shows p+p (left) and d+Au (right) collisions at $\sqrt{s}=200\mathrm{\,GeV}$ for two sets of $p_{T}$.}\label{4.emclead}
\end{figure}

Azimuthal correlations between a forward $\pi^{0}$ and a mid-rapidity $\pi^{0}$ have been studied with the double goal to look at confirmation to the FMS-TPC correlations previously discussed and to perform a deeper analysis by increasing the statistics of reasonably well identified particles. As already mentioned, the BEMC, which we use to reconstruct mid-rapidity neutral pions, is a faster detector than the TPC and we can use the fast stream of data to compute FMS-BEMC correlations. On the other hand, the quality of the reconstructed pions is not as good as for the TPC tracks, so more care needs to be taken while estimating the background components. The coincidence probability for a pair of pions in forward and mid-rapidity per trigger event has been measured for both leading and inclusive production. In order to compare it with the $\pi^{0}$-$h^{\pm}$ correlation, the same set of kinematical cuts has been applied. Forward pions are reconstructed by pairing photons in the FMS fiducial volume. Neutral pion candidates are selected within the pseudo-rapidity range $2.8<\eta^{(FMS)}<3.8$ with requirements on energy asymmetry $Z_{\gamma\gamma}<0.7$ and invariant mass $(0.07<M_{\gamma\gamma}<0.30) \mathrm{\,GeV/c^{2}}$. At mid-rapidity, neutral pions are similarly reconstructed by considering pairs of clusters within the BEMC acceptance $|\eta^{BEMC}_{\gamma}|<0.9$. In addition to this, a quality cut on the single cluster is applied in order to reduce the hadronic background component. As discussed in Section \ref{3.backgroundtreatment}, clusters are required to have a shower shape profile so that at least 90\% of the energy is contained within the peak tower. Additional cuts are applied on the pion candidate's pseudo-rapidity ($|\eta^{BEMC}|<0.9$), energy asymmetry ($Z_{\gamma\gamma}<0.7$) and invariant mass ($(0.07<M_{\gamma\gamma}<0.30) \mathrm{\,GeV/c^{2}}$). Figure \ref{4.emclead} shows the resulting azimuthal correlations between a forward and a mid-rapidity neutral pion, in the case where both particles are selected as the leading ones in the respective detectors (left) and where correlations are calculated inclusively over all candidates. In order to compare FMS-BEMC and FMS-TPC correlations, the same two sets of $p_{T}$ cuts have been applied. As indicated in the figure, for the top row the cuts are $p_{T}^{(FMS)}>2.5\mathrm{\,GeV/c}$ and $p_{T}^{(FMS)}>p_{T}^{(BEMC)}>1.5\mathrm{\,GeV/c}$ while for the bottom row the cuts are $p_{T}^{(FMS)}>2.0\mathrm{\,GeV/c}$ and $p_{T}^{(FMS)}>p_{T}^{(BEMC)}>1.0\mathrm{\,GeV/c}$.
As before, coincidence probabilities per trigger event have been fitted with a constant plus a periodic Gaussian function. The comparison with FMS-TPC correlations, in the same kinematical range, shows that the features of the two kind of correlations ($\pi^{0}$-$\pi^{0}$ and $\pi^{0}$-$h^{\pm}$) are qualitatively comparable. As before, correlations are characterized by a constant (uncorrelated) background from underlying events, which is larger for d+Au interactions. In addition, the correlated back-to-back peak is again evident in p+p and d+Au collisions, in both $p_{T}$ regimes. Statistical uncertainties are highly reduced by the larger amount of fast data available for this analysis.

\begin{figure}
\begin{center}
\includegraphics[height=0.6\textwidth,clip=]{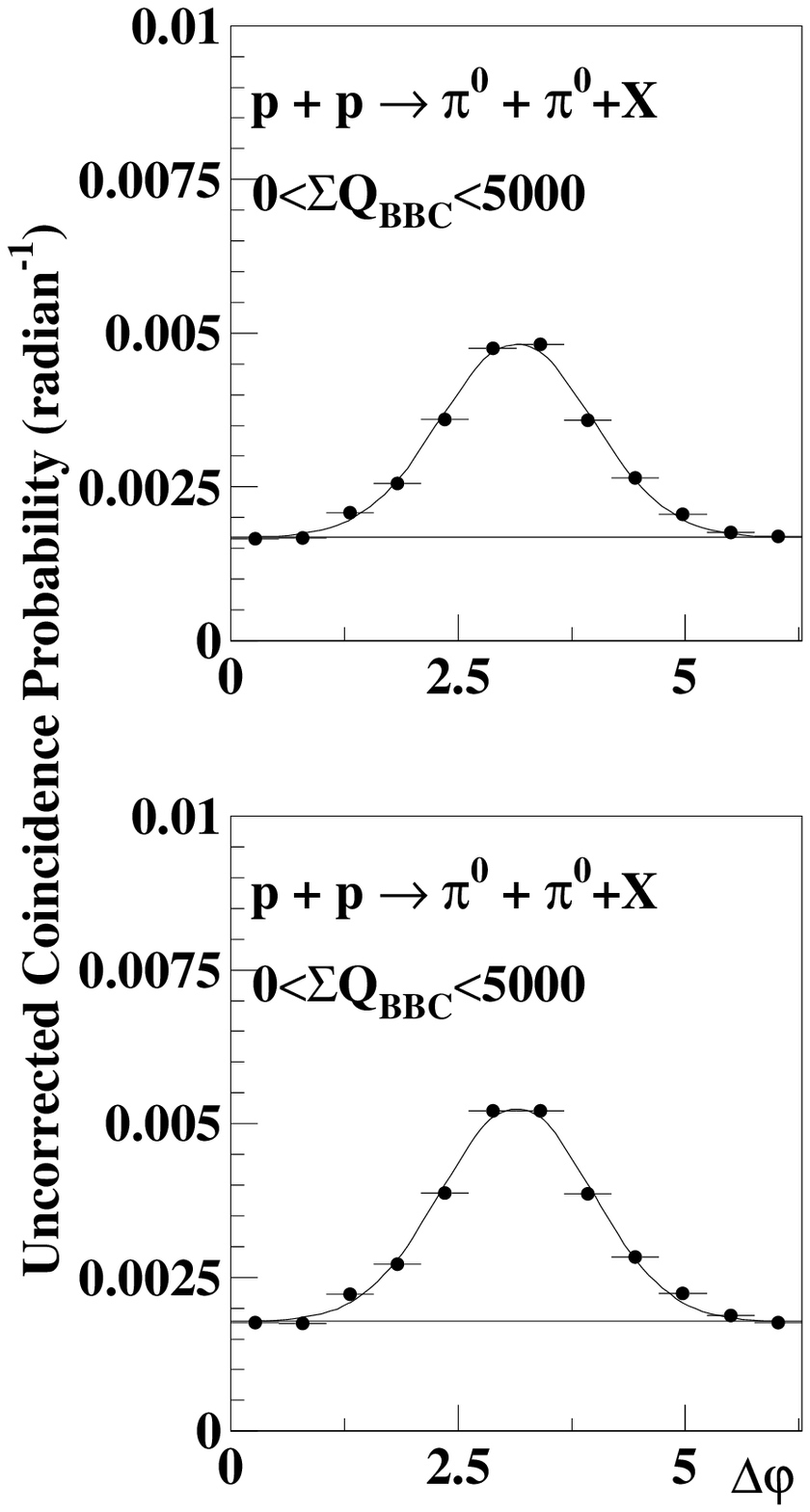} 
\includegraphics[height=0.6\textwidth,clip=]{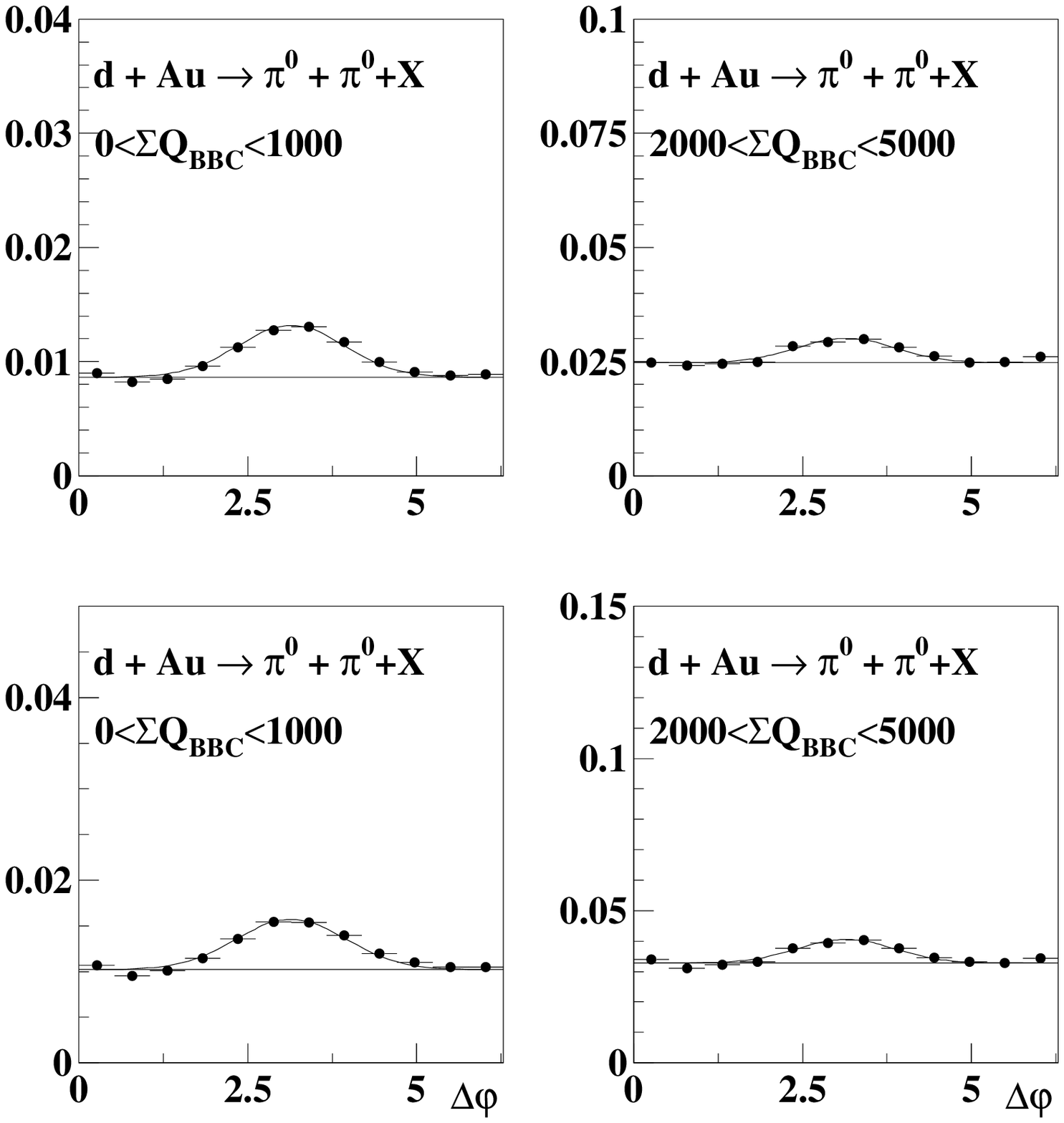} 
\caption{Uncorrected coincidence probability versus azimuthal angle difference between a forward $\pi^{0}$ and a mid-rapidity $\pi^{0}$, with lower $p_{T}$ selection, in p+p (left), peripheral (center) and central d+Au interactions (right). Leading (top) and inclusive (bottom) correlations are shown.}\label{4.multi}
\end{center}
\end{figure}

In order to enhance possible non-linear effects in the nuclear wave-function, we can select a subset of interactions characterized by high multiplicity. Broadening effects due to saturation are in fact expected to appear more significantly when the dense central part of the nucleus is probed. Figure \ref{4.multi} shows a comparison between low $p_{T}$ azimuthal correlations in p+p, peripheral and central d+Au collisions, in the leading (top row) and inclusive (bottom row) analysis (the multiplicity selection does not significantly impact correlations in p+p interactions). In d+Au interactions, although the signal over background ratio appears smaller for higher multiplicity events, there is no hint of significant broadening with respect to peripheral collisions. 

\subsection{Background subtraction}

The reconstruction of mid-rapidity neutral pions is characterized by larger systematic uncertainties than charged tracks. The main source of such uncertainties is the presence of a significant background in the invariant mass spectrum of the pion candidates. The subtraction of a background component from the correlation distributions (background pair - forward pion correlation) is therefore necessary before applying any efficiency correction. In order to estimate the background contribution and to calculate the reconstruction efficiency and the detector acceptance, a simulation of one million minimum bias p+p events have been produced by PYTHIA, using GEANT to simulate the detector responses., as well as a one million d+Au events using HIJING+GEANT. The choice of minimum bias, instead of a simulation of events presenting a forward trigger $\pi^{0}$ within the FMS acceptance, has been made for practical reasons, in order to enhance the statistics at mid-rapidity. The $z$ component of the vertex has been randomly smeared (using a Gaussian noise with $\sigma=30\mathrm{\,cm}$) in simulation, in order to emulate the BBC vertex reconstruction uncertainty. The calibration table used for the simulation has been chosen in order to best represent the average BEMC configuration during the run. 

\begin{figure}
\begin{center}
\includegraphics*[width=0.60\textwidth]{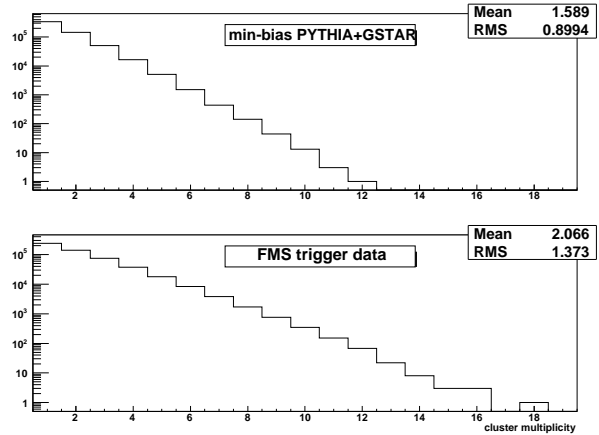} 
\end{center}
\caption{Cluster multiplicity comparison between minimum bias PYTHIA+GEANT simulation (top) and forward triggered data (bottom).}\label{5.multiplicitycomparison}
\end{figure}

\begin{figure}
\includegraphics[width=0.9\textwidth]{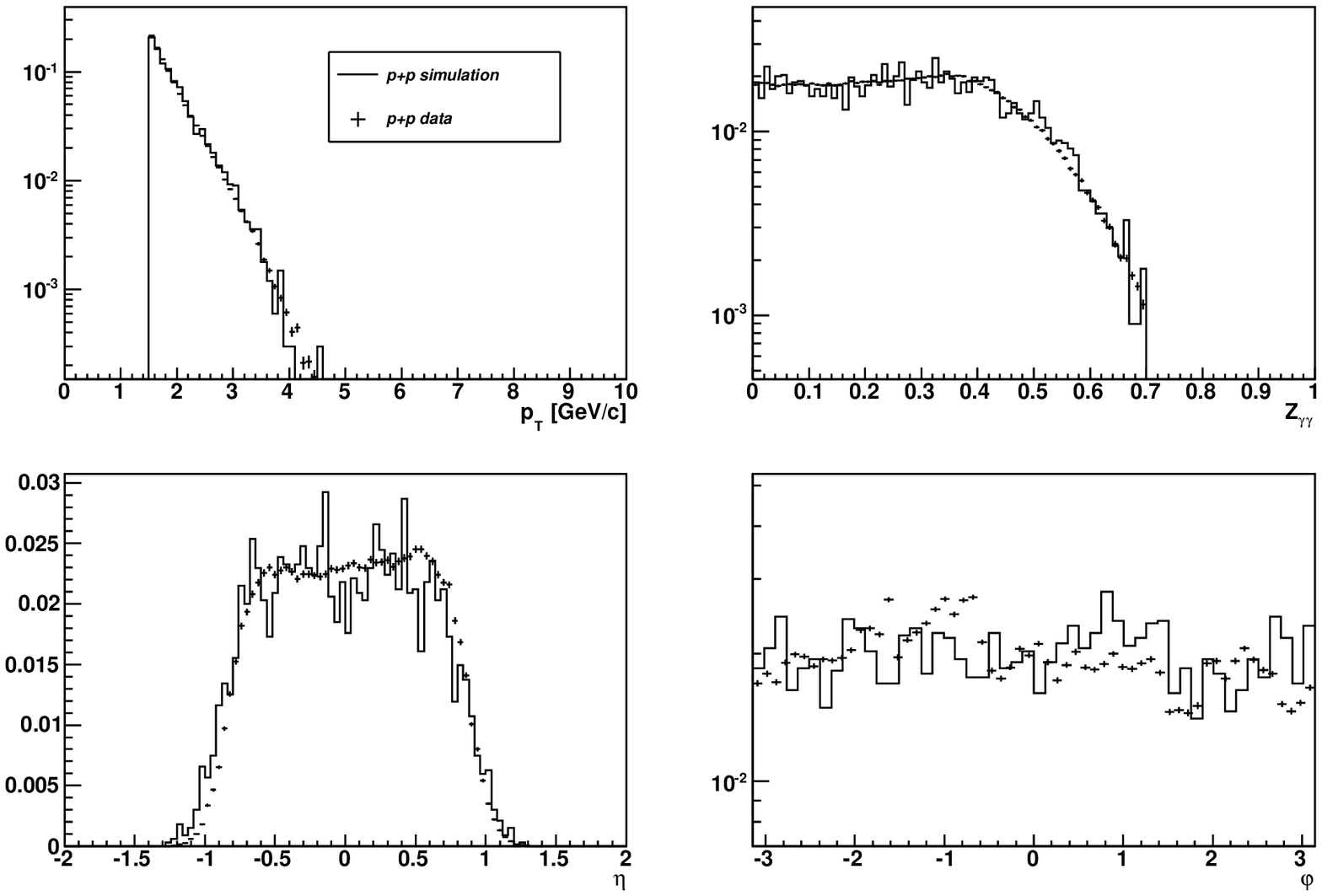} 
\caption{Simulation-Data comparison for $\pi^{0}$ parameters ($p_{T}$, $Z_{\gamma\gamma}$, $\eta$, $\varphi$) in p+p interactions. }\label{5.comparisonsimu}
\end{figure}

\begin{figure}
\includegraphics[width=0.9\textwidth]{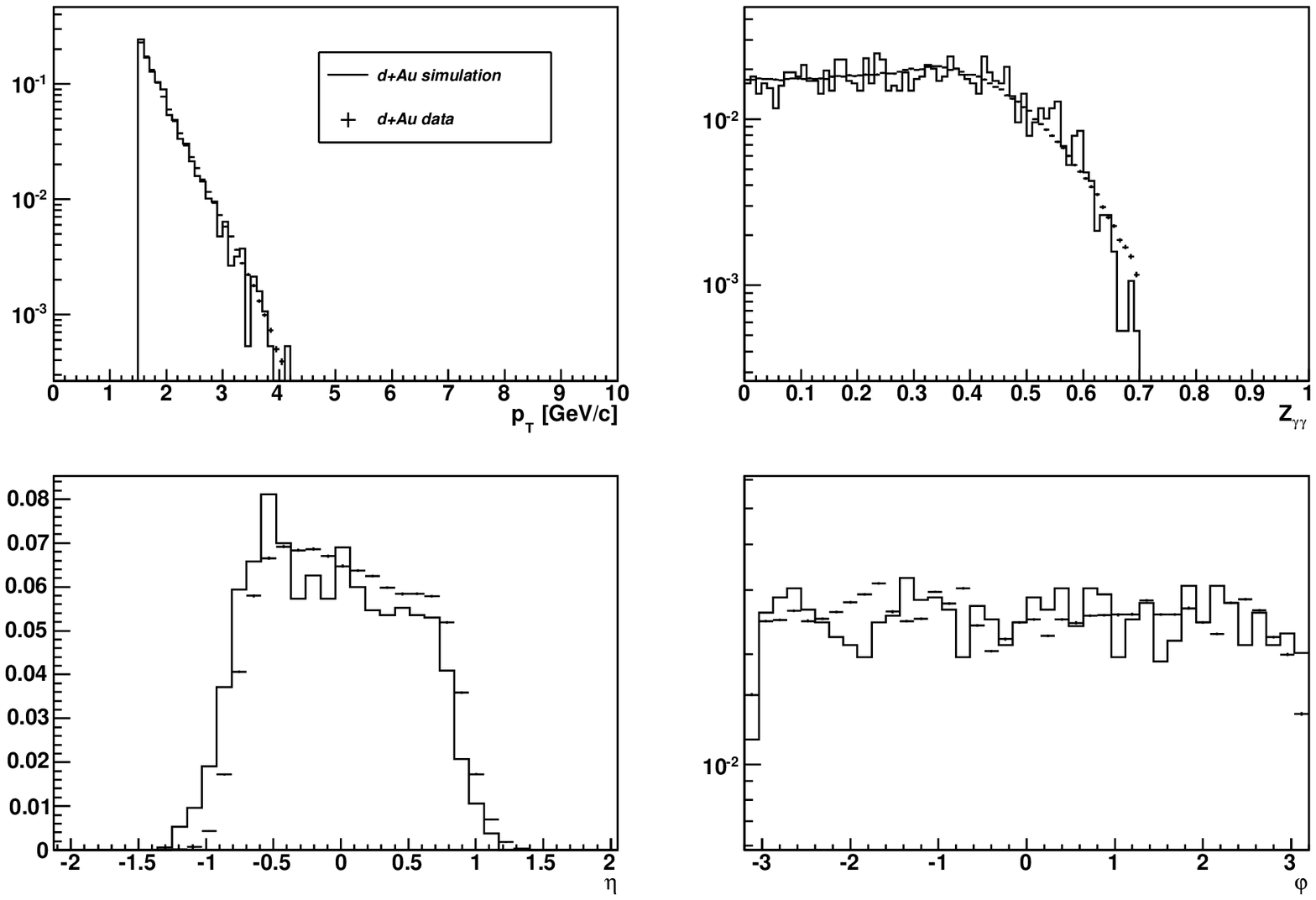} 
\caption{Simulation-Data comparison for $\pi^{0}$ parameters ($p_{T}$, $Z_{\gamma\gamma}$, $\eta$, $\varphi$) in d+Au interactions. }\label{5.comparisonsimudue}
\end{figure}

\begin{figure}
\includegraphics[width=0.9\textwidth]{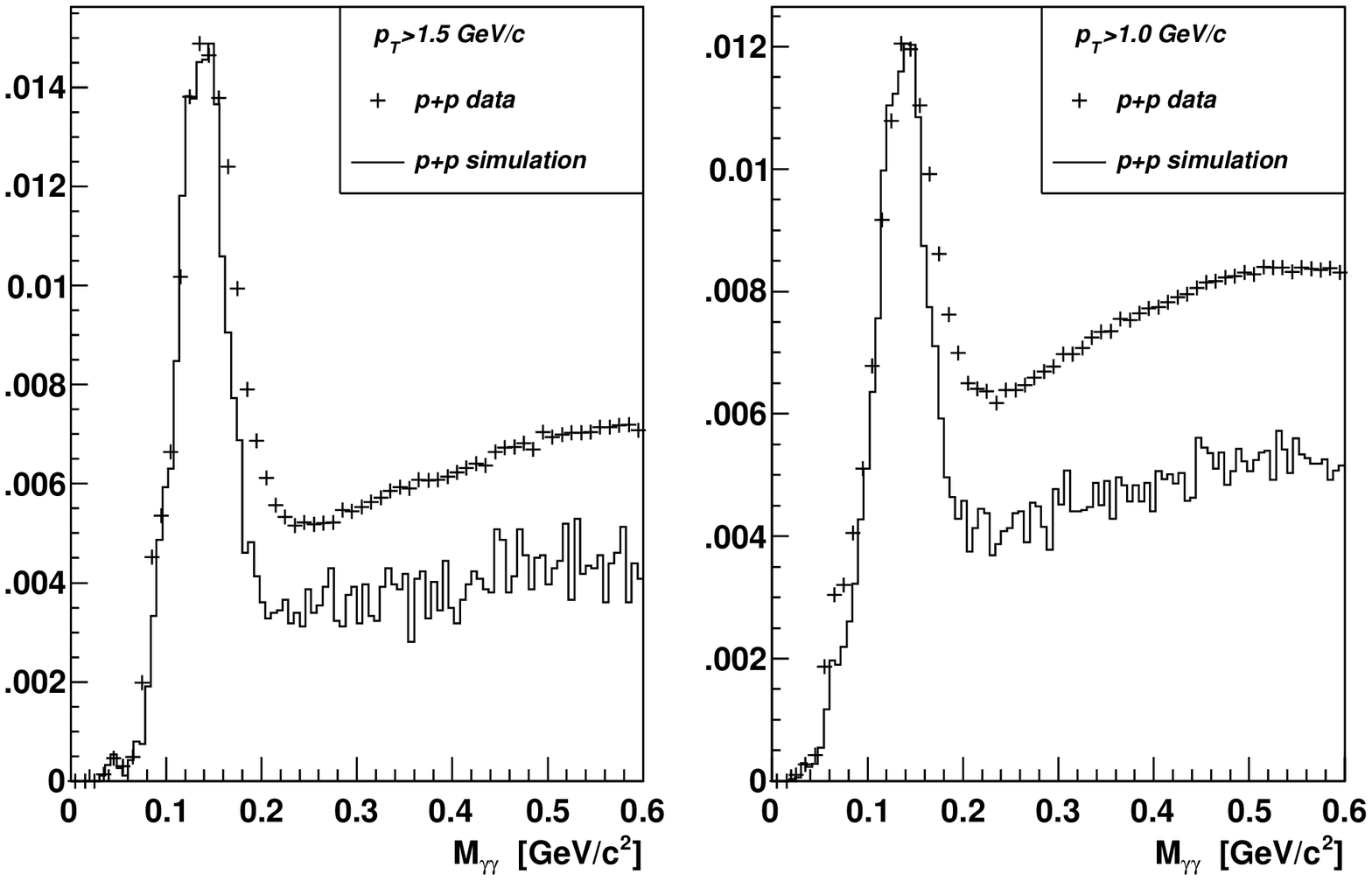} 
\caption{Simulation-Data comparison between invariant mass distributions for BEMC cluster pairs  for different $p_{T}$ selections, in p+p interactions.}\label{5.masse}
\end{figure}

\begin{figure}
\includegraphics[width=0.9\textwidth]{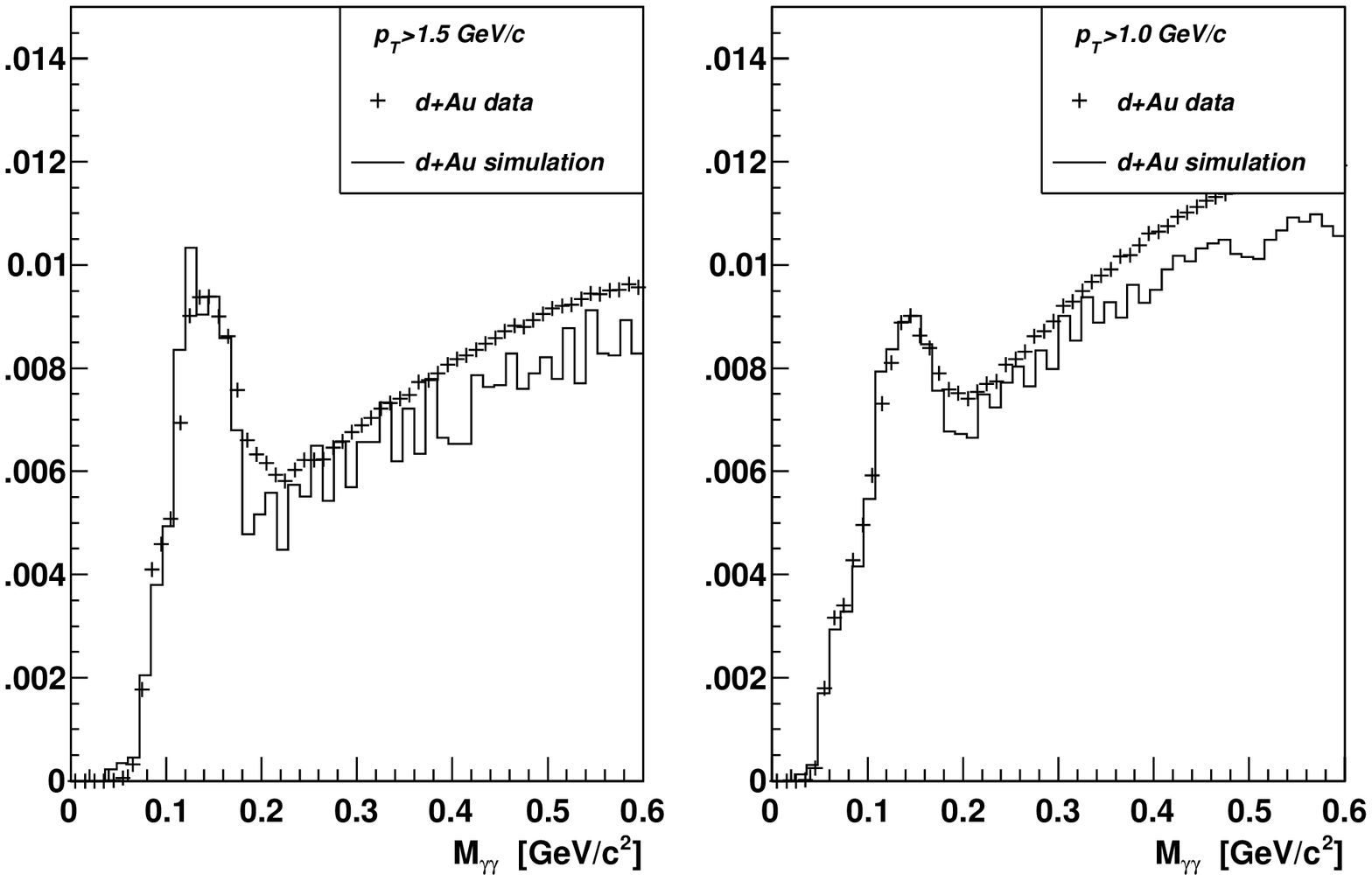} 
\caption{Simulation-Data comparison between invariant mass distributions for BEMC cluster pairs  for different $p_{T}$ selections, in d+Au interactions.}\label{5.massedue}
\end{figure}

The main difference between simulation and data is represented by the higher multiplicity in forward triggered events compared to the minimum bias simulation. This is indeed expected from pure PYTHIA studies and confirmed in the multiplicity comparison between minimum bias simulation and triggered data (Figure \ref{5.multiplicitycomparison}). Despite this fact, the $\pi^{0}$ reconstruction shows good agreement between simulation and data, as illustrated in Figures \ref{5.comparisonsimu} (p+p) and \ref{5.comparisonsimudue} (d+Au interactions). On the other hand, another known and expected difference comes from the invariant mass distribution, and in particular in the background component. As shown in Figures \ref{5.masse} (pp) and \ref{5.massedue} (d+Au interactions), while the $\pi^{0}$ peak appears to be consistent in width with data, the hadronic background is substantially underestimated in simulation. This means that we cannot estimate the signal/background ratio directly from simulation. We will need instead to scale the background component to fit the data or, alternatively, use the information of the shape to estimate how the background behaves below the $\pi^{0}$ peak.

\begin{figure}
\includegraphics[height=0.6\textwidth,clip=]{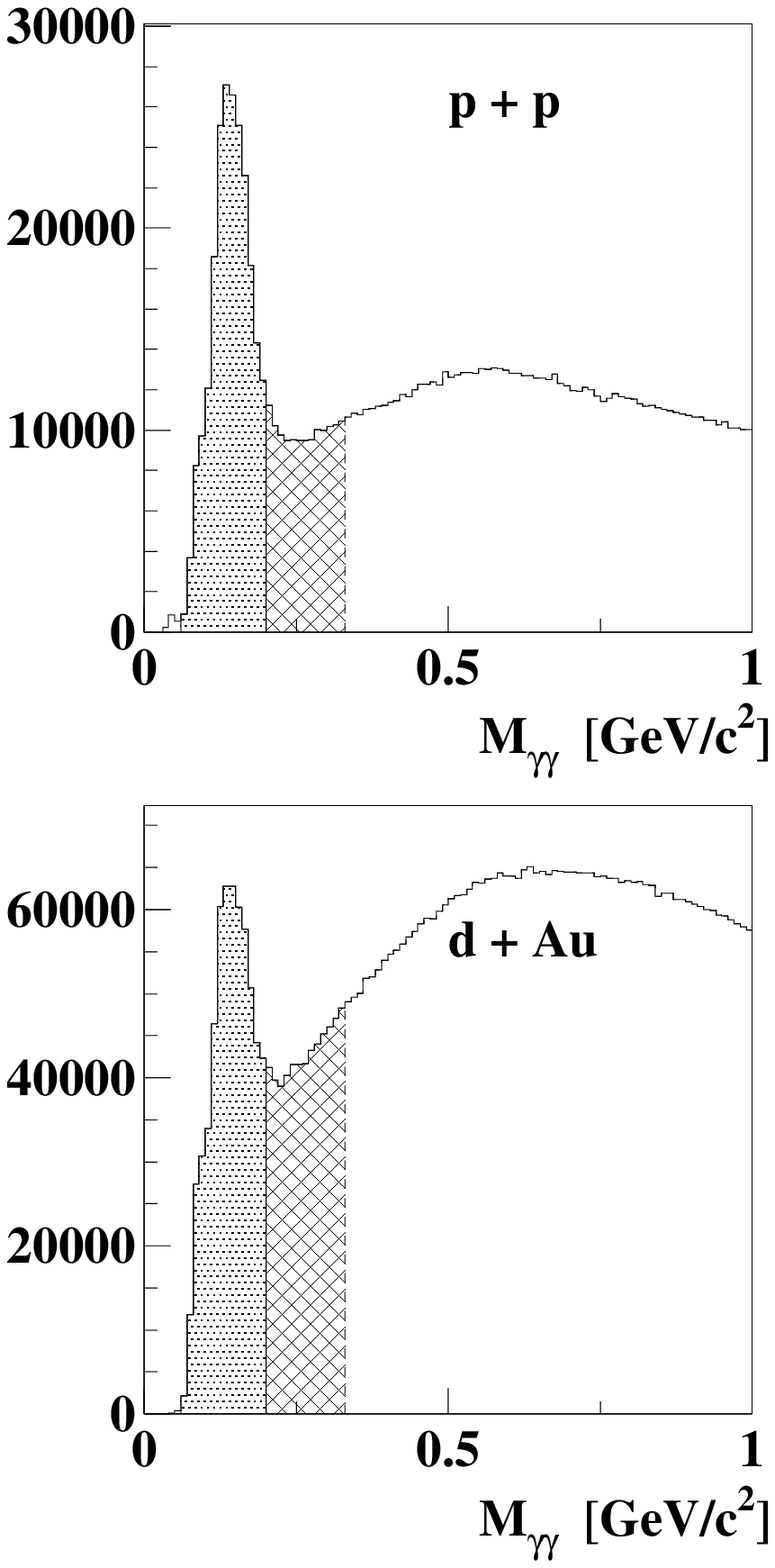} 
\includegraphics[height=0.6\textwidth,clip=]{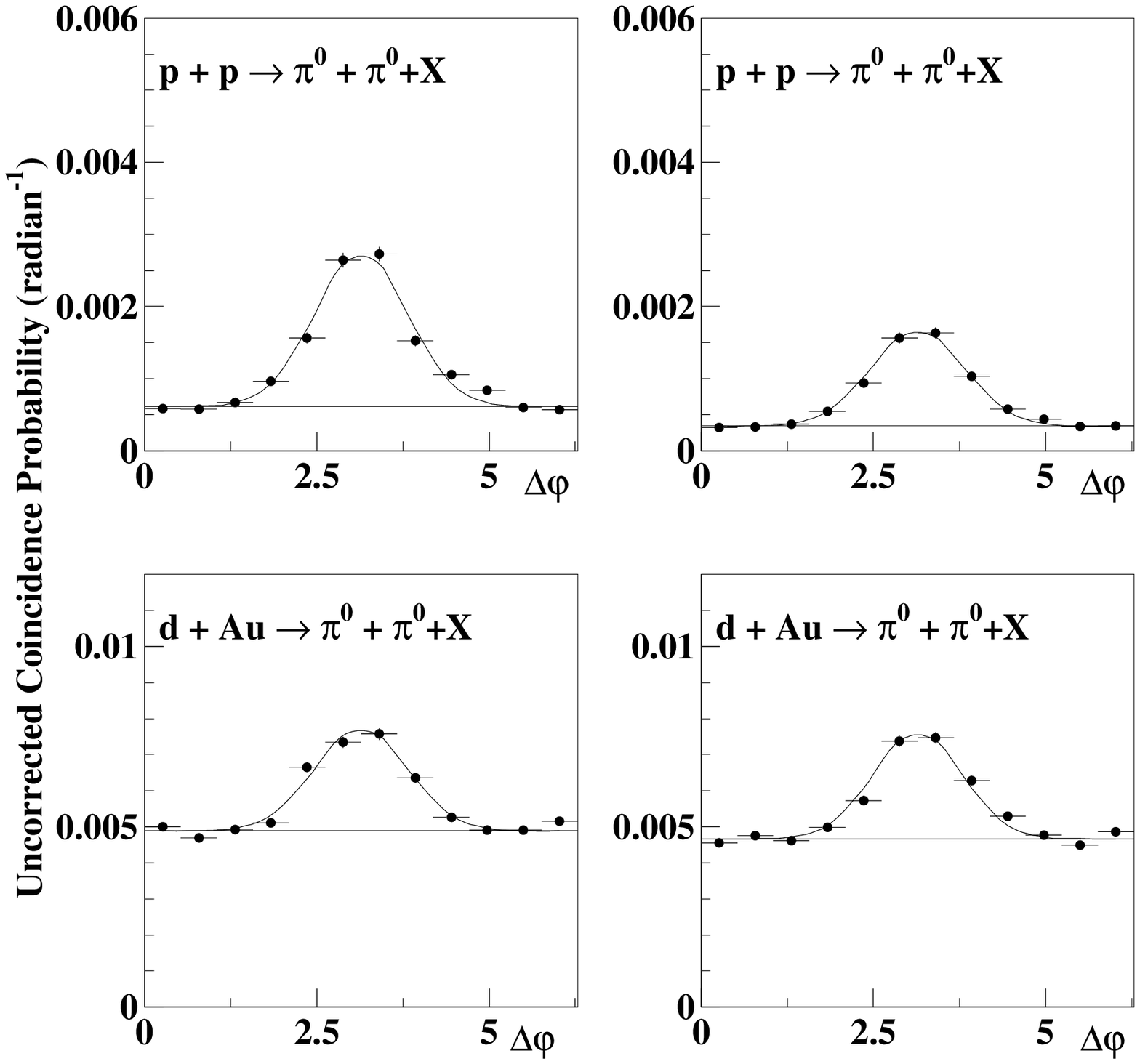} 
\caption{Off-peak analysis. Comparison of azimuthal correlations between a forward $\pi^{0}$ and a mid-rapidity photon pair, where the pair is selected within the $\pi^{0}$ mass window (center) and slightly off-peak (right), as indicated by the invariant mass spectra (left). On the top (bottom) row: p+p (d+Au) interactions. Particles are selected using the higher $p_{T}$ cut.}\label{5.emcoff}
\end{figure}

The first step towards a corrected coincidence probability  is represented by the study of the background below the $\pi^{0}$ peak and its effect on azimuthal correlations. From simulations studies, it is clear that the main background component comes from pairs of a photon with a high energy charged hadron. Such pairs likely originate within the same jet; therefore, they provide a good characterization of the jet itself and they are expected to be as correlated in $\Delta\varphi$ as neutral pions. In order to check this, correlations have been computed between a forward $\pi^{0}$ and a mid-rapidity photon pair selected in the invariant mass window $(0.20<M_{\gamma\gamma}<0.33)\mathrm{\,GeV/c^{2}}$, relatively far from the nominal value of the $\pi^{0}$ mass. Figure \ref{5.emcoff} shows the criteria to select the photon pairs and the inclusive correlations between a forward pion and a mid-rapidity pair. Pairs selected off the pion peak appear very similar compared to correlated pion candidates. Also the signal widths, as extrapolated from the Gaussian fit and indicated in figure, appear to be consistent between the two selections. We can therefore conclude that the hadronic background underlying the $\pi^{0}$ peak in the invariant mass spectrum contains similar information to the $\pi^{0}$ which does not affect our conclusions.

The off-peak correlation is also used for extracting the background component from the ``peak'' region. The coincidence probability can be in fact disentangled into two components: the $\pi^{0}$-$\pi^{0}$ (signal-signal) component and the $\pi^{0}$-pair (signal-background) component. In order to obtain the first one in the (BEMC) $\pi^{0}$ mass window (peak area) we can subtract from the measured  ``on-peak'' correlation the $\pi^{0}$-background correlation estimated from the off-peak region, where the background is by far the largest component. This can be done using the following formula:

\begin{equation}\label{5.eq.back}
\left.\frac{dN^{\pi^{0}\pi^{0}}}{d(\Delta\varphi)}\right|_{P}=\left[\left.\frac{dN}{d(\Delta\varphi)}\right|_{P}-\left.\frac{dN}{d(\Delta\varphi)}\right|_{O}\cdot\frac{\left.N^{b}\right|_{P}}{\left.N^{b}\right|_{O}}\right]\cdot\left[1-\frac{\left.N^{s}\right|_{O}}{\left.N^{s}\right|_{P}}\cdot\frac{\left.N^{b}\right|_{P}}{\left.N^{b}\right|_{O}}\right]^{-1}
\end{equation}

\begin{figure}
\includegraphics[width=0.99\textwidth, clip=]{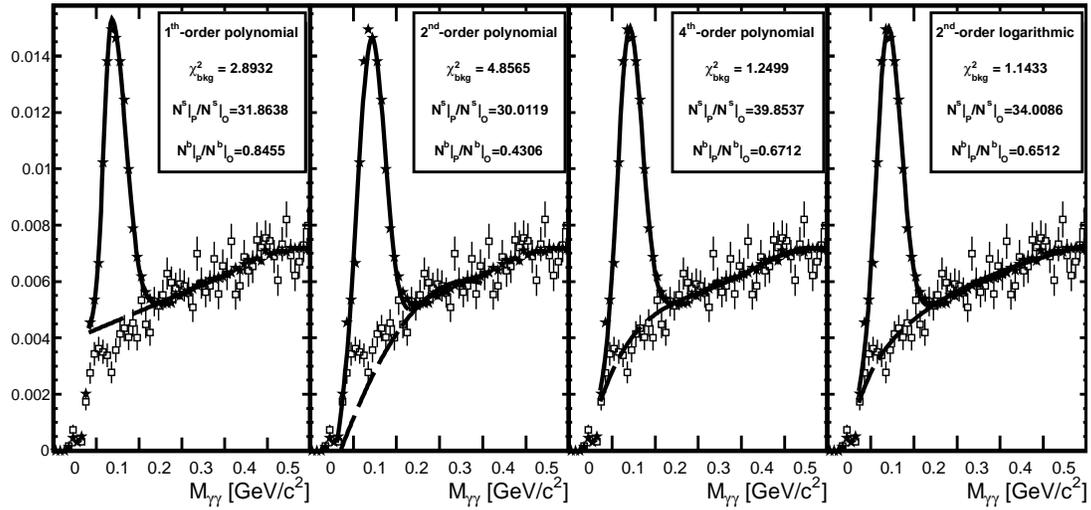} 
\caption{Different fits for the background component are used to fit $M_{\gamma\gamma}$ in data (stars), together with an ``asymmetric Gaussian'' function used for the peak. The background component from simulation (squares) has been scaled up with a factor 1.7 and superimposed to the off-peak background in data. The comparison ($\chi^{2}$) between the scaled simulated background and the background component of the fit to data in the $\pi^{0}$ peak region is indicated. Relative proportions of signal and background components between mass region and off-mass are also indicated.}\label{5.fitsdatadue}
\end{figure}

In this equation, the azimuthal correlations between two neutral pions (signal-signal correlation: $dN^{\pi^{0}\pi^{0}}/d(\Delta\varphi)|_{P}$), where the mid-rapidity pion is selected in the $\pi^{0}$ mass region (P), is obtained by subtracting from the measured ``on-peak'' correlation ($dN/d(\Delta\varphi)|_{P}$) the $\pi^{0}$-background correlation measured ``off-peak'' ($dN/d(\Delta\varphi)|_{O}$), scaled with the ratio of number of background pairs in the peak region ($N^{b}|_{P}$) and off-peak ($N^{b}|_{O}$) in order to reproduce the amount of background present in the $\pi^{0}$ mass window. The scaling factor on the right-hand side in Eq.\ref{5.eq.back} accounts for the possibility that the tail of the $\pi^{0}$ peak extends into the off-mass region. When the $\pi^{0}$ peak is fully contained into the on-peak region ($N^{s}|_{P}=\mathrm{max}$) and no signal pairs are to be found in the off-peak region ($N^{s}|_{O}=0$), the off-peak correlation is a purely $\pi^{0}$-background correlation and the scaling factor reduces to one. This formula has the advantage that it is not necessary to know the signal/background ratio to perform the subtraction. In fact, the only knowledge required is the shape of the signal and background components or, more precisely, the relative percentage of the two distribution (independently) in the two mass regions. These relative proportions ($N^{s}|_{P}/N^{s}|_{O}$ for the signal and $N^{b}|_{P}/N^{b}|_{O}$ for the background) can be estimated in different ways. One can either integrate or fit the different components of the $M_{\gamma\gamma}$ spectrum in simulation, or, alternatively, one can fit the global $M_{\gamma\gamma}$ distribution obtained from data, using the fitting functions from simulation as guideline. Figure \ref{5.fitsdatadue} shows four different fitting functions tested on data (stars) and compared with the background as obtained from simulation (squares). Since the background component is underestimated in simulation, it has been scaled up to fit the data in the off-peak region\footnote{The hadronic and combinatorial component of the background have been scaled by a factor $1.7$, chosen by performing the Kolmogorov test between simulation and data.}. The peak has been fitted using an ``asymmetric Gaussian'' function (i.e. two half Gaussians sharing the mean value but with different variances), while the background has been fitted with polynomial of $\mathrm{1^{th}}$, $\mathrm{2^{nd}}$ and $\mathrm{4^{th}}$ order and with a $\mathrm{2^{nd}}$ order logarithmic function, the last of which provides the best $\chi^{2}$ when compared to the (scaled) background from simulation \cite{Jory}. The background estimated from integrating simulated spectra, as well as from this last fit on data, will be used to do the subtraction, while the other fits will be used to estimate the systematic uncertainty of the final (corrected) yield. 

\subsection{Efficiency correction}

\begin{figure}
\begin{center}
\includegraphics[width=0.48\textwidth, height=0.30\textwidth]{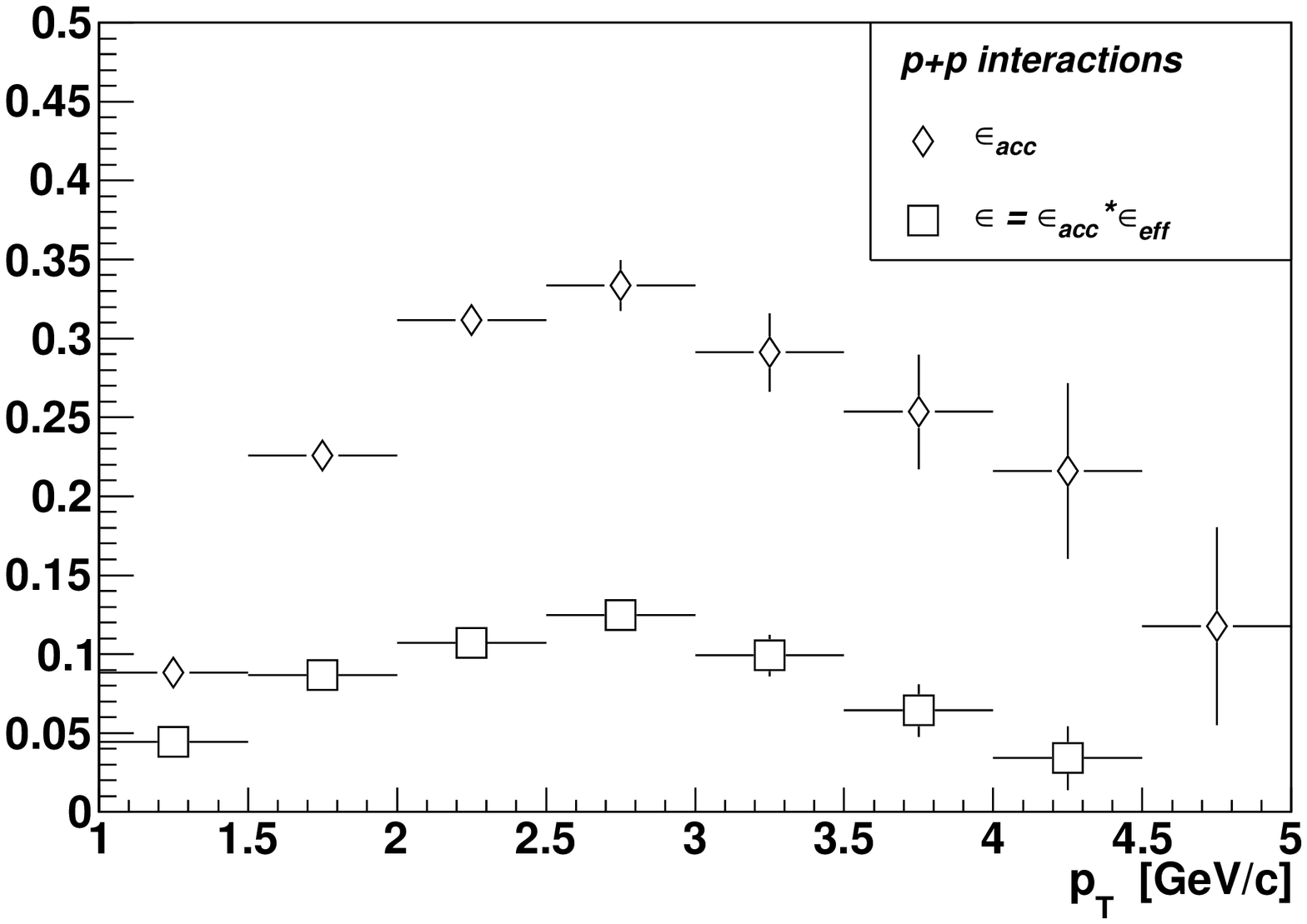} 
\includegraphics[width=0.48\textwidth, height=0.30\textwidth]{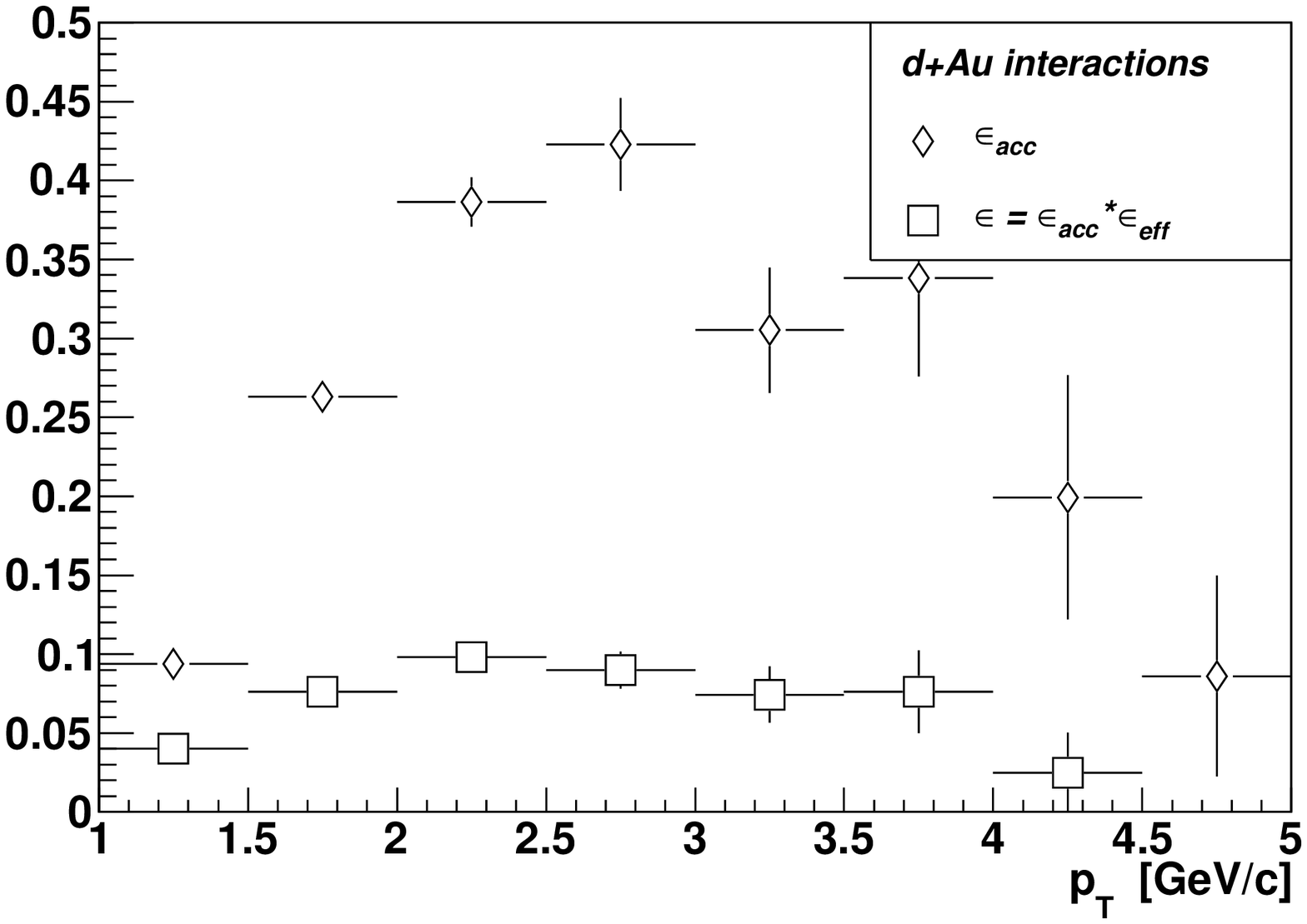} 
\end{center}
\caption{BEMC acceptance and reconstruction efficiency for $\pi^0$ candidates in p+p (left) and d+Au interaction (right) as function of $p_{T}$.}\label{5.efficiency}
\end{figure}

\begin{figure}
\begin{center}
\includegraphics[width=1\textwidth, height=0.50\textwidth]{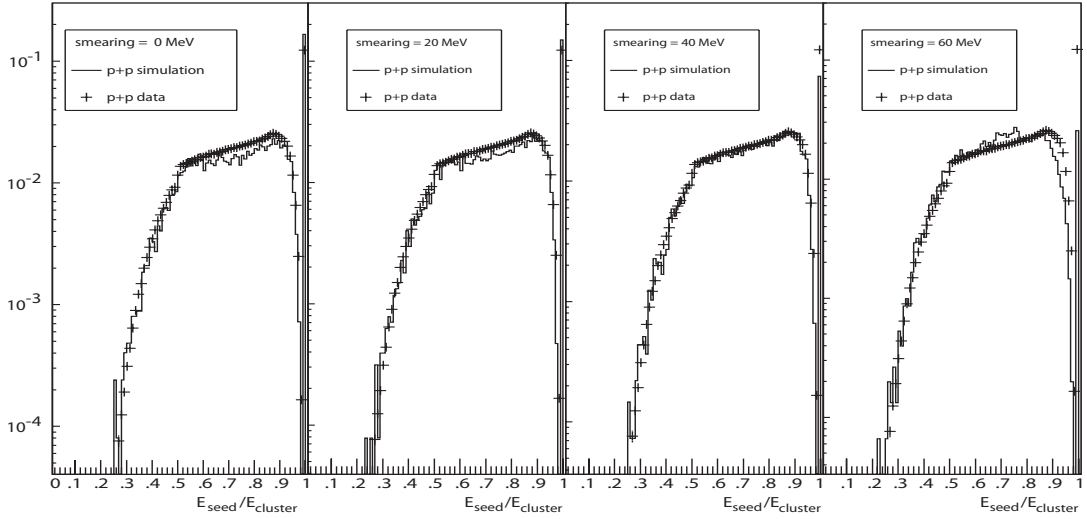} 
\end{center}
\caption{Simulation-Data comparison between the variable $x=E_{seed}/E_{cluster}$ for different values of the tower energy smearing for p+p interactions (higher $p_{T}$ selection). Note the significant difference at $x=1$ (single tower clusters).}\label{5.quality}
\end{figure}

Once the prescription to subtract the background has been determined, it is possible to apply an efficiency correction to the azimuthal correlation. The reconstruction efficiency and detector acceptance is calculated by performing an association analysis on the reconstructed pair of clusters. Each cluster is associated with a simulated track pointing to it. When more than one track is pointing to the same cluster, the one contributing with the highest energy fraction to the total cluster energy is selected. The detector acceptance $\epsilon_{acc}$ is calculated as the ratio between reconstructed and simulated neutral pions with pseudo-rapidity $|\eta|<0.9$. The number of reconstructed pions is incremented when a pair of clusters, both within the detector acceptance, are associated with the two decaying photons from the same $\pi^{0}$. The reconstruction efficiency $\epsilon_{rec}$ takes into account, in addition, the quality cut applied to the pion candidate to reduce the background, namely the asymmetry cut $Z_{\gamma\gamma}<0.7$ and the ``quality cut'' which requires the peak tower to carry at least $90\%$ of the cluster energy.
Figure \ref{5.efficiency} shows the $p_{T}$ dependence of the acceptance $\epsilon_{acc}$ and the total efficiency $\epsilon=\epsilon_{acc}\times\epsilon_{rec}$ in p+p and d+Au interactions. The acceptance (and thus the efficiency) is lower at low $p_{T}$, due to a higher background contamination, and at large $p_{T}$ because of the opening angle cut applied by the clusterfinder. The efficiency proves to be independent on the multiplicity of the event, which is the main difference between triggered data and minimum bias simulation. However, the distribution of the variable $x=E_{seed}/E_{cluster}$, on which the quality cut is applied, appears to be different in simulation and data. In order to make the two distribution comparable, a Gaussian energy smearing has been applied independently to all BEMC towers before the clustering procedure. Figure \ref{5.quality} shows the effect of different smearing values on the quality variable. This provides access to systematic uncertainties on the value of the efficiency, calculated as the average of the values of the efficiency in the two most plausible cases (second and third panel in Figure \ref{5.quality}), while the other cases provide the limits for the systematic uncertainty. The efficiency, when integrated over the whole $p_{T}$ range, turns out to be $\epsilon_{pp}=0.0974\pm0.0032\mathrm{(stat)}\pm0.0067\mathrm{(syst)}$ (and $\epsilon_{dAu}=0.0854\pm0.0031\mathrm{(stat)}\pm0.0039\mathrm{(syst)}$) for pions with $p_{T}>1.5\mathrm{\,GeV/c}$, $\epsilon_{pp}=0.0570\pm0.0001\mathrm{(stat)}\pm0.0013\mathrm{(syst)}$ (and $\epsilon_{dAu}=0.0475\pm0.0011\mathrm{(stat)}\pm0.0007\mathrm{(syst)}$) for pions with $p_{T}>1.0\mathrm{\,GeV/c}$.

\subsection{Systematics}
Once the BEMC reconstruction efficiency is estimated, it is possible to correct the coincidence probability between a forward $\pi^{0}$ and a mid-rapidity $\pi^{0}$. The two-step process consists on subtracting first the $\pi^{0}$-background contribution from the azimuthal correlation, using Equation \ref{5.eq.back}, and then scaling it for the $\pi^{0}$ efficiency. As for the FMS-TPC correlations, the correlation obtained is fitted with a constant background plus a periodic Gaussian centered at $\Delta\varphi=\pi$, so that the correlated yield can be extracted as the area of the Gaussian function. We estimated the shape of signal and background in the $\pi^{0}$ invariant mass by fitting the histogram components from simulation and we scale the correlation so obtained by an overall efficiency factor, integrated over $p_{T}$, indicated in the previous section. The correlated yield for p+p (and d+Au) calculated in this way reads:      
$A_{pp}=0.023\pm0.002$ ($A_{dAu}=0.025\pm0.004$) for higher $p_{T}$ cuts ($p_{T}^{FMS}>2.5\mathrm{\,GeV/c}$, $p_{T}^{FMS}>p_{T}^{EMC}>1.5\mathrm{\,GeV/c}$)  and $A_{pp}=0.059\pm0.003$ ($A_{dAu}=0.067\pm0.014$) for lower $p_{T}$ cuts ($p_{T}^{FMS}>2.0\mathrm{\,GeV/c}$, $p_{T}^{FMS}>p_{T}^{EMC}>1.0\mathrm{\,GeV/c}$). However, due to the larger uncertainties affecting the FMS-BEMC measurement, a number of additional checks have been performed in order to extract the correlated yield. These include different efficiency and background estimates, checks on data sanity and kinematical cuts. They are listed below, while the values for the correlated yield in all cases are summarized at the end (Table \ref{5.syssum}. This first reference measurement is referred as number zero.).

\begin{description}
\item[1. Transverse momentum weighted efficiency.]
The $p_{T}$ distribution of mid- rapidity pions is well described by simulation (Figure \ref{5.comparisonsimu}). Therefore we can use an overall efficiency factor, integrated over the whole $p_{T}$ range. However, a different approach can be to estimate the efficiency for single $p_{T}$ bins and to weight them using the $p_{T}$ distribution from data. This leads again to a single efficiency factor, which does not differ much from the value used in the reference measurement. 

\item[2. Transverse momentum dependent efficiency and background.]
A similar approach consists in performing the efficiency correction (and, before that, the background subtraction) for each single $p_{T}$ bin. In this way the background component is estimated (as before, from fit on simulation) in each bin and the efficiency correction is applied accordingly. The total azimuthal correlation is eventually obtained by summing all these components. 

\item[3. Cluster ``quality cut''.]
In order to test the consistency of the variable $E_{seed}/E_{cluster}$, efficiency and (fitted) background have been estimated using the stricter condition $E_{seed}/E_{cluster}>95\%$ (a). 
Analogously we calculated the correlated yield for $\pi^{0}$ correlations where this cut has not been applied to the cluster selection (b).
  
\item[4. Energy asymmetry.]
A check on the energy asymmetry cut has been performed by requiring a stricter ($Z_{\gamma\gamma}<0.5$) condition. 

\item[5. Background correction with different $M_{\gamma\gamma}$ windows.]
The larger uncertainty in measurement of the $\pi^{0}$-$\pi^{0}$ correlation is represented by the quantification of the background component in the BEMC $M_{\gamma\gamma}$ spectrum. For this reason, a series of different methods to estimate the background below the $\pi^{0}$ peak have been pursued. At first, we can use a different invariant mass window to define the off-mass pair in the usual formula \ref{5.eq.back}. Instead of estimating the correlated background from correlations with a pair in the region $(0.20<M_{\gamma\gamma}<0.33)\mathrm{\,GeV/c^{2}}$, we use the further region $(0.33<M_{\gamma\gamma}<0.46)\mathrm{\,GeV/c^{2}}$. In this case, the efficiency and the background shape from simulation stay the same as in the reference measurement.  

\item[6. Background correction with $M_{\gamma\gamma}$ windows based on resolution.]
A different background estimate can be achieved by select $M_{\gamma\gamma}$ windows based on the peak mean value and its resolution. Data show that the $\pi^{0}$ peak in the invariant mass spectrum in data ($\mu=(0.139+0.031-0.026)\mathrm{\,GeV/c^{2}}$ for hight $p_{T}$ p+p interaction)\footnote{The asymmetry in the signal width is due to the ``asymmetric Gaussian'' used for the fit.} is broader than in simulation ($\mu=(0.131+0.032-0.020)\mathrm{\,GeV/c^{2}}$ for the same set). Moreover, its position is centered at a slightly different mean value. In order to make the selection consistent between data and simulation, cluster pairs in the invariant mass spectra have been selected in two regions $\left[\mu-2.7\sigma_{-};\mu+1.9\sigma_{+}\right]$ for the peak and $\left[\mu+1.9\sigma_{+};\mu+6.1\sigma_{+}\right]$ for the off-mass selection. 
 
\item[7. Background estimated from integrals in simulation.]
Instead of estimating signal and background in the invariant mass spectrum by fitting the different component in simulation, one can take directly the integrals of such components in the $M_{\gamma\gamma}$ regions of interest. The difference between fit and histograms is particularly significant for low mass values in the hadronic background and the tail of the signal peak in the off-mass region ($M_{\gamma\gamma}>0.20\mathrm{\,GeV/c^{2}}$).  

\item[8. Background estimated from fit on data.]
Alternatively, one can estimate the signal and the background shapes by fitting the invariant mass spectrum obtained from data. As shown in Figure \ref{5.fitsdatadue}, the fitting function (on data) that best represents the (scaled) background shape from simulation is a second-order logarithmic function (a). Another fitting function that provides a reasonable comparison with simulated background is the fourth-order polynomial (b). As an limit case, we can fit the invariant mass on data using either a first-order (c) or a second-order polynomial (d). In the first case the background is overestimated in comparison with simulation. In the second case (but only for higher $p_{T}$ selections) the background is underestimated.

\end{description}

\begin{table}
\begin{tabular}{c c}
\begin{tabular}{| c | c  | c |}
\hline Check & $A_{pp}$ & $A_{dAu}$\\ \hline \hline
0 & 0.023$\pm$0.002 & 0.025$\pm$0.004 \\ \hline
1 & 0.024$\pm$0.002 & 0.026$\pm$0.004 \\ \hline
2 & 0.023$\pm$0.007 & 0.026$\pm$0.007 \\ \hline
3a &  0.027$\pm$0.002 & 0.023$\pm$0.004 \\ \hline
3b &  0.023$\pm$0.001 & 0.025$\pm$0.004 \\ \hline
4 & 0.027$\pm$0.002 & 0.022$\pm$0.005 \\ \hline
5 & 0.023$\pm$0.002 & 0.025$\pm$0.003 \\ \hline
6 & 0.025$\pm$0.002 & 0.028$\pm$0.004 \\ \hline
7 & 0.022$\pm$0.002 & 0.023$\pm$0.004 \\ \hline
8a &  0.022$\pm$0.002 & 0.024$\pm$0.003 \\ \hline
8b &  0.022$\pm$0.002 & 0.025$\pm$0.004 \\ \cline{2-3}
8c &  0.018$\pm$0.002 & 0.020$\pm$0.003 \\ \cline{2-3}
8d &  0.027$\pm$0.002 & 0.033$\pm$0.004 \\ \hline
\end{tabular}
&
\begin{tabular}{| c | c  | c |}
\hline Check & $A_{pp}$ & $A_{dAu}$\\ \hline \hline
0 & 0.059$\pm$0.003 & 0.067$\pm$0.014 \\ \hline
1 & 0.064$\pm$0.004 & 0.067$\pm$0.014 \\ \hline
2 & 0.066$\pm$0.019 & 0.067$\pm$0.031 \\ \hline
3a &  0.066$\pm$0.005 & 0.070$\pm$0.014 \\ \hline
3b &  0.060$\pm$0.003 & 0.060$\pm$0.003 \\ \hline
4 & 0.060$\pm$0.004 & 0.061$\pm$0.005 \\ \hline
5 & 0.056$\pm$0.003 & 0.070$\pm$0.010 \\ \hline
6 & 0.067$\pm$0.003 & 0.067$\pm$0.014 \\ \hline
7 & 0.060$\pm$0.003 & 0.067$\pm$0.014 \\ \hline
8a &  0.056$\pm$0.003 & 0.063$\pm$0.004 \\ \hline
8b &  0.053$\pm$0.005 & 0.067$\pm$0.014 \\ \cline{2-3}
8c &  0.055$\pm$0.003 & 0.067$\pm$0.010 \\ \cline{2-3}
8d &  0.048$\pm$0.004 & 0.063$\pm$0.011 \\ \hline
\end{tabular}
\end{tabular}
\caption{Summary of the systematic analysis for the extraction of the correlated yield in p+p and d+Au interactions for higher (left panel) and lower (right panel) $p_{T}$ selections.}\label{5.syssum}
\end{table}

This procedure provides access to the systematic uncertainty on the value of the correlated yield. The main contribution to such uncertainty is carried by the background estimation. The value for the correlated yield and its statistical errors are taken from the reference measurement. The systematic uncertainty range is conservatively defined by the two outermost values found in the study. This leads to the following values for the correlated yield: $A_{pp}=0.023\pm0.002^{+0.003}_{-0.006}$ ($A_{dAu}=0.025\pm0.004^{+0.008}_{-0.005}$) for higher $p_{T}$ cuts, $A_{pp}=0.059\pm0.003^{+0.009}_{-0.010}$ ($A_{dAu}=0.067\pm0.014^{+0.008}_{-0.003}$) for lower $p_{T}$ cuts. 

%%%%%%%%%%%%%%%%%%%%%%%%%%%%%%%%%

\section{FMS-FMS correlations}

\begin{figure}
\begin{center}
\includegraphics[width=0.60\textwidth]{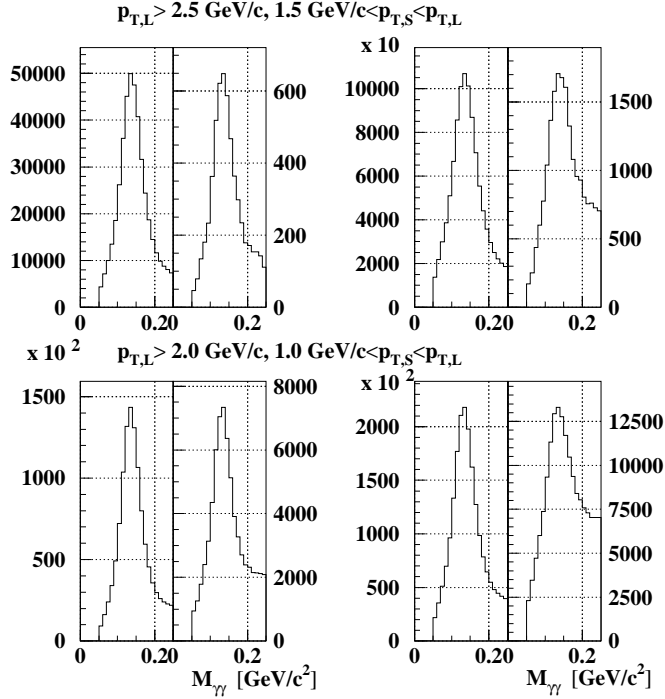} 
\end{center}
\caption{Invariant mass distributions for photon pairs with higher (top row) and lower $p_{T}$ cuts (bottom row) in p+p (left) and d+Au collisions (right). Each quadrant shows invariant mass for leading and sub-leading (associated) pion candidates.}\label{4.fms}
\end{figure}

\begin{figure}
\begin{center}
\includegraphics[width=0.60\textwidth]{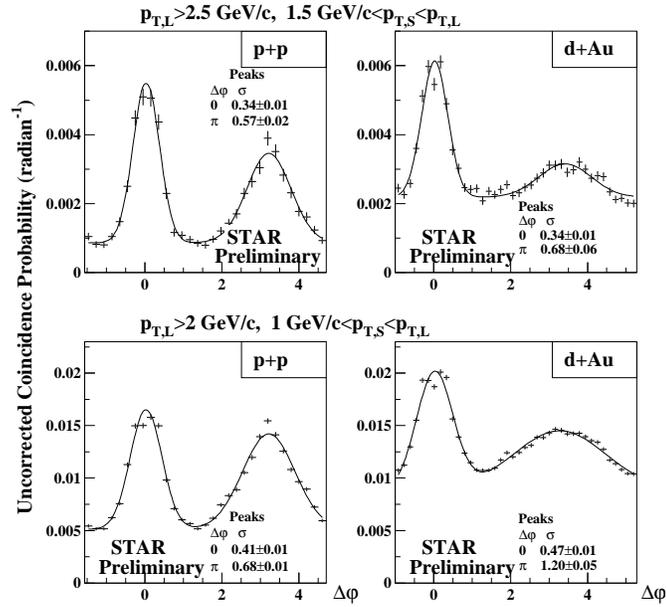} 
\end{center}
\caption{Uncorrected coincidence probability versus azimuthal angle difference between two forward pions in p+p (left) and d+Au collisions (right) for two $p_{T}$ selections. Data are fitted with a constant plus two Gaussian functions.}\label{4.fmsdue}
\end{figure}

As we mentioned before, the saturation scale can be approached by requiring both particles to be detected in the forward region. The large acceptance of the FMS and its fine granularity allows us to reconstruct multiple pion candidates and therefore to compute di-pion correlations in the forward region, where the lowest $x$ value of the probed gluons is accessed. The coincidence probability for di-pion production has been measured imposing similar kinematical cuts than for forward - mid-rapidity correlations. Neutral pions are reconstructed from pairs of photon clusters in the FMS fiducial volume $2.5<\eta<4.0$. Good pion candidates are required to have an energy asymmetry $Z_{\gamma\gamma}<0.7$ and an invariant mass in the range $(0.07<M_{\gamma\gamma}<0.25)\mathrm{\,GeV/c^{2}}$. The pair with the largest $p_{T}$ is selected as the leading (trigger) $\pi^{0}$ and it's azimuthal coordinate is compared inclusively to those of all other (associated) pion candidates. As before, two sets of $p_{T}$ cuts have been used to explore the $p_{T}$ dependence of possible saturation effects. For consistency, the two classes of cuts reflect the ones used before: higher $p_{T}$ selection requires $p_{T}^{trg}>2.5\mathrm{\,GeV/c}$ and $p_{T}^{trg}>p_{T}^{asc}>1.5\mathrm{\,GeV/c}$ while the lower cut requires $p_{T}^{trg}>2.0\mathrm{\,GeV/c}$ and $p_{T}^{trg}>p_{T}^{asc}>1.0\mathrm{\,GeV/c}$. The invariant mass distributions of leading and associated pion candidates for these two classes of events, in both p+p and d+Au collisions, are shown in Figure \ref{4.fms}. The corresponding azimuthal correlations are shown in Figure \ref{4.fmsdue}.

The first clear difference we can notice by comparing FMS-FMS and FMS-BEMC (or FMS-TPC) correlations is the presence of a second peak, centered around $\Delta\varphi=0$. The so called \emph{near-side peak} is the contribution to the coincidence probability coming from correlations of particles originating in the same jet. Since they belong to the same jet, their azimuthal directions are similar and the difference $\Delta\varphi$ is close to zero. This peak is not expected to undergo modification due to saturation effects. On the contrary, the \emph{away-side} peak is expected to broaden and disappear. As before, data are fitted with a constant plus two Gaussian functions centered at $\Delta\varphi=0$ and $\Delta\varphi=\pi$, made periodic at the interval extremes. Currently, no background corrections are available for the FMS $\pi^{0}$ reconstruction. As a consequence, FMS-FMS correlations are uncorrected, as well as the forward-midrapidity correlations for what concerns the FMS. However, the FMS correction is not expected to be significantly different between p+p and d+Au interactions.

Systematics studies have been performed for this analysis and results are reported in \cite{Braidot:2010ig}. The impact of the combinatorial background has been studied by looking at azimuthal correlations between pairs of cluster with an invariant mass in the range $(0.25<M_{\gamma\gamma}<0.45)\mathrm{\,GeV/c^{2}}$, relatively distant from the nominal value of the $\pi^{0}$ mass. Such correlations present away and near-side peaks qualitatively consistent with the ``on-peak'' correlations. A small impact is expected for jet-like correlations where the combinatorial background is produced between pairs in the same jet.

The sensitivity of di-pion correlations to the gluon distribution function has been systematically studied. The latest available HIJING versions present only a limited choice of parton distribution functions, all of which predate HERA discovery of the rapid growth of the gluon density at low-$x$. The dependence on the PDF has therefore been studied using PYTHIA 6.222. Azimuthal correlations in p+p interactions were found to be consistent (both in correlated yield and peak width) with gluon distributions that include a rapid rise of the gluon density at low-$x$.

The effect of multiplicity on the correlations has been studied in embedded PYTHIA plus GEANT events. This was done to study the possibility that the larger multiplicity which characterize d+Au interactions, compared to p+p, could cause the loss of correlation seen in these results.
Azimuthal correlations between two forward pions have been simulated using pure PYTHIA events and PYTHIA events embedded into minimum bias peripheral and central d+Au interactions. The larger multiplicity in d+Au affects primarily the $\pi^{0}$ reconstruction, causing a higher combinatoric background. The features of the near-side peak are reproduced consistently in p+p and d+Au interactions. The away-side peak in embedded events does not show any sign of broadening or suppression. We conclude the additional multiplicity in d+Au collisions does not affect the shape of the azimuthal correlation.

%\include{chapter_5}

%%% Chapter heading commands %%%

\chapter{Results and discussion}

In this last chapter, we quantitatively compare the azimuthal correlations described in Chapter \ref{chapter_results}. This is done with emphasis on the dependence of possible saturation effects on the transferred momentum $Q^{2}$ of the interaction (characterized through the dependence on the transverse momentum $p_{T}$ of the detected particles), the longitudinal momentum fraction $x$ of the gluon field probed in the nuclear medium (through the pseudo-rapidity $\eta_{asc}$ of the associate particle) and the density of the target (by characterizing the centrality of the collision based on the multiplicity of the event).  

\begin{figure}\begin{center}
\includegraphics[width=1.0\textwidth]{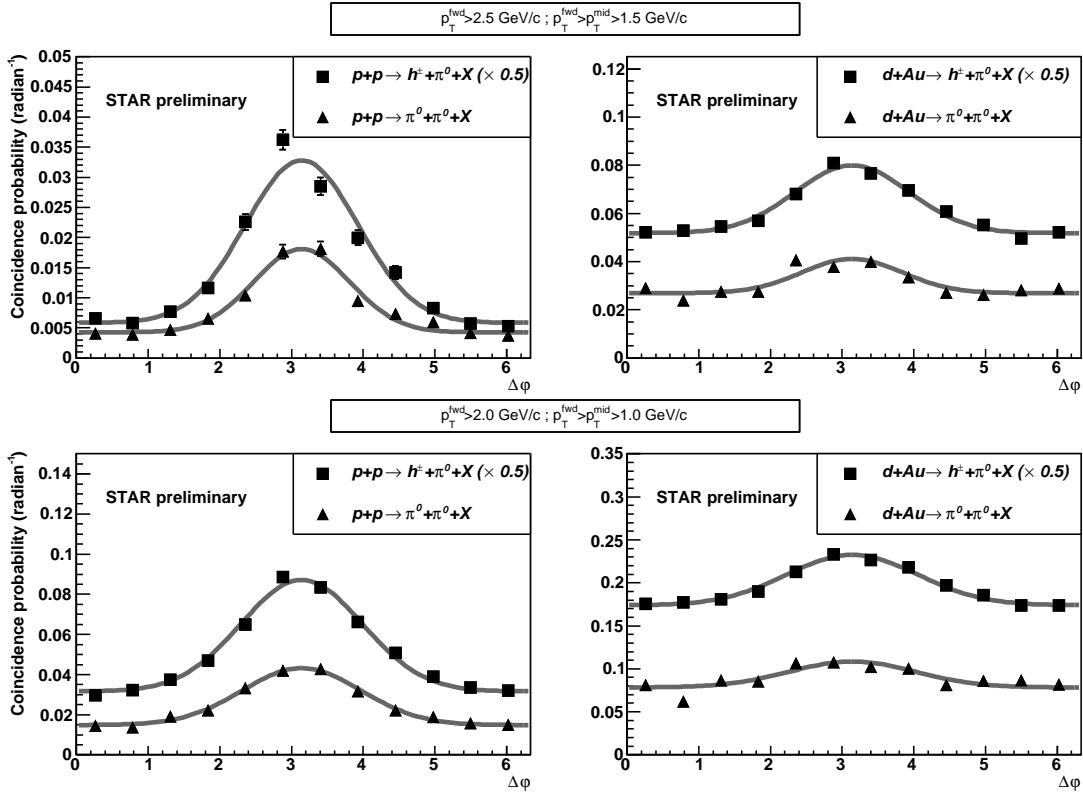} 
\caption{Coincidence probability between a forward $\pi^{0}$ and either a mid-rapidity charged hadron (squares) or a mid-rapidit $\pi^{0}$ (triangles) in p+p (left) and d+Au interactions for two $p_{T}$ selections. Data are corrected for BEMC and TPC background and efficiencies. FMS-TPC correlations have been scaled of 0.5.}\label{6.midSummary}\end{center}
\end{figure}

As a first step, the measurements of the coincidence probability between a forward $\pi^{0}$ and a mid-rapidity particle are studied and compared after being corrected for background and efficiency. Figure \ref{6.midSummary} shows the azimuthal correlations between a forward $\pi^{0}$ in the FMS and either a mid-rapidity $\pi^{0}$ in the BEMC (triangles) or a charged track in the TPC (squares) in the same kinematic regime. FMS-TPC correlations have been corrected for luminosity effects, by subtracting the constant (uncorrelated) pile-up contribution, and for the TPC efficiency. In the FMS-BEMC correlations the (correlated) background contribution to the BEMC invariant mass spectra have been subtracted from the correlations before applying the BEMC efficiency correction. In order to easily compare their features, FMS-TPC correlations have been scaled by 0.5. Correlations are shown for p+p and d+Au interactions using two different $p_{T}$ thresholds. A function composed of a constant plus a periodic Gaussian has been used to fit the data. The results of the fit are listed on Table \ref{6.midSumTable}.

The Gaussian fit on the correlated contribution provides the following numbers for the standard deviation: for higher $p_{T}$ cuts, the signal width goes from $\sigma_{pp}=0.783\pm0.036$ in p+p to $\sigma_{dAu}=0.796\pm0.036$ in d+Au, with a difference of $\Delta\sigma=0.013\pm0.051$; for lower $p_{T}$ cuts, the signal width goes from $\sigma_{pp}=0.844\pm0.025$ in p+p to $\sigma_{dAu}=0.914\pm0.025$ in d+Au, with a difference of $\Delta\sigma=0.070\pm0.035$. Statistical errors are too large to allow us to claim any significant broadening in going from p+p to d+Au. The main conclusions is again that the back-to-back contribution to the coincidence probability is still clearly visible and not significantly affected by modifications in the gluon field of the gold nucleus.  

The results of the fit on the correlations are summarized in Table \ref{6.midSumTable}. The comparison of the signal width from the fit shows good agreement between $\pi^{0}$-$h^{\pm}$ (FMS-TPC) and $\pi^{0}$-$\pi^{0}$ (FMS-BEMC). The clear back-to-back correlation peak indicates that we are not yet probing the saturation region. The ratio between correlated yields (the integral of the Gaussian in the range $\left[0,2\pi\right]$) in $\pi^{0}-\pi^{0}$ and $\pi^{0}-h^{\pm}$ correlations is consistently equal to $R_{A}\simeq4.5$ for the higher $p_{T}$ selection and $R_{A}\simeq4.0$ for the lower $p_{T}$ selection, both in p+p and d+Au (this will be further discussed in Section \ref{sec6.ratio}). The ratio between the uncorrelated background components $b$ is significantly similar between p+p and d+Au, and reasonably smaller for lower $p_{T}$ range ($R_{b}\simeq3.6$) than for higher cuts ($R_{b}\simeq4.3$). This reflects the larger uncertainty in quantifying the background contribution in the BEMC than in the TPC, especially at low $p_{T}$. FMS-BEMC interactions present, for what concerns the correlated signal width $\sigma$, a slightly narrower peak than for FMS-TPC correlations. This is due to the choice of relatively high energy thresholds that allows a better $\pi^{0}$ reconstruction in the BEMC by effectively selecting higher $p_{T}$ pion candidates (thus leading to a narrower peak)\footnote{An early step of the analysis, using lower energy threshold in the BEMC pion reconstruction can be found in \cite{Braidot:2009ji}.}. 

In both correlations the background component is larger in d+Au than in p+p correlations, and for lower $p_{T}$ particles, as expected in events with higher multeplicity. On the contrary, the correlated yield does not appear to change significantly from p+p to d+Au interactions. Analogously, the correlated peak width, given by the standard deviation $\sigma$ of the Gaussian fitting function, does not significantly differ from p+p to d+Au interactions. The large statistical errors do not allow in fact any conclusion on the apparent slight broadening from p+p to d+Au, not even at low $p_{T}$ where the effect seems larger. The conclusion is that, at these rapidities, the back-to-back contribution to the coincidence probability is still clearly visible in both $p_{T}$ selections and mostly not affected by modifications in the gluon field of the gold nucleus.

\begin{table}\begin{center}\begin{tabular}{|c|c|c|c|c|c|c|}
\hline
$\mathrm{p_{T}}$ cuts & int. & correlation & $b$ & $A$ & $\sigma$\\
\hline\hline
\multirow{4}{*}{higher} & \multirow{2}{*}{p+p} & $\mathrm{\pi^{0}+h^{\pm}}$ & $0.012\pm 0.001$ & $0.106\pm0.005$ & $0.783\pm0.038$\\
\cline{3-6}
&  & $\mathrm{\pi^{0}+\pi^{0}}$ & $0.004\pm 0.001$ & $0.023\pm0.002$ & $0.669\pm0.055$\\
\cline{2-6}
&\multirow{2}{*}{d+Au} & $\mathrm{\pi^{0}+h^{\pm}}$& $0.104\pm 0.001$ & $0.112\pm0.006$ & $0.796\pm0.036$\\
\cline{3-6}
& & $\mathrm{\pi^{0}+\pi^{0}}$ & $0.027\pm 0.001$ & $0.025\pm0.003$ & $0.697\pm0.075$\\
\hline
\multirow{4}{*}{lower} & \multirow{2}{*}{p+p}& $\mathrm{\pi^{0}+h^{\pm}}$ & $0.063\pm 0.001$ & $0.236\pm0.008$ & $0.846\pm0.025$\\
\cline{3-6}
& & $\mathrm{\pi^{0}+\pi^{0}}$ & $0.015\pm 0.001$ & $0.059\pm0.003$ & $0.829\pm0.045$\\
\cline{2-6}
& \multirow{2}{*}{d+Au} & $\mathrm{\pi^{0}+h^{\pm}}$ & $0.348\pm 0.002$ & $0.269\pm0.009$ & $0.915\pm0.025$\\
\cline{3-6}
&  & $\mathrm{\pi^{0}+\pi^{0}}$ & $0.078\pm 0.002$ & $0.067\pm0.009$ & $0.880\pm0.111$\\
 \hline
\end{tabular}\caption{Summary of the values of uncorrelated background ($\mathrm{b}$), signal yield ($\mathrm{A}$) and width ($\mathrm{\sigma}$) from the fit to Figure \ref{6.midSummary}.}\label{6.midSumTable}\end{center}\end{table}

\begin{table}\begin{center}
\begin{tabular}{|c|c|c|c|c|c|}
\hline correlations & peak & $ p_{T} $ cuts & $\sigma_{pp}$ & $\sigma_{dAu}$ & $\Delta\sigma$ \\ \hline\hline 
\multirow{4}{*}{FMS-FMS} & \multirow{2}{*}{near-side} & higher & $0.34\pm0.01$ & $0.34\pm0.01$ & $0.00\pm0.01$ \\ 
& & lower & $0.41\pm0.01$ & $0.47\pm0.01$ & $0.06\pm0.01$\\ \cline{2-6}
& \multirow{2}{*}{away-side} & higher & $0.57\pm0.02$ & $0.68\pm0.06$ & $0.11\pm0.06$ \\
& & lower & $0.68\pm0.01$ & $1.20\pm0.05$ & $0.52\pm0.05$\\ \hline
\end{tabular}\caption{Summary of the values of the standard deviation of a Gaussian fit to the $\Delta\varphi$ distributions in Figure \ref{4.fmsdue}.}\label{6.fmssum}\end{center}
\end{table}

The situation changes when we consider azimuthal correlations between two forward neutral pions (FMS-FMS). The results of the width from the double Gaussian fit on the distribution of Figure \ref{4.fms}, for p+p and d+Au interactions and for two $p_{T}$ selections (compatible with the ones used for forward-midrapidity correlations) are summarized in Table \ref{6.fmssum}. 
This shows that the near-side peak in d+Au stays unchanged when compared to p+p interactions. On the contrary, the away-side contribution presents a significant broadening, even stronger when the lower $p_{T}$ selection is applied. The presence of the back-to-back peak in d+Au collisions tells us that, at these $p_{T}$ and $\eta$ regimes and by considering multiplicity averaged events,  we are not yet into the saturation region. We are instead probing a transitional region were its first effects start to appear in the form of broadening of the away-side peak.

\begin{figure}
\begin{tabular}{c c c}
\includegraphics[height=0.315\textwidth, clip=]{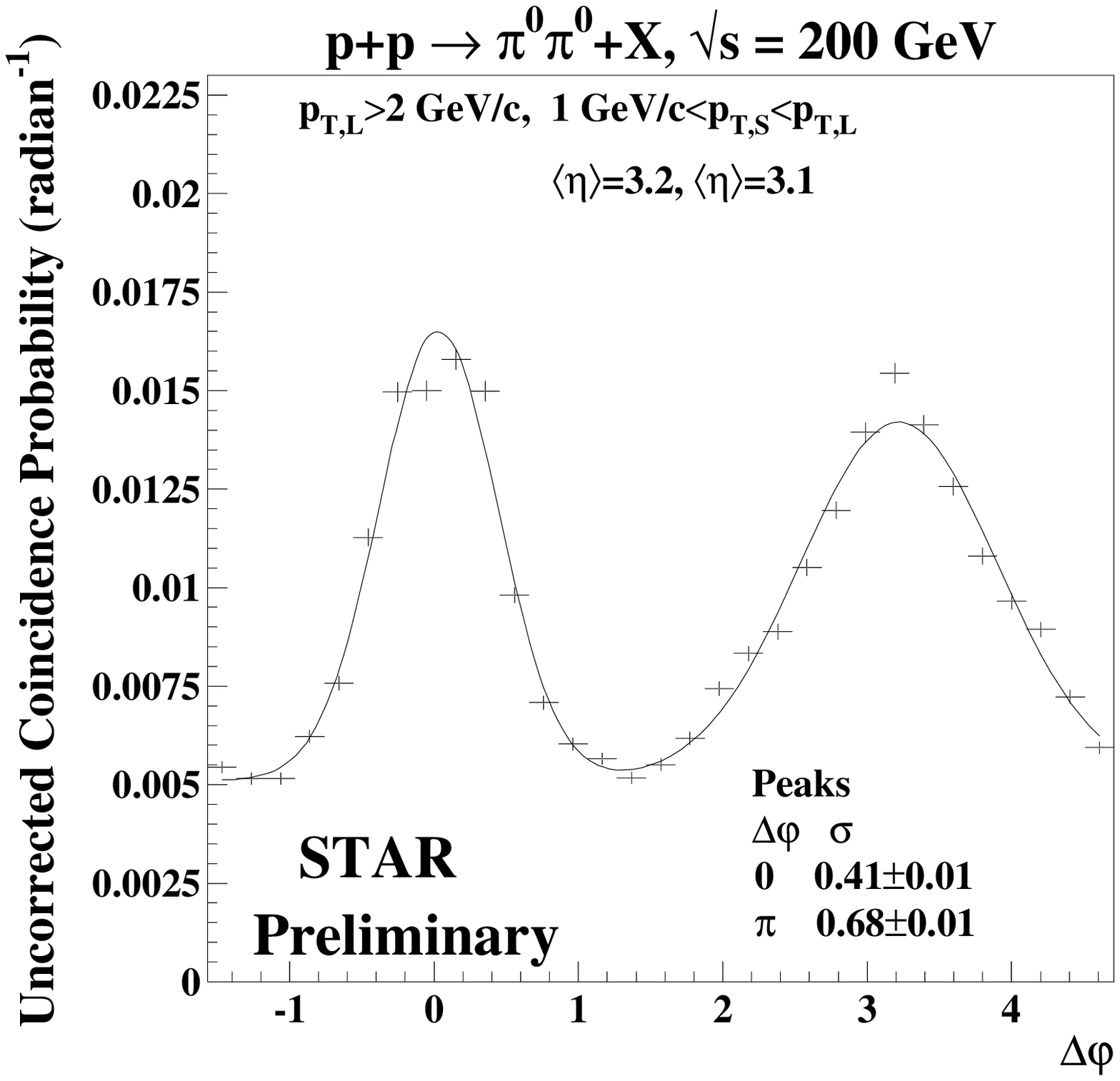} &
\includegraphics[height=0.315\textwidth, clip=]{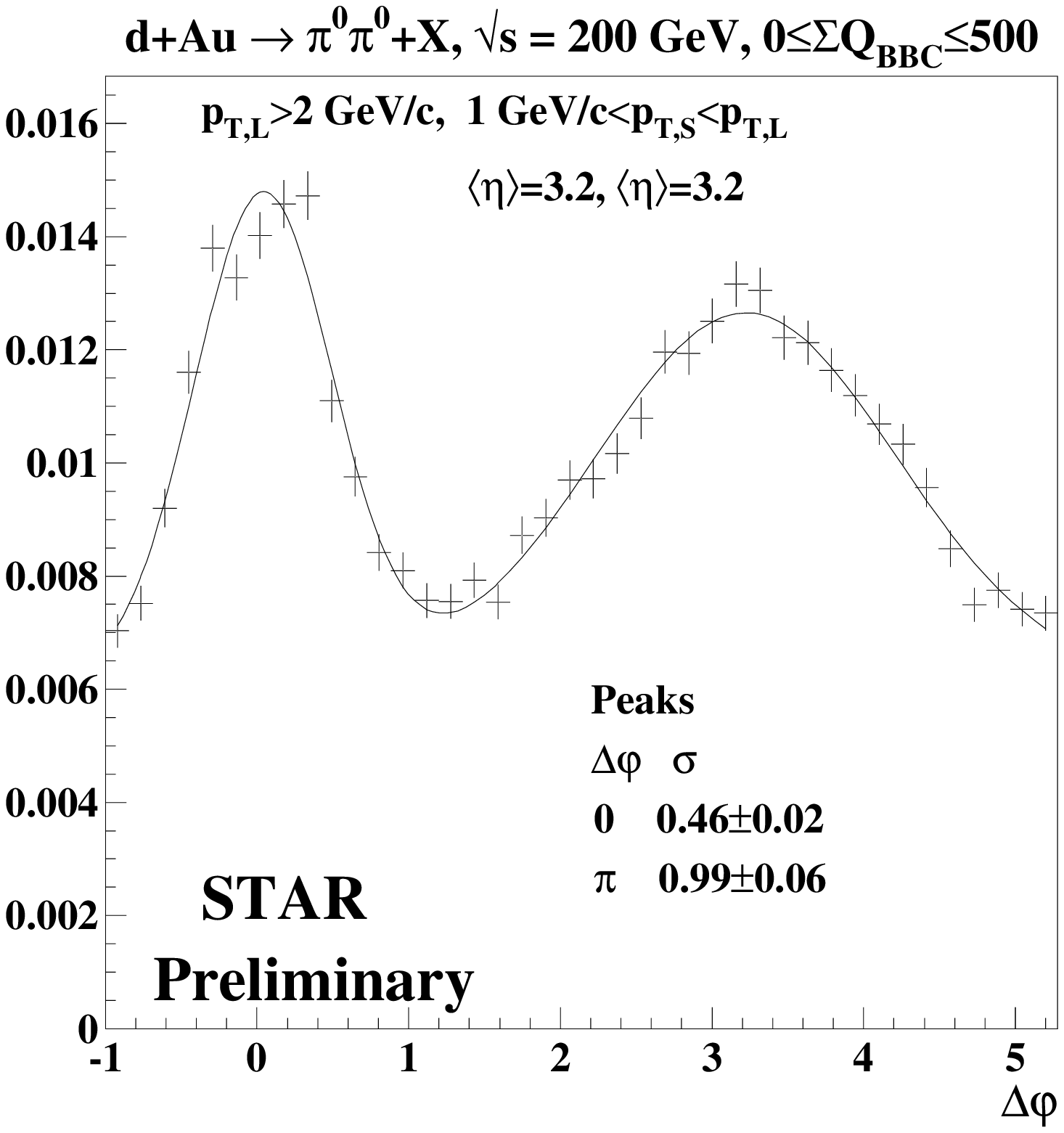} &
\includegraphics[height=0.315\textwidth, clip=]{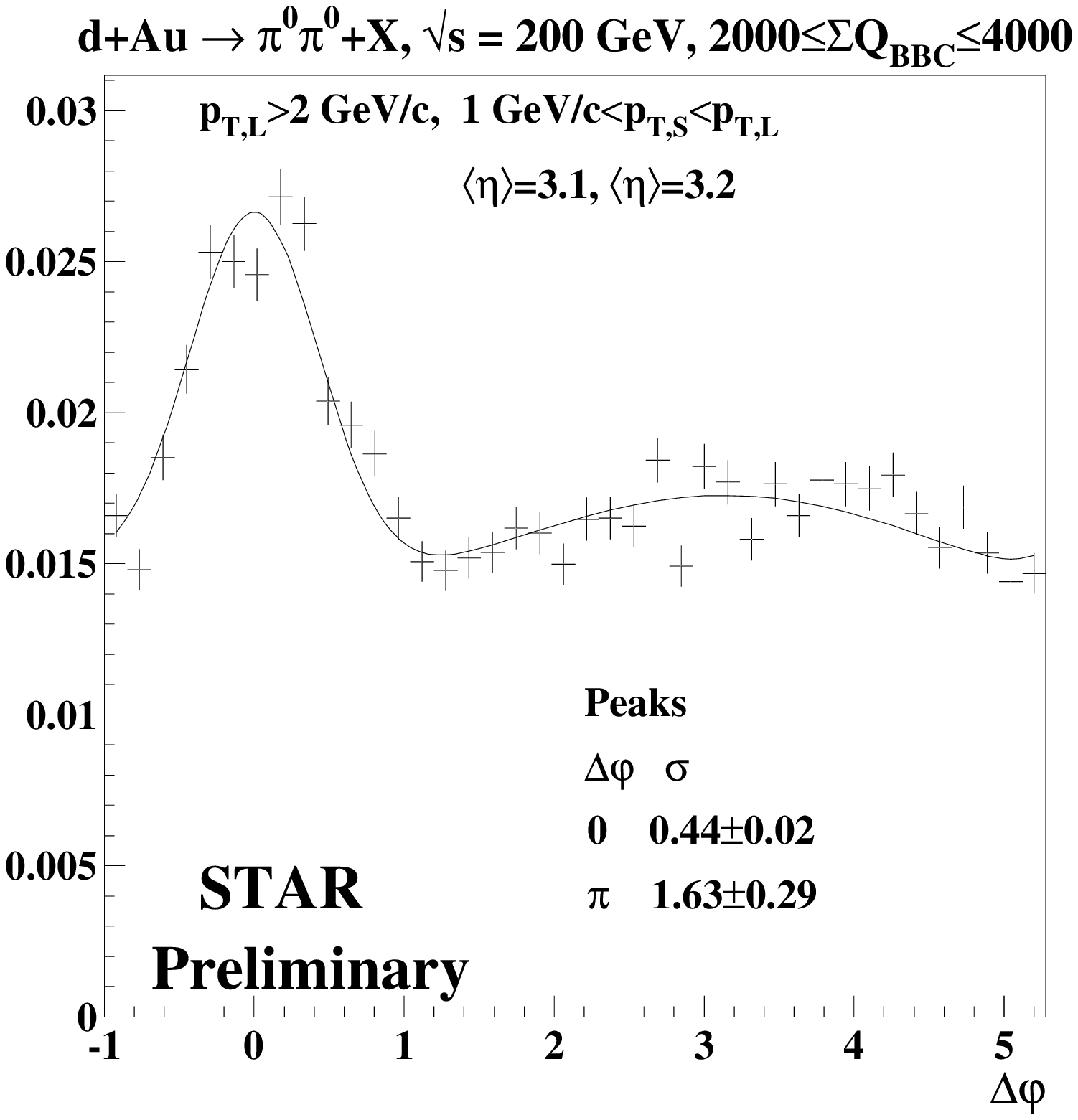}
\end{tabular}
\caption[]{Uncorrected coincidence probability versus azimuthal angle difference between two forward neutral pions in p+p collisions (left) compared to peripheral (center) and central d+Au collisions (right). Data are shown with statistical errors and are fitted with a constant plus two Gaussian functions.}
\label{4.fmscentral}
\end{figure}

As a step forward, the effect from possible saturation can be enhanced by selecting more central d+Au events. Saturation effects should in fact be enhanced when the denser part of the nucleus is probed. This is done, as for the BEMC-FMS correlations, by cutting on the East BBC charge multiplicity. In this case, the two multiplicity classes are defined by $0<\sum{Q_{BBC}}<500$ (peripheral) and  $2000<\sum{Q_{BBC}<4000}$ (central collisions), corresponding to $\langle b \rangle=(6.8\pm1.7)$ fm and $\langle b \rangle=(2.7\pm 1.3)$ fm respectively, as determined from a HIJING 1.383 simulation. A comparison between azimuthal correlations in p+p, peripheral and central d+Au interactions, using the lower $p_{T}$ selection (where the strongest effect is expected), is shown in Figure \ref{4.fmscentral}. As before, the near-side peak is mostly unchanged from p+p to peripheral and central d+Au collisions: the width from the Gaussian fit is $\sigma_{pp}=0.41\pm0.01$ for p+p collisions, $\sigma_{dAu}^{(peri)}=0.46\pm0.02$ in peripheral d+Au collisions and $\sigma_{dAu}^{(cent)}=0.44\pm0.02$ in central d+Au collisions. 
On the contrary, the away-side peak present striking differences: the signal width significantly broadens from $\sigma_{dAu}^{(peri)}=0.99\pm0.06$ in peripheral collisions to $\sigma_{dAu}^{(cent)}=1.63\pm0.29$ in central collisions. As expected, azimuthal correlations in peripheral d+Au collisions present a marked away-side peak which is more similar to the correlation in p+p interactions ($\sigma_{pp}=0.68\pm0.01$). This is due to the fact that the peripheral regions of the Gold nucleus are less dense and a smaller nuclear effect is expected. The more central parts are instead very dense and this is were the largest effect from saturation is expected to be seen. 
This is indeed what has been observed: there is a strong broadening of the away-side peak in d+Au collisions with respect to p+p (and to peripheral d+Au) interactions, while the near-side peak stays mostly unchanged.

%\begin{table}\begin{center}
%\begin{tabular}{|c|c|c|c|}
%\hline correlations & centrality & $I_{AA}$ (high $p_{T}$) & $I_{AA}$ (high $p_{T}$) \\ \hline\hline 
%FMS-TPC & averaged & 1.06$\pm$0.08 & 1.14$\pm$0.05 \\\hline
%FMS-BEMC & averaged & 1.09$\pm$0.16 & 1.14$\pm$0.16 \\\line
%FMS-FMS & averaged & & \\\line
%FMS-FMS & peripheral & & \\\line
%FMS-FMS & central & & \\\line
%\end{tabular}\caption{}\label{6.sommario}\end{center}
%\end{table}

\section{Charge particle to neutral pion ratio}\label{sec6.ratio}
The comparison of coincidence probabilities between a forward $\pi^{0}$ and either a mid-rapidity neutral pion or a charged track in the same kinematical range is expected to reflect the relative abundance of such particles. Although this is not the topic of this thesis work and it is not expected to impact directly the correlation results so far discussed, it can provide an useful insight in understanding the environment in which the measurements are performed. The ratio between charged particles and neutral pions at mid-rapidity can be estimated from the ratio of their coincidence probability with a forward pion. This is equal, in p+p interactions, to $X=4.42\pm0.01^{+0.01}_{-0.01}$ for higher $p_{T}$ cuts and $X=4.1\pm0.01^{+0.01}_{-0.01}$ for lower $p_{T}$ cuts. Here the statistical uncertainties come mostly from $\pi^{0}-h^{\pm}$ while the systematic errors are purely from $\pi^{0}-\pi^{0}$ correlations. The fact that the charged to neutral ratio decreases at lower $p_{T}\lesssim1.0\mathrm{\,GeV/c}$ values is expected (among others) from early STAR measurements. However the absolute value of  the ratio does is consistently higher than the number $R\sim3.2$ quoted in literature (see for example \cite{Zhang:2003jr}).

\begin{figure}\begin{center}
\includegraphics[width=1\textwidth,  clip=]{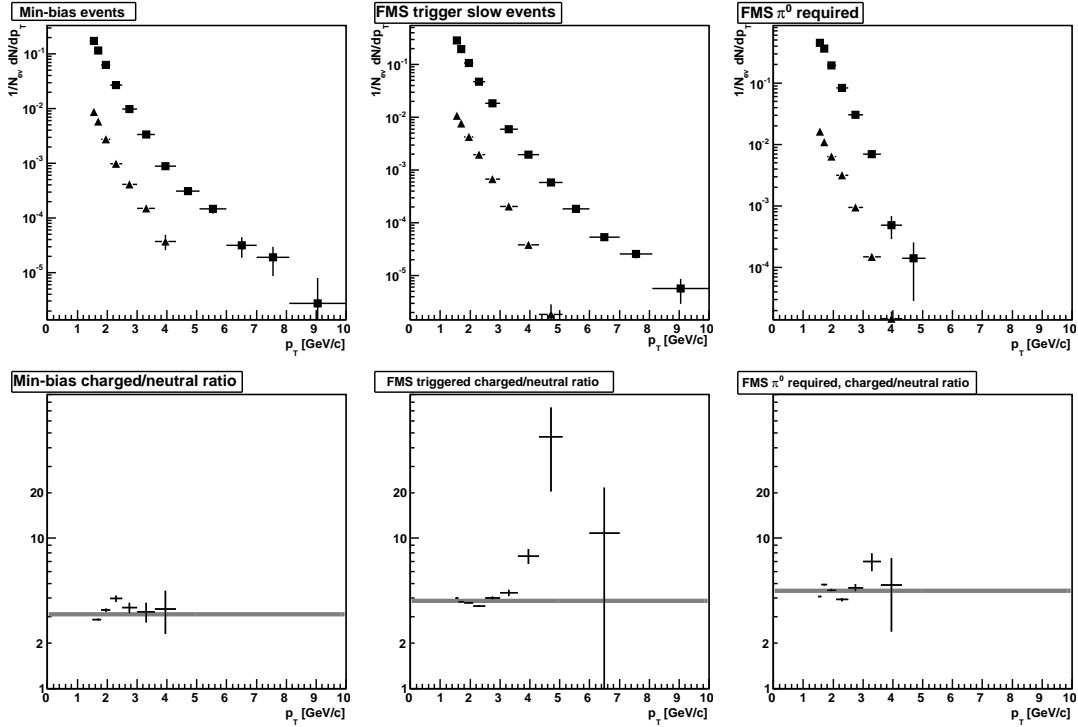} 
\caption{Top row: uncorrected transverse momentum spectra for charged particles (squares) and neutral pion (triangles) for minimum bias (left), forward triggered (center) and forward triggered events requiring a full $\pi^{0}$ reconstruction in the FMS (right). Bottom row: (background and efficiency) corrected ratio between the charged particles and neutral pions spectra. The grey line indicates the average over the whole $p_{T}$ range.}\label{6.MB}\end{center}
\end{figure}

In order to study the origin of such discrepancy, the reconstruction algorithm used for the correlation analysis has been applied to a minimum bias set of data, similar to the one used for this measurement. The top row of plots in Figure \ref{6.MB} shows the $p_{T}$ spectra of charged particles (squares) and neutral pion candidates (triangles) as used for the correlation analysis (p+p interaction, $p_{T}>1.5\mathrm{\,GeV/c}$) for minimum bias data (left), forward triggered events(central) and forward triggered events were the reconstruction of a $\pi^{0}$ with $p_{T}>2.5\mathrm{\,GeV/c}$ is required (right). The right-most panel reflect exactly the spectra of the particles used in the correlation analysis. These last spectra appear to be softer than the previous ones (in particular, the tail at high-$p_{T}$ disappears). This is due to the requirement $p_{T}^{fwd}>p_{T}^{mid}$ imposed to the analysis. The charged particle ``raw'' spectra have been corrected for luminosity (pile-up) effects and efficiency. The $\pi^{0}$ candidates ``raw'' spectra have been corrected for the hadronic background (singularly estimated from each invariant mass distribution) and for the efficiency, obtained in all cases from minimum bias simulation. The charge to neutral ratios of the spectra so corrected are shown on the bottom row of Figure \ref{6.MB}, together with the value of the ratio averaged over the $p_{T}$ range (in grey). 

The comparison shows indeed an increase of the charge particle to neutral pion ratio from minimum bias data (were the value is $R=3.11$, consistent with the literature), forward triggered data without ($R=3.83$) and with the requirement of a forward $\pi^{0}$ ($R=4.48$, consistent with the correlation measurements). The source of such systematic difference is not clear. Possibilities include uncertainties in estimating background or efficiency, or unpredicted physical aspects. Background levels in $\pi^{0}$ reconstruction are evaluated independently from the invariant mass spectra of the three data sets. We have seen that the background correction carries the largest systematic uncertainty. However, once the method for estimating the background in the $M_{\gamma\gamma}$ spectrum is decided (for example, by choosing the logarithmic fitting function), the measurement of signal over background ratio, performed independently in the three cases, does not differ significantly. The efficiency correction is calculated from a minimum-bias simulation and indiscriminately applied to the three data sets. A set of simulated triggered events may provide different efficiency values to be applied to triggered data points. From a preliminary test in minimum bias p+p simulation, there is no sign of dependence of the efficiency from multiplicity (the only quantity that changes significantly from minimum-bias to triggered events), nor do the ``quality cut'' distributions, on which the efficiency is calculated. However, the BEMC efficiency in d+Au is slightly lower than in p+p interactions. A multiplicity dependent efficiency could indeed explain such observation and eventually bring the charged particle to neutral ratio between minimum bias and forward triggered events to a lower value. Another observation is that the (minimum bias) efficiency correction seems to be more reliable when applied to minimum bias data than in forward triggered events. In particular, the charged particle to neutral ratio, once corrected for the minimum bias efficiency, appears to be independent of $p_{T}$ for minimum bias events, while this is not happening at the same level for forward triggered events (at high $p_{T}$).  

\section{Theory comparison}

Comparisons between measured azimuthal correlations and theoretical calculations have been explored \cite{Braidot:2010zh}. Measurements of correlations between two neutral pions in the forward region (FMS-FMS correlations) have been compared to a di-pion coincidence probability calculation \cite{Albacete:2010pg} within the CGC framework. In \cite{Albacete:2010pg}, the forward di-hadron production yield is obtained by calculating the scattering between a fast valence quark from the deuteron with the saturated gluon field of the nuclear target. In this picture, the scattering process is initiated by the valence quark, which splits into a back-to-back quark-gluon pair. By interacting with the low-$x$ color field, these partons acquire transverse momentum of the order of the saturation scale of the nucleus $Q_{S}^{2}=0.4\mathrm{\,GeV^{2}}$. The angular correlation between the two outgoing partons survives as long as their $p_{T}$ is much larger than the saturation scale. When it becomes comparable to $Q_{S}$, the correlation between the two partons weakens and eventually disappears.
The only constraint to this calculation is represented by the starting point of the BK evolution equation in $x$ and the correspondent value of the saturation scale. These two inputs are taken from the analysis of the single-inclusive hadron production \cite{Albacete2010bs}, calculated using the same approach \cite{Marquet200741}. This constraint is also used to normalize the coincidence probability, while the uncorrelated background has been subtracted from the data, since the underlying event component was not calculated.

\begin{figure}
\begin{tabular}{c}
\includegraphics[width=0.9\textwidth]{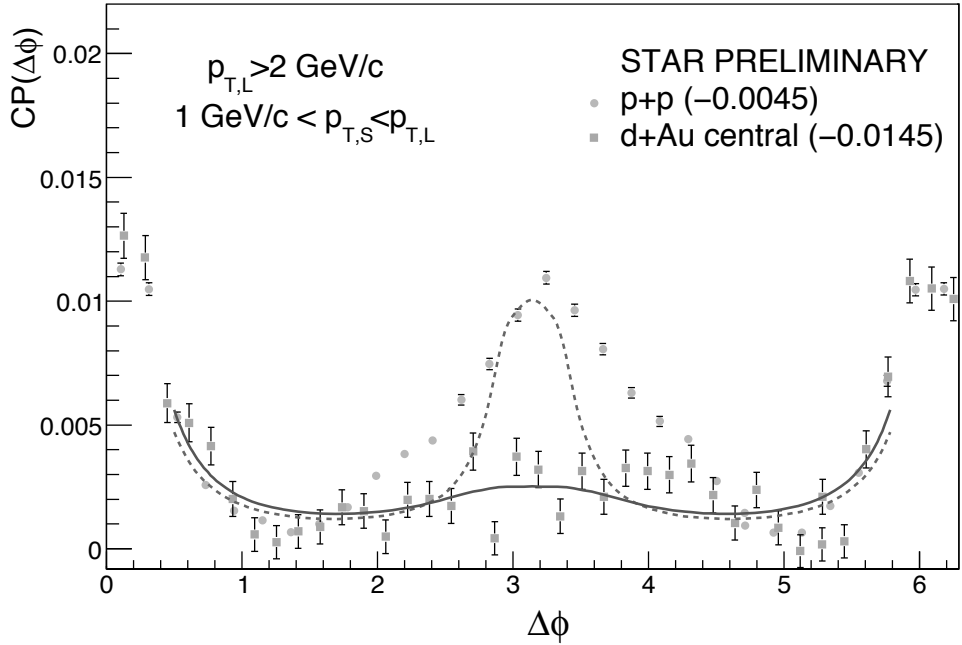} 
\end{tabular}
\caption{Background subtracted azimuthal correlations between two forward pions in p+p and central d+Au collisions, superimposed with a CGC calculation (dashed line for p+p and solid line for d+Au interactions). Figure from \cite{Albacete:2010pg}.}\label{6.cyrille}
\end{figure}  
 
Figure \ref{6.cyrille} shows the CGC calculation for central d+Au collisions together with the corresponding measurement. The calculation qualitatively describes the disappearance of the away-side peak, as well as the tails of the near-side peak\footnote{The CGC calculation does not extend to $\Delta\varphi\sim0$ because of the assumption of independent parton fragmentation.}. The same approach has been applied to p+p interactions (where the saturation scale for the proton has been estimated to be $Q_{S}^{2}=0.2\mathrm{\,GeV^{2}}$), even if its applicability in this case is questionable. Here, although some discrepancies remain in the description of the width of the correlated yield, a well defined back-to-back peak is expected.  

\begin{figure}
%\begin{tabular}{c c}
\includegraphics[height=0.261\textwidth, clip=]{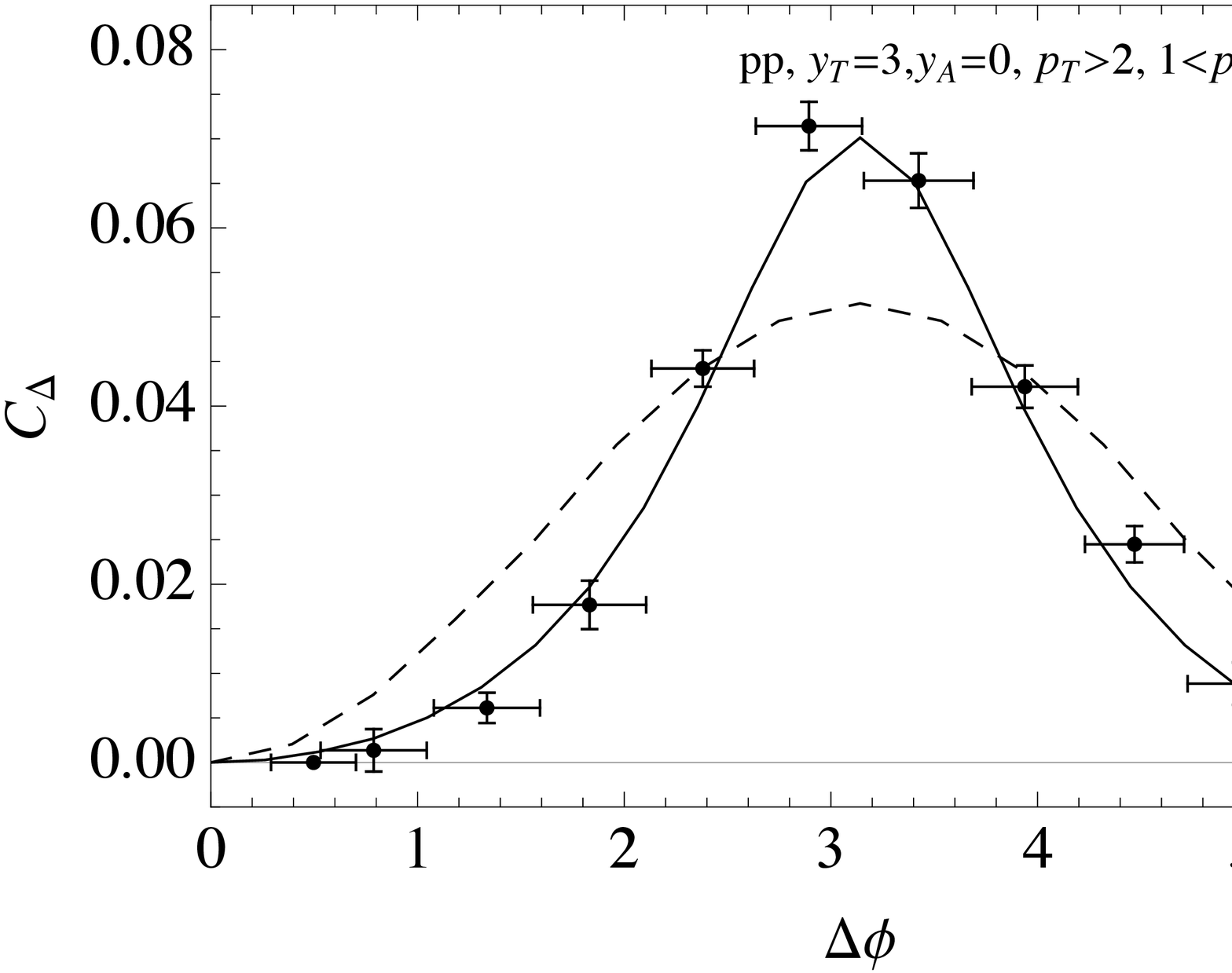} 
\includegraphics[height=0.262\textwidth, clip=]{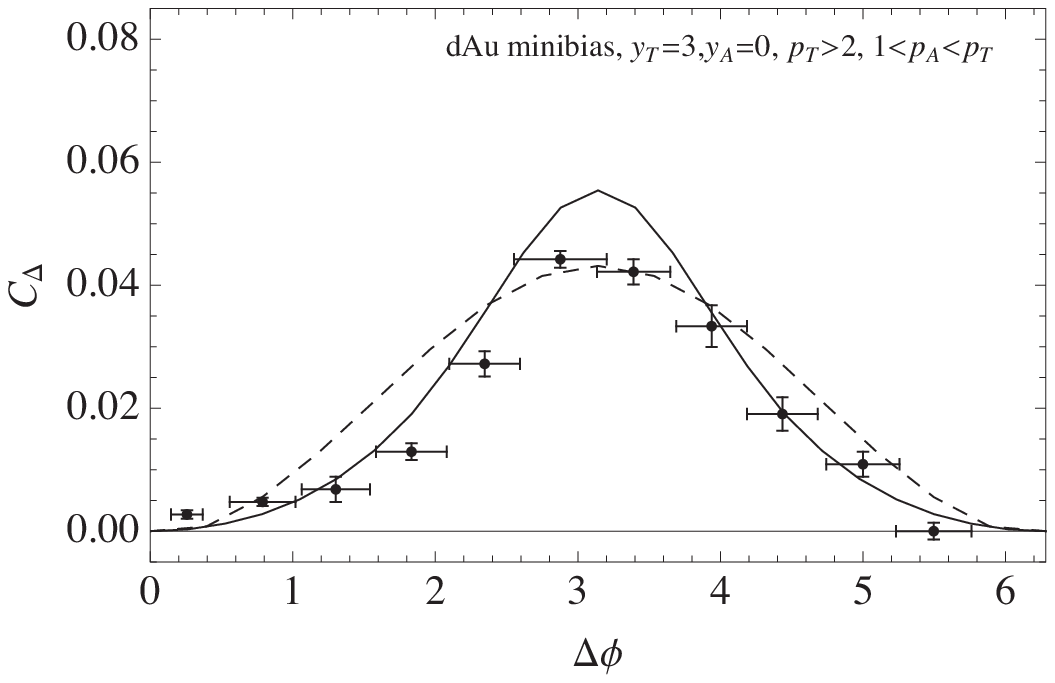} \\
\includegraphics[height=0.3\textwidth]{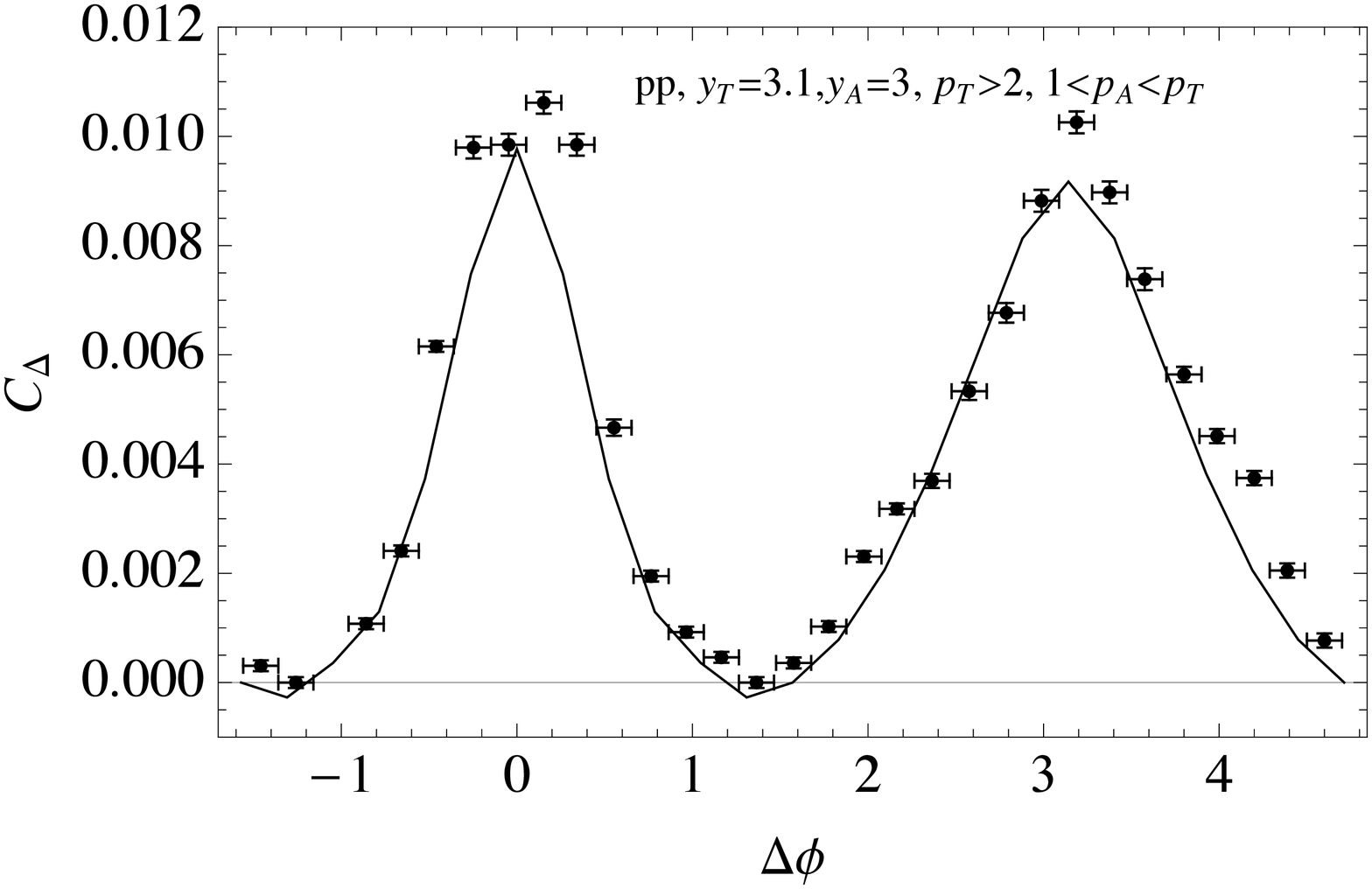} 
\includegraphics[height=0.303\textwidth]{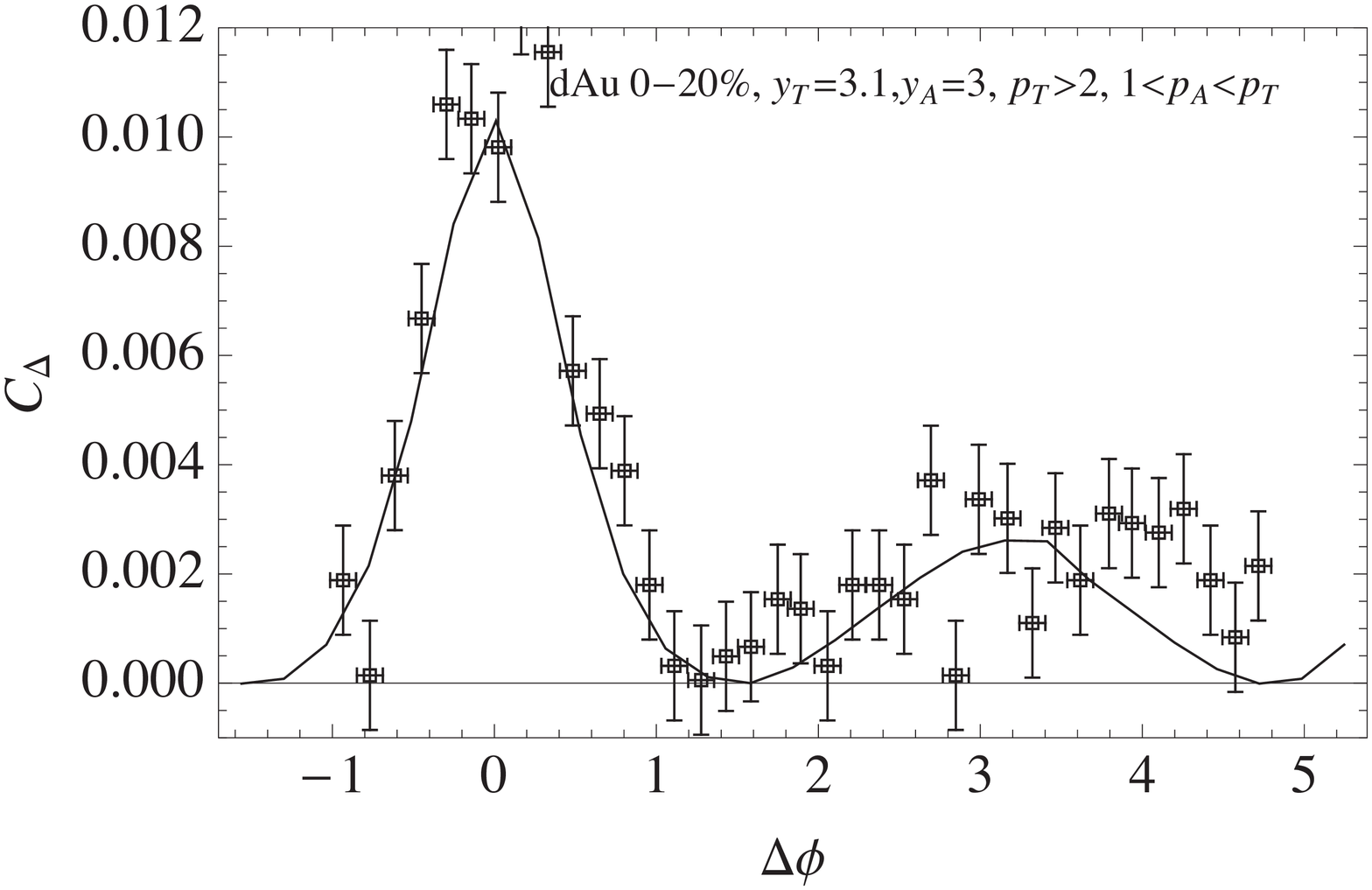} \\
%\end{tabular}
\caption{Azimuthal correlations between a forward $\pi^{0}$ and a second $\pi^{0}$ either at mid-rapidity (top row) or in the forward region (bottom row). Data are superimposed with a CGC calculation using ``$k_{T}$-factorization''. Figures from \cite{Tuchin:2009nf}. The dashed curve includes evolution (gluon emission) for the outgoing partons.
}\label{6.kiril}
\end{figure}  

Another approach \cite{Kharzeev:2004if}, based on ``$k_{T}$-factorization''  instead of the dipole model, has been used to compute azimuthal correlations within the CGC framework and compare them to these data. In \cite{Tuchin:2009nf}, the azimuthal correlations have been calculated by factorizing out the $2\rightarrow4$ elementary process (here only gluons initiate the process\footnote{This means that the interaction between a valence quark in the deuteron and the low-$x$ gluon in the nucleus, believed to be the leading process, are not considered in this model.}) and the unintegrated gluon distribution (defined as function of the saturation scale $Q_{S}$). Although generally $k_{T}$-factorization is not appropriate to describe processes in the gluon saturation region, this approach has shown reasonable agreement  in the description of many observed quantities, providing exact results for the single and (at least at RHIC energies) double inclusive production. The calculations have been normalized to fit the correlated peak in p+p data, since this approach is known to miss the overall normalization (this is essentially the only free parameter of the model). Figure \ref{6.kiril} shows a comparison between these calculations and measured azimuthal correlations between a forward $\pi^{0}$ and a second $\pi^{0}$, either at mid-rapidity (top row) or in the forward region (bottom row). The comparison shows that the calculation, once normalized, describes with qualitatively good agreement the signal widths, both in p+p (which description is possible using $k_{T}$-factorization) and d+Au interactions. While the correlation peak is reproduced both in p+p and d+Au for forward-mid-rapidity correlations, with no sign of significant broadening, for forward-forward correlations the peak almost disappears for central d+Au interaction. At the same time, the near-side correlation is reproduced with a width consistent with data, both in p+p and d+Au.

\chapter{Conclusions and outlook}

Two particle azimuthal correlations are a powerful tool for characterizing the transitional region between dilute and saturated partonic systems. Early RHIC measurements of inclusive particle production in p+p collisions at $\sqrt{s_{NN}}=200\mathrm{\,GeV}$ \cite{PhysRevLett.93.242303, Adams:2006uz} show general agreement with perturbative QCD. A strong suppression of forward particles is instead observed in d+Au interactions, suggesting the onset of parton saturation. Two particle correlations allow us to probe more selectively the broad range of longitudinal momentum fraction $x$ (averaged over in inclusive production). 

Gluon saturation is a feature of dense relativistic hadrons, firstly nuclei. At very high energy, non-linear contributions need to be included in the hadronic wave-function in order to soften the otherwise divergent rise of the gluon density and to restore unitarity. In a collision involving a saturated target, the leading back-to-back ($2\rightarrow 2$) contribution to the partonic scattering is replaced by the monojet ($2\rightarrow 1$) contribution. The transverse momentum of the jet produced by a large momentum parton from the probe is balanced here by many gluons in the target, which recoil collectively. In this scenario, the back-to-back peak is expected to significantly broaden when saturation sets in and eventually to disappear. 

In order to determine if the saturation region is accessible at RHIC energies, the STAR collaboration is pursuing a systematic plan of measurements of azimuthal correlations. Correlations of particles produced in interaction between dilute systems (p+p) are compared with those in which one of the two projectile is a relativistic nucleus (d+Au). In this way, initial state features of the nuclear structure are studied without significant contributions from final state effects (QGP), typical of heavy ion interactions (Au+Au). The low-$x$ component of the nuclear gluon field is accessed by selecting events which present a neutral pion reconstructed in the forward region. The associated particle is selected with a transverse momentum down to $1\mathrm{\,GeV/c}$ over a broad rapidity range, allowing us to gradually lower the momentum fraction $x$ of the struck gluon.

In this work, forward neutral pions, reconstructed using the FMS, are correlated with mid-rapidity neutral pions (using the BEMC) or charged particles (using the TPC), as well as with a second forward $\pi^{0}$ in the FMS. The analysis performed shows no significant broadening in the back-to-back peak when the associated particle is reconstructed at mid-rapidity. On the contrary, 
when both particles are reconstructed in the forward region (that is when the lowest $x$ value is probed) the correlated peak in d+Au is significantly broader than in p+p, while the near-side peak stays unchanged. The broadening effect, as expected from saturation models, is larger when particles are reconstructed with a lower transverse momentum. This corresponds to measuring interactions with a lower transferred momentum $Q^{2}$. In the same way, the broadening is maximum when the dense (central) part of the nucleus is probed: selecting central d+Au collisions leads to a strong reduction (almost disappearance) of the correlated peak. In contrast, peripheral d+Au interactions reveal a correlated peak similar to the peak in p+p. Theoretical expectations for the coincidence probability in interactions involving a saturated target, described with two different approaches (dipole model and $k_{T}$-factorization) within Color Glass Condensate (CGC) framework, show qualitative agreement with our measurements.

The results in this thesis show that gluon saturation occurs at a scale that is relevant for RHIC energies. However, initial state effects are visible only in special kinematic conditions such as forward di-pion production at low $p_{T}$, particularly in high multiplicity (central) events. For other (mid-rapidity) correlation measurements, there are no hints of saturation at the values of $p_{T}$ here considered. This places RHIC kinematical regime in the transitional region between dilute and saturated systems. The analysis presented in this thesis can be improved and extended by  probing more exhaustively the boundaries of the saturation region. This includes a measurement of correlations between forward pions (FMS) and mid-rapidity charged particles (TPC) with a transverse momentum lower than $1\mathrm{\,GeV/c}$. In addition, FMS-BEMC correlations will highly benefit from a larger data sample which would allow a significant comparison with FMS-TPC correlations in central d+Au collisions, after a consistent background subtraction in the BEMC pion reconstruction. Another important piece of the analysis, currently ongoing, is the measurement of azimuthal correlations between a forward pion and an associated particle in the intermediate pseudo-rapidity region $1.0<\eta<2.0$, using the STAR Endcap Electromagnetic Calorimeter (EEMC).

\begin{figure}
\includegraphics[width=0.9\textwidth]{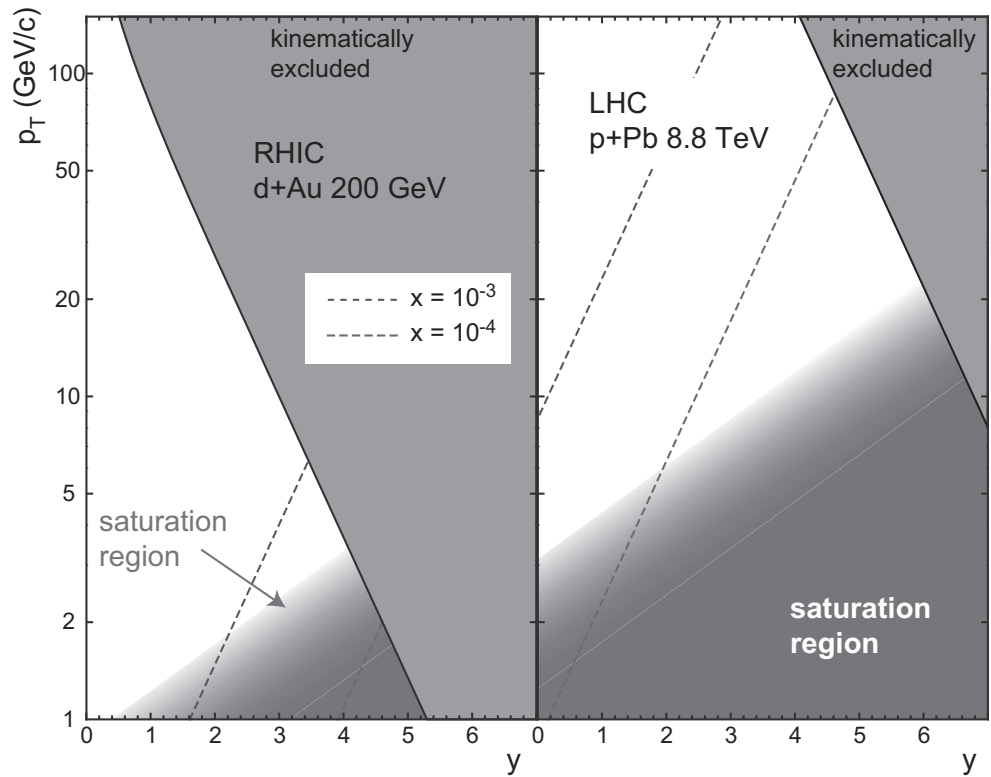} 
\caption{Pictorial representation of the kinematic acceptance at RHIC and LHC. The region of the rapidity-transverse momentum space where saturation effects are expected is indicated.}\label{6.thomas}
\end{figure}  
 
The results here presented are very encouraging with respect to the expectations for LHC. The kinematical range available at the energies of the new hadron collider greatly exceeds that of RHIC, as illustrated in Figure \ref{6.thomas}. The large pseudorapidity acceptance will allow measurement of momentum fraction down to $x\approx10^{-6}$. The saturation scale $Q_{S}(x,A)$ is therefore predicted to be at LHC approximatively 2-3 times higher than at RHIC. As a consequence, saturation effects are expected already at low rapidity, unlike at RHIC. This implies that a good understanding of the nuclear structure in the initial state of the interaction is crucial for a correct description of the final state effects, especially in a regime where saturation can be significant in many heavy ion measurements (and, possibly, in p+p interactions as well). For this reason, the contribution coming from RHIC is important since it provides a first insight of saturation in a regime where discrimination between initial and final states is still clean. These considerations illustrate however the great potential of the LHC experiments for what concerns low-$x$ and saturation physics. In particular, the ALICE Collaboration is planning a substantial upgrade in the forward region, with the installation of a new forward calorimeter which will exploit the larger kinematical range in measurements similar to those described in this thesis.

\appendix
\appendix

\chapter{Run-3 results emulation}

%\begin{figure}
%\begin{tabular}{c c}
%\includegraphics[width=0.45\textwidth, bb =16 6 543 160]{figures/appendix/paperBW2.eps} 
%\includegraphics[width=0.45\textwidth, bb =28 0 500 147]{figures/appendix/emulation.eps} 
%\end{tabular}
%\caption{On the left: coincidence probability versus azimuthal angle difference between a forward $\pi^{0}$ and a leading mid-rapidity charged track with $p_{T}>0.5\mathrm{GeV/c}$. Figure taken from \cite{Adams:2006uz}. On the right: uncorrected coincidence probability reproducing 2003 conditions using data from 2008.}\label{4.emu}
%\end{figure} 

\begin{figure}
\begin{center}
\includegraphics[width=0.75\textwidth]{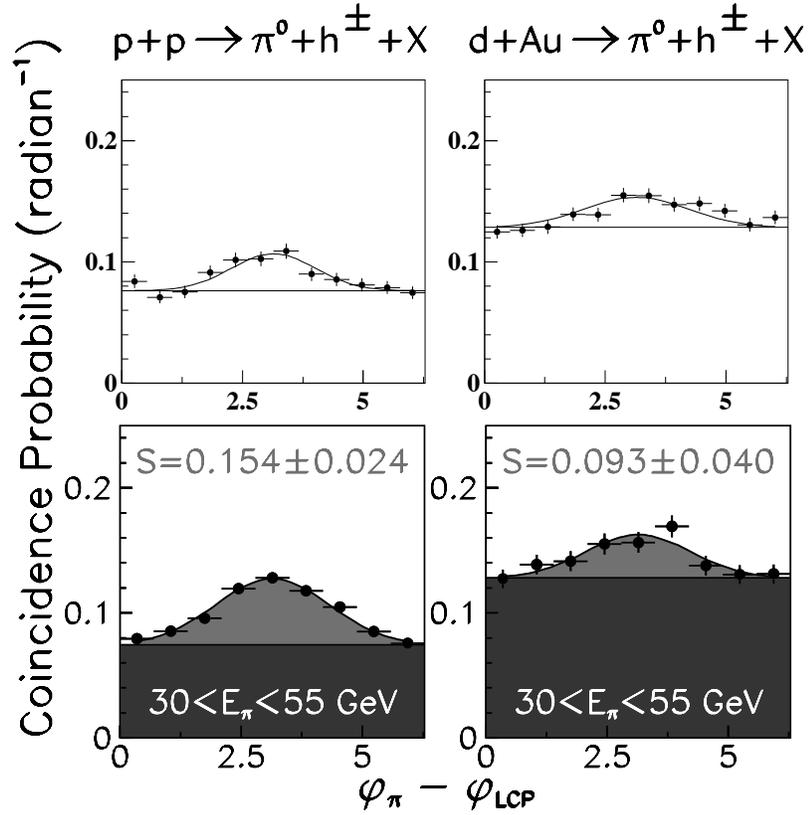} 
\end{center}
\caption{Coincidence probability versus azimuthal angle difference between a forward $\pi^{0}$ and a leading mid-rapidity charged track with $p_{T}>0.5\mathrm{GeV/c}$. Bottom part taken from \cite{Adams:2006uz}. Upper part: uncorrected coincidence probability reproducing 2003 conditions using data from 2008.}\label{4.emu}
\end{figure} 

An interesting testbed for the hardware and software FMS configuration used in Run-8 is to being able to reproduce published STAR results from Run-3, using the FPD detector \cite{Adams:2006uz}. Figure \ref{4.emu} compares published azimuthal correlations from Run-3 with Run-8 data. The new analysis tries to reproduce as  faithfully as possible the kinematical conditions used for Run-3. The forward $\pi^{0}$ is reconstructed using a limited portion of the FMS matrix in order to emulate the FPD acceptance. In addition, pions are required to satisfy the following cuts: $3.8<\eta^{(FMS)}<4.1$, $Z_{\gamma\gamma}<0.7$, $(0.07<M_{\gamma\gamma}<0.30)\mathrm{\,GeV/c^{2}}$. The charged tracks are reconstructed within the range $|\eta^{(TPC)}|<0.75$ and required to have at least 25 TPC hits and a minimum transverse momentum of $p_{T}^{(TPC)}>0.5\mathrm{\,GeV/c}$. Correlations between the forward $\pi^{0}$ and the leading charged track are calculated in energy bins of the trigger pion. Despite the fact that there are still inevitable differences in the two measurements (in particular, the trigger algorithms were different) and that the low energy bin does not present sufficient statistics in this trigger configuration, the emulation in the higher energy bin looks encouraging. All the features of the azimuthal correlations in the energy bin $(30<E_{\pi^{0}}<55)\mathrm{\,GeV/c^{2}}$ are in fact qualitatively reproduced.

\bibliographystyle{epjc}
\bibliography{Thesis.bbl}

\chapter*{Samenvatting}
\addcontentsline{toc}{chapter}{Samenvatting}
\setlength{\epigraphwidth}{.3\textwidth}
\epigraph{Tout y parlerait\\\`{A} l'\^{a}me en secret\\Sa douce langue natale.}{\textit{L'invitation au voyage}\\\textsc{Charles Baudelaire}}

Dit proefschrift bevat een verzameling van metingen van azimutale correlaties
in p+p en d+Au botsingen van $\sqrt{s_{NN}}=200\mathrm{\,GeV}$ met de STAR
detector van RHIC. De vergelijking van azimutale correlaties in botsingen met
een systeem van lage dichtheid (p+p) met het geval waar een van de
projectielen een hoge dichtheid heeft (d+Au) geeft inzicht in de structuur van
de relativistische nucleus en de effecten in het begin van de evolutie na de
botsing. Bij hele hoge energie ondergaat de relativistische nucleus Lorentz
contractie langs de longitudinale as en is het net een dunne ``pannenkoek''
van quarks en gluons. Als de dichtheid hoog genoeg is, beginnen de nucleaire
compenenten te interacteren. Een recente beschrijving van de structuur van een
nucleus wordt beschreven door een raamwerk van prescripties en vergelijkingen
die de non-lineaire componenten van de golf-functie van de nucleus bevat,
bekend als de Color Glass Condensate (CGC).

De nieuwe Forward Meson Spectrometer (FMS) die geinstalleerd is in het STAR
experiment, geeft de mogelijkheid om pion correlaties bij lage $p_{T}$ in de
voortgaande richting the doen. De verwachting is dat saturatie wordt bereikt
als de laagste momentum fractie $x$ in het nucleaire gluon veld wordt
bekeken. Dit wordt bereikt door beide deeltjes in de voorwaartse richting te
reconstrueren. In dit geval zijn de effecten aan het begin van de evolutie
duidelijk zichtbaar, vooral in centrale d+Au botsingen, in de vorm van een
verbreding of verdwijning van de tegenoverliggende piek behorend bij de pQCD
$2\rightarrow2$ scatter. Aan de andere kant, wanneer het geassocieerde deeltje
bij een lagere pseudo-rapiditeit wordt geselecteerd (en daardoor bij hogere
momentum fractie $x$ in het nucleaire gluon veld), wordt geen significante
verbreding geobserveerd. Dit laat zien dat het saturatie proces plaatsvindt
binnen het kinematisch regime van RHIC in het transitie gebied tussen het
minder dichte en gesatureerde systeem. Het levert een aanmoedigende eerste
stap richting lage-$x$ studies bij LHC waar een significante saturatie wordt
verwacht voor veel zware ionen metingen.

\newpage
\chapter*{Acknowledgments}
\addcontentsline{toc}{chapter}{Acknowledgments}

These past four (and a half) years have been for me pretty intense. \\
I enjoyed many happy moments that changed my life for the better and, inevitably, I had to face a few setbacks. The most significant of the latter made it not possible for my father Roberto to enjoy this moment. I do, however, know exactly what he would have felt today, what he would have said and how he would have glanced at me and this makes the loss a bit easier to bear. I will be always very thankful for the support he constantly granted me and the honest trust he showed for my judgments and my choices. 

There are obviously many other people that I want to thank for the part they played on my Ph.D.
First of all my promoters Thomas Peitzmann and Eric Laenen, for driving me through these four years with suggestions and encouragements, and my co-promoter Andr\'{e} Mischke. A humongous thanks goes then to Marco van Leeuwen, the inspiration for, and the solution to, all the physics (and coding) questions that came to my mind. Thanks also to Raimond Snelling that has always been able to find a crack in my confidence dyke holding on saturation physics. I would like to thank also the whole Utrecht group (in particular Ton van den Brink, for fixing four times my iPowerAdaptors) and all the fellow Ph.D. students in Utrecht and Amsterdam with whom I shared these four years. A special thanks goes to Martijn Russcher (for teaching me how to survive a Ph.D. in The Netherlands and at BNL), Marta Verweij (for helping me with \emph{de Nederlandse taal}) and Cristian G. Ivan \cite{Ivan2010ev} (for trying very hard to keep me from my duties; admittedly, meeting not much opposition).

Another big piece of help came from Brookhaven. I am very grateful for many things to Leslie Bland, leader of the FMS group and thoughtful host of thanksgiving dinners. A supel thanks goes to Akio Ogawa, for sharing his knowledge and teaching me all the secrets of the STAR code with tireless perseverance and proverbial patience. I will also heartily remember all the other FMS/STAR people I met at BNL and, in particular, Hank Crawford and my buddy Len Eun. I am greatly thankful to the many experts I had the opportunity to talk with that helped to clear my mind about saturation: Daniel Boer, Raju Venugopalan, Larry McLerran, Al Mueller.

Finally, I would like to thank my family for their support during these four years in the Low Countries.  Thanks in particular to my mother Anna Maria and my brother Mattia. And, of course, the warmest acknowledgment goes to Eleonora. This time in Amsterdam gave us the possibility to live together, to get married and strengthen our relationship, to buy a house and fill it with every new (ultimate) gadget. I was very lucky to have my wife beside me every time I needed support or a kick to finish the job, and I am very fond of all the moments we shared and all the goals we achieved together.

\newpage

  \thispagestyle{empty}%              % Empty header styles

%\cleardoublepage
%\printnomenclature

\end{document}